%% file: thesis.tex
\documentclass[12pt, a4paper,twoside]{report}
\usepackage[english,ngerman]{babel} 
\usepackage[utf8]{inputenc}
\usepackage{graphicx,amsmath,amsfonts,amssymb,rotating,bm}
\usepackage{ngerman} 
\usepackage{psfrag,fancyhdr,appendix,subfigure}
\usepackage[onehalfspacing]{setspace}
\usepackage[font=small,format=plain,labelfont=bf,up,
width=.8\textwidth,labelsep=space]{caption}
\usepackage{lettrine,longtable,fix-cm,tikz,natbib,afterpage}

\newcommand{\matha}{{\sc Mathematica\ }}
\newcommand{\dd}{\,\mathrm d}
\renewcommand{\vec}{\bm}
\newcommand{\mat}[1]{\mathbf{\mathsf{#1}}}
\newcommand{\unit}[1]{\ensuremath{\, \mathrm{#1}}}
\newcommand{\comment}[1]{}
\newcommand{\eq}[1]{eq. (\ref{#1})}
\newcommand{\ellmax}{\ell_\mathrm{max}}
\newcommand{\calN}{\mathcal N}
\newcommand{\FOM}{\mathop{\mathrm{FOM}}}

\usepackage[hyperfootnotes=false,linkcolor=blue,
urlcolor=blue,citecolor=blue]{hyperref} 

\begin{document}
\pagenumbering{roman}
\input{title}
\input{header}

\cleardoublepage
\tableofcontents
\cleardoublepage

\newcounter{mypage}
\setcounter{mypage}{\value{page}}
\pagenumbering{arabic}
\setcounter{page}{\value{mypage}}

\input{quote}
\input{chapter1}

\input{chapter2}

\input{chapter3}
\input{chapter4}
\input{chapter6}

\newpage
\renewcommand{\nochapterzero}[1]{}
\bibliographystyle{apsrmp}
\bibliography{manual,library}
\appendix

\input{coupled}

\cleardoublepage

\input{thanks}

\end{document}

%% file: title.tex
\pagestyle{empty}
\begin{center}
  \renewcommand{\baselinestretch}{2.00}
  \LARGE
  Fakultät für Physik und Astronomie\\
  \Large
  Ruprecht-Karls-Universität Heidelberg
  \par\vfill\normalfont
  Diplomarbeit\\
  im Studiengang Physik\\
  vorgelegt von\\
  Adrian Vollmer\\
  geboren in Ochsenfurt\\[2cm]
  \textit{2011}
\end{center}
\cleardoublepage

\comment{
\begin{center}
  \renewcommand{\baselinestretch}{2.00}
  \LARGE\bfseries
    Zu erwartende Einschränkungen der Ent\-wicklung des Hubble Parameters
    sowie der Wachs\-tums\-funk\-tion durch zukünftige Ver\-messung\-en von
    schwachen Gravitationslinsen\\
  \par
  \vfill
  \Large\normalfont
  Die Diplomarbeit wurde von\\ Adrian Vollmer\\
  ausgeführt am\\
  Institut für theoretische Physik\\
  unter der Betreuung von\\
  Herrn Prof. Luca Amendola
\end{center}\par
\vspace{5\baselineskip}
\cleardoublepage

\renewcommand{\baselinestretch}{1.00}\normalsize

\begin{center}
  \renewcommand{\baselinestretch}{2.00}
  \LARGE
  Department of Physics and Astronomy\\
  \Large University of Heidelberg
  \par\vfill\normalfont
  Diploma thesis\\
  in Physics\\
  submitted by\\
  Adrian Vollmer\\
  born in Ochsenfurt\\[2cm]
  \textit{2011}
\end{center}
\cleardoublepage
}

\begin{center}
  \renewcommand{\baselinestretch}{2.00}
  \LARGE\bfseries
    Forecasting Constraints on the Evolution of the Hubble Parameter and the
    Growth Function by Future Weak Lensing Surveys \\
  \par
  \vfill
  \Large\normalfont
  This diploma thesis has been carried out by\\ Adrian Vollmer\\
  at the\\
  Institute for Theoretical Physics\\
  under the supervision of\\
  Prof. Luca Amendola
\end{center}\par
\cleardoublepage
\renewcommand{\baselinestretch}{1.00}\normalsize

\selectlanguage{ngerman}
{\small
\paragraph{Zusammenfassung.}\hspace{.1cm} Die kosmologische Information, die
im Signal des schwa\-che-Gravitationslinsen-Effekts verborgen ist, lässt
sich mit Hilfe des Potenzspektrums der sogenannten Konvergenz analysieren.
Wir verwenden den Fish\-er-In\-for\-ma\-tion-For\-ma\-lis\-mus mit dem
Konvergenz-Potenzspektrum als Observable, um abzuschätzen, wie gut
zukünftige Vermessungen schwacher Gravitationslinsen die Expansionrate und
die Wachstumsfunktion als Funktionen der Rotverschiebung einschränken
können, ohne eine bestimmtes Modell zu deren Parametrisierung zu Grunde zu
legen. Hierfür unterteilen wir den Rotverschiebungsraum in Bins ein und
interpolieren diese beiden Funktionen linear zwischen den Mittelpunkten der
Rotverschiebungsbins als Stützpunkte, wobei wir für deren Berechnung ein
Bezugsmodell verwenden.  Gleichzeitig verwenden wir diese Bins für
Potenzspektrum-Tomographie, wo wir nicht nur das Potenzspektrum in jedem Bin
sondern auch deren Kreuzkorrelation analysieren, um die extrahierte
Information zu maximieren.

Wir stellen fest, dass eine kleine Anzahl von Bins bei der gegebenen
Präzision der photometrischen Rotverschiebungsmessungen ausreicht, um den
Großteil der Information zu erlangen. Außerdem stellt sich heraus, dass die
prognostizierten Einschränkungen vergleichbar sind mit derzeitigen
Einschränkungen durch Beobachtungen von Clustern im Röntgenbereich. Die
Daten von schwachen Gravitationslinsen alleine könnten manche modifizierte
Gravitationstheorien nur auf dem $2\sigma$ Niveau ausschließen, aber wenn
man den Prior von Vermessungen der kosmischen Hintergrundstrahlung
berücksichtigt, könnte sich dies zu einem $3\sigma$ Niveau verbessern.}

\vfill

\selectlanguage{english}
{\small
\paragraph{Abstract.} The cosmological information encapsulated within a
weak lensing signal can be accessed via the power spectrum of the so called
convergence. We use the Fisher information matrix formalism with the
convergence power spectrum as the observable to predict how future weak
lensing surveys will constrain the expansion rate and the growth function as
functions of redshift without using any specific model to parameterize 
these two quantities. To do this, we divide redshift space into bins and
linearly interpolate the functions with the centers of the redshift
bins as sampling points, using a fiducial set of parameters. At the same
time, we use these redshift bins for power spectrum tomography, where we
analyze not only the power spectrum in each bin but also their
cross-correlation in order to maximize the extracted information.

We find that a small number of bins with the given photometric redshift
measurement precision is sufficient to access most of the information
content and that the projected constraints are comparable to current
constraints from X-ray cluster growth data. This way, the weak lensing data
alone might be able to rule out some modified gravity theories only at the
$2\sigma$ level, but when including priors from surveys of the
cosmic microwave background radiation this would improve to a $3\sigma$
level.}

\newpage

%% file: header.tex
\setlength{\headheight}{15pt}
\pagestyle{fancy}
\fancyhf{}
\renewcommand{\chaptermark}[1]{\markboth{#1}{}}
\renewcommand{\sectionmark}[1]{\markright{#1}{}}

\newcommand{\nochapterzero}[1]{\ifnum\value{chapter}>0#1\fi}

\setcounter{tocdepth}{2}
\fancyhead[LE]{\textsf{\textbf\thepage\hfill\footnotesize{
\nochapterzero{Chapter \thechapter. }\nouppercase{\leftmark}}}}
\fancyhead[RO]{\textsf{\nochapterzero{\footnotesize{\thesection\ %
\nouppercase{\rightmark}}}\hfill\textbf\thepage}}
\fancypagestyle{plain}{%
\fancyhf{} 
\renewcommand{\headrulewidth}{0pt}
\renewcommand{\footrulewidth}{0pt}}

%% file: quote.tex
\cleardoublepage
\thispagestyle{plain}

\vspace*{\stretch{1}}
\begin{center}
\begin{minipage}{.5\columnwidth}
\lettrine[lines=1]{C}{}osmologists are often in error, but never in doubt.

\begin{flushright}\textit{Contributed to Lev Landau\\Before 1962}\end{flushright}

\vspace*{1cm}

\lettrine[lines=1]{I}{} am certain that it is time to retire Landau's quote.

\begin{flushright}\textit{Michael Turner\\
 Physics Today 2001, \\
 Vol. 54 Issue 12, p. 10}\end{flushright}
\end{minipage}
\end{center}
\vspace{9em} 
\vspace*{\stretch{1}}
\clearpage{\pagestyle{plain}\cleardoublepage}

%% file: chapter1.tex
\chapter{Introduction}

\section{Motivation}

It is a truth universally acknowledged that any enlightened human being in
possession of a healthy curiosity must be in want of an understanding of the
world around him. In the past few decades, two mysterious phenomena have
emerged in cosmology that challenge our understanding of the Universe in the
post provoking way: dark matter and dark energy. The ``stuff'' that we
observe, that makes up the galaxies and stars, the protons and the neutrons,
the planets and the moons, the plants and the animals and even ourselves,
everything we ever knew and all of what we thought there ever was, makes up
only a meager 4\% of the cosmos. At first glance, this presents a huge
problem for cosmology. But in the grand picture, it might be an opportunity
for all of physics to make progress. Now, for the first time in history,
cosmology is inspiring particle physics via dark matter for the hunt of a
new particle, bringing physics to a full circle. At the same time, dark
energy is causing embarrassment in the quantum field theory department by
deviating from the predicted value by 120 orders of magnitude.
Closing all gaps in our knowledge about the Universe on a large scale is
therefore necessary in order to understand physics and our world as a whole.

Experiments are the physicist's way of unlocking the Universe's secrets, but
they are incredibly limited in the field of cosmology. The studied object is
unique and cannot be manipulated, only be observer through large telescopes
either on the ground or in space. There is no shortage of theories that
could explain some of the issues that are troubling cosmologists these days,
but they all look very similar and it is hard to tell them apart by looking
at the sky. Subtle clues need to be picked up to succeed, and for this we
need better and better surveys. To shed light onto the dark Universe,
several missions are being planned, designed, or are already in progress.
In the design stage in particular, it is interesting to see what kind of
results one can expect depending on the experiment's specifications.

A lot of dark energy models are degenerate in the sense that some of their
predictions match or are indistinguishable by reasonable efforts. The
most popular ones predict critical values for some parameters, so unless we
are provided with a method that allows for infinite precision in our
measurements, we will never know, for instance, whether the Universe is flat
or has a tiny, but non-zero curvature that is too small to be detectable, or
whether dark energy is really a cosmological constant or is time-dependent
in an unnoticeable way. We hope to break at least some of the degeneracies
by investigating how growth of structure in the matter distribution evolved
throughout the history of the Universe.  

Information about how the matter used to be distributed is naturally
obtained by analyzing far away objects, since the finite speed of light
allows us to look as far into the past as the Universe has been transparent.
But the nature of dark matter is hindering us from getting a full picture of
the matter distribution, because only 20\% of all matter interacts
electromagnetically. The only sensible way to detect all of matter is via
its influence on the geometry of space-time itself. Distortions of
background galaxies will reveal perturbations in the fabric of the cosmos
through which their light has passed, unbiased in regard of matter that
happens to be luminous or not. In particular, the signal coming from the
so-called weak lensing effect is sensitive to the evolution of both dark
energy and growth related parameters.  To make the most out of this
phenomenon, we ``slice up'' the Universe into shells in redshift space and
study the correlation of these redshift bins, a technique which has been
dubbed \textit{weak lensing tomography}.\footnote{Tomos ($\tau\acute o\mu o
\varsigma$): Greek for slice, piece.} 

\section{Objective}

By assuming the validity of Einstein's general relativity, all observations
are biased towards the standard model and the growth of structure might not
have been measured correctly. Instead, we want to allow for the possibility
of modified gravity. Our goal is  to investigate how future weak lensing
surveys like {\sc Euclid} can constrain the expansion rate $H(z)$ and the
growth function $G(z)$ without assuming a particular model parameterizing
those two quantities. We first select a suitable number of redshift bins in
which we divide a given galaxy distribution function.  Starting from a
fiducial model with a list of cosmological parameters that take on
particular values determined by Wilkinson Microwave Anisotropy Probe
(WMAP)\footnote{A probe of the cosmic microwave background radiation, see\\
\url{http://map.gsfc.nasa.gov}.}, we then take the values of $H$ and $G$ in
each bin as a constant, treat them as additional cosmological parameters,
and rebuild these two functions as linear interpolations between supporting
points in the redshift bins. 

Using these new functions, we calculate the weak lensing power spectrum in
each bin as well as their cross-correlation spectra based on the non-linear
matter power spectrum, which we take from fitting formulae found by other
groups. The weak lensing power spectrum is then used in the powerful Fisher
matrix formalism, which allows us to estimate the uncertainty of all
cosmological parameters, including the values of $H$ and $G$ at the chosen
redshifts. With these forecast error bars, we can compare competing theories
of modified gravity to our simulated results of next generation weak lensing
surveys.

\section{Symbols and notation}

We use units where the speed of light equals unity. Vectors are lower case
and  printed in bold (e.g. $\vec x$), while matrices are upper case and
printed sans serifs (e.g. $\mat F$). The metric has signature $-$+++.
Derivatives are sometimes written using the comma convention: $\partial
\phi/\partial x^i = \phi_{,i}$. The logarithm to base ten is denoted by
$\lg$, and the logarithm to base $e$ by $\ln$. The Fourier transform of a
function $f(x)$ is written with a tilde: $\tilde f(k)$. Due to the
one-to-one correspondence of the redshift and the scale factor, $a=1/(1+z)$,
we may denote the dependence on either one of those quantities
interchangeably, for instance it is obvious that $H(a) = H(1/(1+z))$, so we
might as well consider $H(a)$ as a function of $z$ and write $H(z)$.
Sometimes, the value of redshift dependent quantities like $\Omega_m(z)$
as of today is given as just $\Omega_m$, for instance. A list of commonly
used symbols follows.

\begin{longtable}{lp{11cm}}
$a$ & Scale factor\\
$\chi$ & Complex ellipticity\\
$C_{ij}$ & Covariance matrix\\
$\delta_{ij}$ & Kronecker delta\\
$\delta_\mathrm{D}$ & Dirac delta function\\
$\delta_m(z)$ & Matter density contrast\\
$\Delta_z$ & Photometric redshift error\\
$E(z)$ & Dimensionless Hubble parameter, $E(z)=H(z)/H_0$\\
$F_{ij}$ & Fisher matrix\\
$f(z)$ & Growth rate\\
$G$ & Newton's gravitational constant\\
$G(z)$ & Growth function\\
$\gamma$ & Growth index\\
$\gamma_\mathrm{int}$ & Intrinsic galaxy ellipticity\\
$\gamma_i$ & Shear, $i=1,2$\\
$\gamma_\mathrm{ppn}$ & Parameterized post-Newtonian parameter\\
$h_i,g_i$ & Values of the logarithm of $H(z),G(z)$ at redshift $z_i$, $i=1,\dots,\calN$\\
$H_0$ & Hubble constant\\
$h$ & Dimensionless Hubble constant, $h=H_0 / 100 \unit{km/s/Mpc}$\\
$J_{ij}$ & Jacobian matrix\\
$k$ & Wave number\\
$\kappa$ & Convergence\\
$\ell$ & Multipole\\
$\mathcal L$ & Likelihood, Lagrangian\\
$n(z)$ & Galaxy distribution function\\
$n_i(z)$ & Galaxy distribution function for the $i$-th redshift bin\\
$n_i$ & Average galaxy density in the $i$-th redshift bin\\
$n_s$ & Scalar perturbation spectral index\\
$n_\vartheta$ & Angular galaxy density\\
$\calN$ & Number of redshift bins\\
$\omega_x$& Reduced fractional density for component $x$: $\omega_x=\Omega_x h^2$\\ 
$\Omega_x(a)$ & Fractional density for component $x$ as a function of the
scale factor or redshift ($x=m,b,c$ for matter, baryons, cold dark matter)\\
$\Omega_m$ & Fractional matter density today\\
$P_m(k)$ & Matter power spectrum\\
$P(\ell)$ & Convergence power spectrum\\
$p$ & Pressure\\
$P(A|B)$ & Probability of $A$ given $B$\\
$\Phi,\Psi$ & Scalar gravitational potentials\\
$r(z)$ & Comoving distance\\
$\rho$ & Density\\
$\sigma_8$ & Fluctuation amplitude at $8\unit{Mpc}/h$\\
$\sigma(x)$ & Uncertainty of the quantity $x$\\ 
$\tau$ & Reionization optical depth\\
$t$ & Cosmic time\\
$\vec \theta$ & Vector in parameter space; angular position\\
$w$ & Equation-of-state ratio\\
$W(z)$ & Window function\\
$z$ & Redshift\\
$z_m$ & Median redshift\\
$z_i,\hat z_i$ & Endpoint, center of the $i$-th redshift bin\\
\end{longtable}

%% file: chapter2.tex
\chapter{Theoretical preliminaries} 

This section outlines the basic theory behind dark energy by giving a brief
historical introduction. Furthermore, the basics of weak lensing and the
Fisher information matrix formalism are explained, as well as the concept of
power spectra, an immensely important tool not only in the remainder of this
thesis, but in all of modern cosmology. A much more detailed treatment of
the topics presented in this section can be found in many excellent standard
text books, such as~\citet{Carroll2003} for the derivation of cosmology from
general relativity,~\citet{Weinberg2008} and~\citet{Amendola2010} (hereafter
referred to as A\&T) for an up to date reference to modern cosmology, dark
energy and useful statistical tools, \citet{Peebles1980} also for the
statistics and~\citet{Bartelmann2001} for a comprehensive treatment of weak
gravitational lensing.

\section{Dark energy: A historical summary}

A short while after Albert Einstein derived his famous field equations he
proposed the idea of inserting a cosmological constant originally in order
to allow for a non-expanding Universe, because he felt that a world that
was spatially closed and static was the only logical 
possibility \citep{Einstein1917}. This view was the general consensus at the
time, considering that all visible stars seemed static and other galaxies
had not been discovered yet. From the first and
second Friedmann equations
\begin{align}
\left( \frac{\dot a}{a} \right)^2 & = \frac{8\pi G}{3}\rho
-\frac{k}{a^2} \label{eq:2-friedmann1}\,,\\
\frac{\ddot a}{\dot a} & = -\frac{4\pi G}{3}(\rho+3p)\,,
\label{eq:2-friedmann2}
\end{align}
we can easily see, that for a static solution with $\ddot a=\dot a=0$, we
need a component in the Universe that bears a negative pressure
\begin{equation}
p = -\frac{1}{3}\rho
\label{eq:2-pressure}
\end{equation}
with an overall positive spatial curvature that is finely tuned to
\begin{equation}
\frac{k}{a^2} = \frac{8\pi G}{3}\rho.
\label{}
\end{equation}
At this point, we mention that it is often useful to plug the time
derivative of eq.~(\ref{eq:2-friedmann1}) into eq.~(\ref{eq:2-friedmann2})
to arrive at the continuity equation
\begin{equation}
\dot \rho = -2\frac{\dot a}{a}(\rho+p)\,.
\label{eq:2-conti}
\end{equation}
It can be shown that for dust, i.e. non-relativistic baryonic or
dark matter, the pressure vanishes and for radiation or ultra-relativistic
matter the pressure is one third of the energy density. This means that
ordinary forms of energy cannot produce the desired result. Einstein
recognized that we the Lagrangian in the Hilbert-Action can be changed to
\begin{equation}
S = \frac{1}{16\pi G}\int \mathrm d^4 x \sqrt{-g}(R-2\Lambda)
\end{equation}
by introucing a constant $\Lambda$.
We can interpret this term as a form of energy density $\rho_\Lambda =
\Lambda/8\pi G$ that does not 
depend on the scale factor, which leads to $p=-\rho$ according to
eq.~(\ref{eq:2-conti}).
Thus $\rho$ in eqs.~(\ref{eq:2-friedmann1})
and (\ref{eq:2-friedmann2}) can be replaced by $\rho_m+\rho_\Lambda$.
Then a Universe in which $\rho_m=2\rho_\Lambda$ or $\Lambda=4\pi G \rho_m$
satisfies \eq{eq:2-pressure}.

\begin{figure}[htbp]
\begin{center}
\includegraphics[width=9cm]{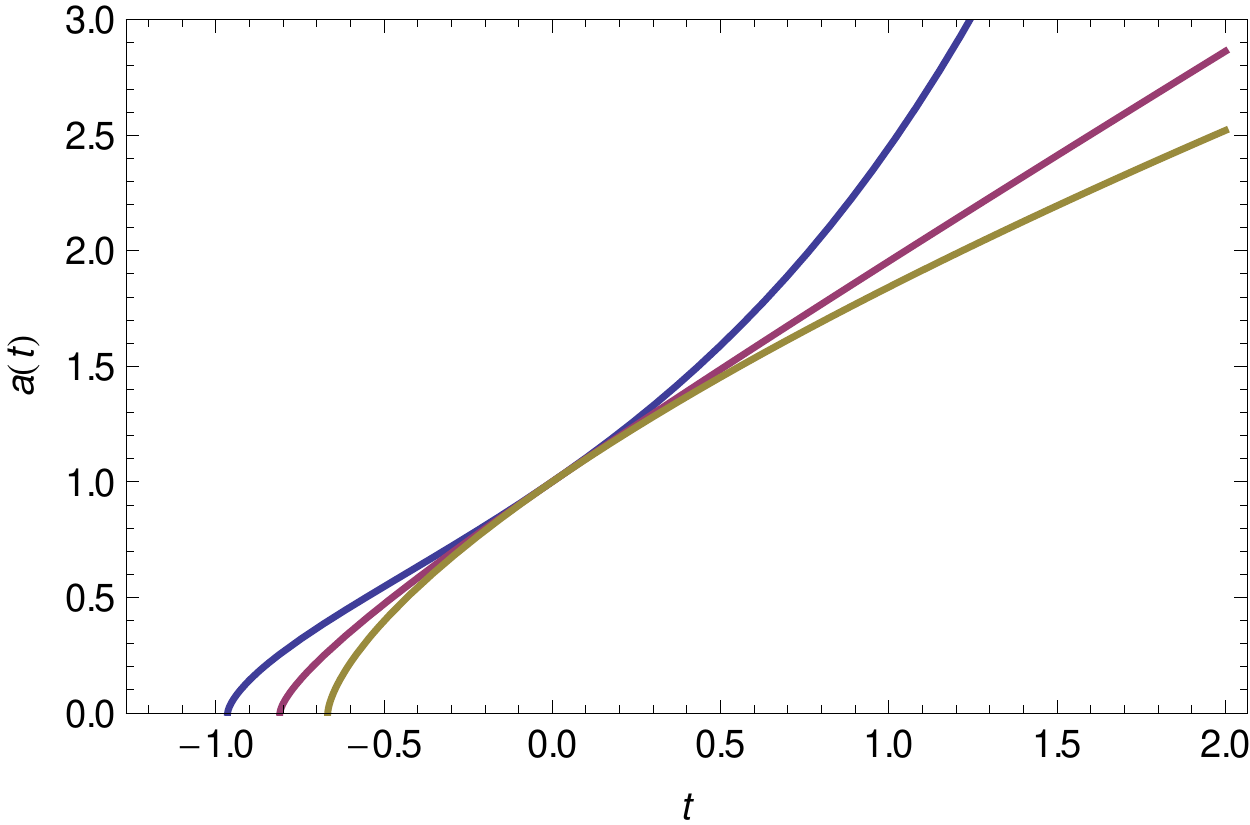}
\includegraphics[width=9cm]{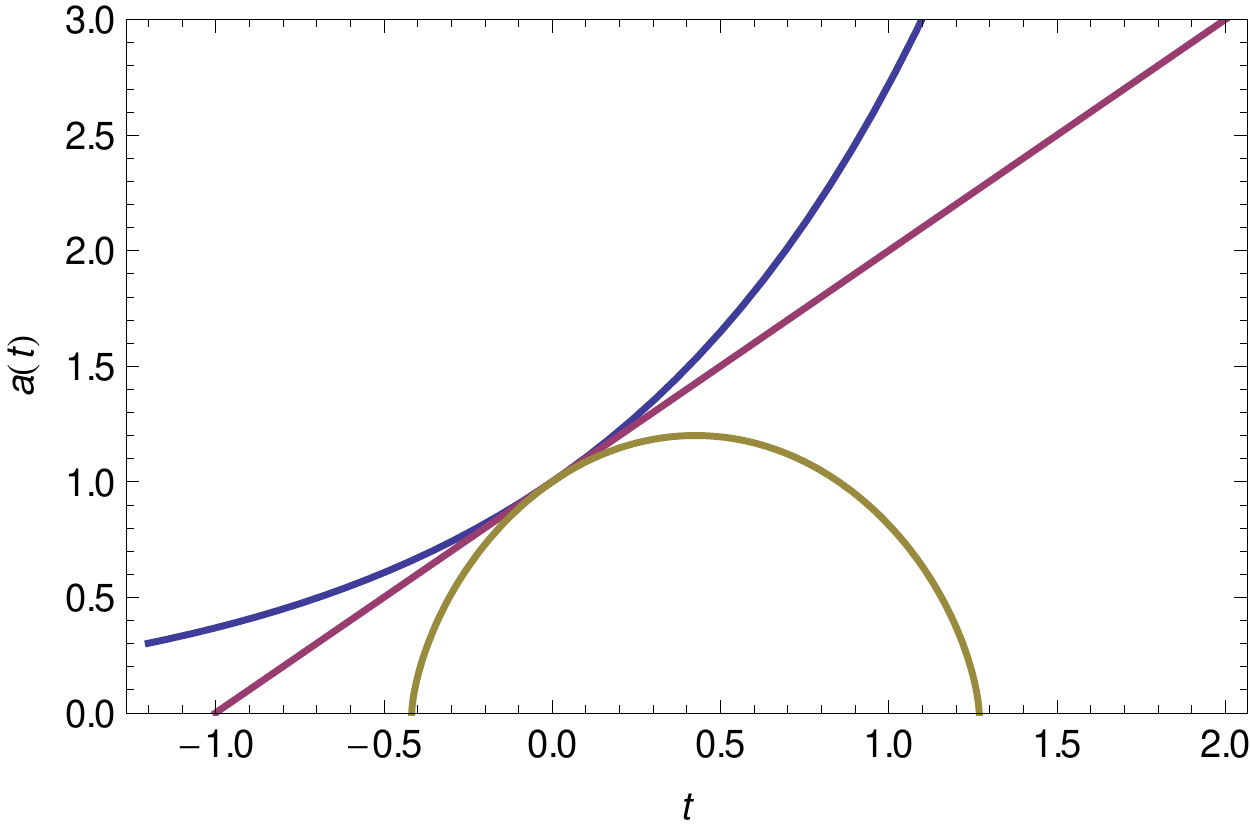}
\end{center}
\caption{Solutions of the Friedmann equation demonstrating the evolution of
the Universe. We have set the Hubble constant $H_0$ to unity for
convenience, which implies that $t=-1$
corresponds to the Hubble time $-13.9\unit{Gyr}$.  All curves pass $t=0$
(today) with $a(0)=\dot a(0)=1$. \textit{Top panel}: From the top: The
expansion history and future for our Universe with $(\Omega_m =
0.3,\Omega_\Lambda=0.7)$, an open matter-only Universe $(0.3,0)$ and the
Einstein--de Sitter model $(1, 0)$.  \textit{Bottom panel}: Some more exotic
solutions: A pure cosmological constant $(0,1)$, the Milne model $(0,0)$ and
a closed Universe $(6,0)$.}
\label{fig:evolscalfac}
\end{figure}

However, Einstein's conservative view of the world did not live up to
experimental facts. Only a short time later, after Edwin Hubble's
revolutionary discovery of apparently receding galaxies in 1923, Einstein
readily discarded the idea of a static Universe. Because an expanding
Universe does not necessarily need a cosmological constant (see
fig.~\ref{fig:evolscalfac}), the introduction of the cosmological constant
was considered a mistake for decades after that \citep{Peebles2003}.

It was not until the early 1960s, when $\Lambda$ had to be revived to
explain a new measurement of $H_0$ by Allen Sandage that was more accurate
than the original one by Hubble by one order of magnitude. After a proper
recalibration of the distance measure, he found $H_0\approx75
\unit{km/s/Mpc}$, causing an incompatibility of what was then concluded
to be the age of the Universe ($7.42 \unit{Gyr}$) with that of the age of
the oldest known star in the Milky Way, which was at the time estimated to
be greater than $15 \unit{Gyr}$ \citep{Sandage1961}. Even though the numbers
were slightly off according to our current knowledge, Sandage was on the
right track and immediately suggested that a cosmological constant could
easily resolve this issue. This was the first hint that the Universe was not
only expanding, but expanding in an accelerated manner.

Several decades later, the highest authority in physics---the
experiment---put an end to all speculations. In the late 1990s
\citet{Riess1998} observed ten supernovae of type Ia that they used as
standard candles to infer their luminosity distance as a function of
redshift and found that they were on average 10\%--15\% farther away than
one would expected in an open, mass dominated Universe. They were able to
rule out the $\Omega_\Lambda=0$ case at the $9\sigma$ confidence level,
leading the way for precision cosmology. Almost simultaneously, an
independent measurement by \citet{Perlmutter1998} confirmed their findings
by analyzing 42 high--redshift supernovae of type Ia. Thus, the notion that
the expansion of the Universe was accelerating was confirmed. Our failure to
understand the nature of this accelerated expansion is reflected in the name
for this phenomenon: \textit{dark energy}. It is important to stress that
the underlying
physical cause of the acceleration does not have to be a new form of energy,
it might as well be a new form of physics that mimics the effects of a
cosmological constant, which is often misunderstood even by physicists
unfamiliar with cosmology. Whether we are faced with a ``physical darkness
or dark physics'', to say it with the words of \citet{Huterer2007}, is the
whole crux of the matter.  Evidence independent of the distance-luminosity
relation for type 1a supernovae which supports the idea of acceleration
has steadily increased since the findings of Riess and Perlmutter. For
instance, the late integrated Sachs-Wolfe effect \citep{Scranton2003}, or
the cosmic microwave background radiation \citep{Komatsu2011} and a dark
energy
component is part of today's standard model of the Universe. But does this
mean that cosmology as a science is done?

On the contrary. Even though the so called flat $\Lambda$CDM model (cold
dark matter with a cosmological constant) with only six free parameters
($\omega_m$,$\omega_b$,$\Omega_\Lambda$,$n_s$,$\sigma_8$,$\tau$)
\footnote{The exact meaning of those parameters will be explained in
section~\ref{survey}.} agrees remarkably well with all observations, most
physicists are still not satisfied with it for a number of reasons. First of
all, the value of the cosmological constant is incredibly small. When we
express $\Lambda$ in natural units, i.e. where the Planck length $l_P =
\sqrt{\hbar G} =1$, we get $\Lambda = 3.5 \times 10^{-122}$. Even worse,
when we interpret the cosmological constant simply as the vacuum energy
predicted from quantum field theory, the value diverges. Because it is
widely believed that new physics are needed at the Planck scale where
neither quantum effects nor gravity are negligible, namely a
theory of quantum gravity \citep{Carroll1992}, one can renormalize the
integral over all modes of a scalar, massive field (e.g.  a spinless boson)
with zero point energy $\hbar \omega/2$ by introducing a cutoff at Planck
length. This generates a finite value, but one that is still 120 orders of
magnitude too large, making it arguably the worst prediction in the history
of physics.  Finding a mechanism that would cause fields to cancel out each
other's contributions to the vacuum energy density would be a challenge, but
a mechanism that leaves a tiny, positive value seems completely out of reach
for now. The most promising resolution might lie in super symmetry,
super gravity or super string theory. This is commonly referred to as the
\textit{cosmological constant problem} \citep{Weinberg1989}.

\begin{figure}[htbp]
\begin{center}
\includegraphics[width=9cm]{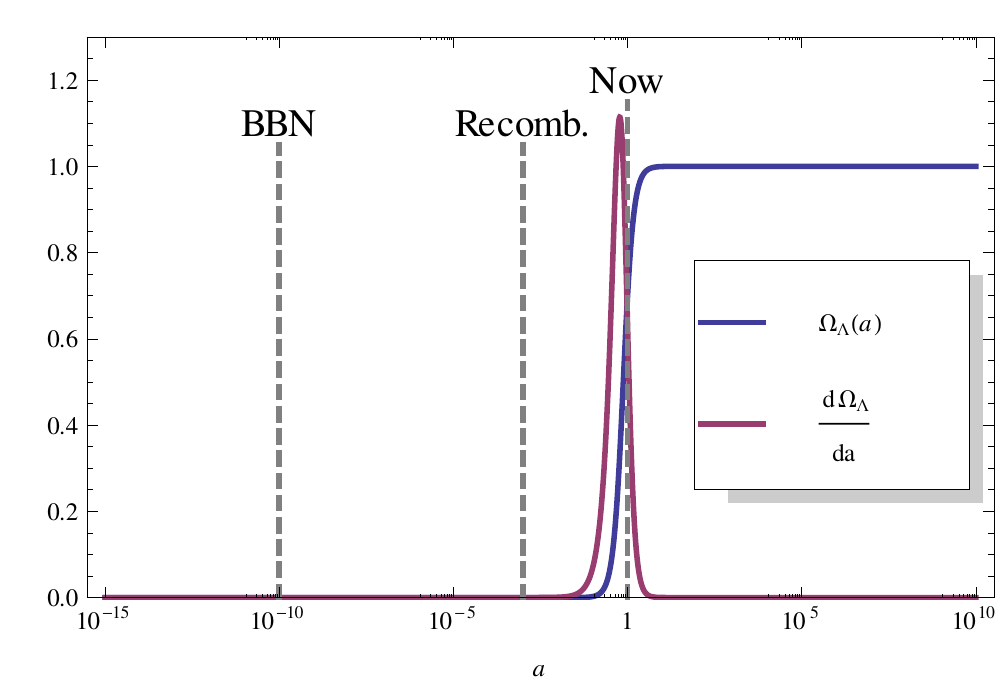}
\end{center}
\caption{Visualization of the coincidence problem. Shown here is the
dimensionless dark energy density $\Omega_\Lambda(a)$ in dependence of the
scale factor $a$ and its derivative on a logarithmic scale. Highlighted with
dashed lines are the times of big bang nucleosynthesis (BBN) at
$a\approx10^{-10}$, recombination at $a\approx 1/1000$ and today at $a=1$.
Dark energy started to dominate right when structures emerged in the
Universe.} 
\label{fig:coinc}
\end{figure}

The second problem is the \textit{coincidence problem}, which describes the
curious fact that we live exactly in a comparably short era in which matter
surrenders its dominance to dark energy. This can be visualized if we
consider the time dependence of the dimensionless dark energy
density\footnote{Radiation energy scales as $a^{-4}$ and was only relevant
at very early stages of the Universe, which is why we shall not consider it
here.}:
\begin{equation}
\Omega_\Lambda(a)=\frac{\rho_\Lambda}{\rho_\mathrm{crit}(a)}=
\frac{\rho_\Lambda}{\rho_m(a)+\rho_\Lambda}=
\frac{1}{\frac{\rho_m(a_0)}{\rho_\Lambda}a^{-3}+1} \approx 
\frac1{\frac37a^{-3}+1},
\end{equation}
because $\rho_m$ scales as $a^{-3}$ while $\rho_\Lambda$ stays constant,
when $a_0$ is the scale factor today, usually normalized to unity.
Plotting this with its derivative on a logarithmic scale (see
fig.~\ref{fig:coinc}) shows the coincidence.

Until now, there is no satisfactory solution to these problems. String
theorists argue that we might live in one of $10^{500}$ realizations of the
Universe, corresponding to the number of false vacua that are allowed by
string theory \citep{Douglas2003}. This approach comes down to an anthropic
argument, stating that it is no surprise that the cosmological constant has
the value that it has, since any other value might be realized in a
different Universe, but would not lead to the growth of structure required
for sentient life that can ponder the value of natural constants
\citep{Susskind2005}.  This argument is controversial amongst physicists. As
one of the leading cosmologists Paul Steinhard puts it in a \textsc{Nature}
article \citep{Brumfiel2007}: ``Anthropics and randomness don't explain
anything.''

Alternative theories include the postulation of Quintessence, a new kind of
energy in the form of a scalar field with tracker solutions mimicking a
cosmological constant \citep{Zlatev1999}, or a modification of Einstein's
theory of general relativity, where, in the most popular case, the Ricci
scalar in general relativity is replaced as the Lagrangian by some function
$f(R)$ \citep{Felice2010}. Unfortunately, current observational constraints
are insufficient when it comes to discriminating between competing theories.

\section{Weak lensing}

\begin{figure}[htbp]
\begin{center}
\includegraphics{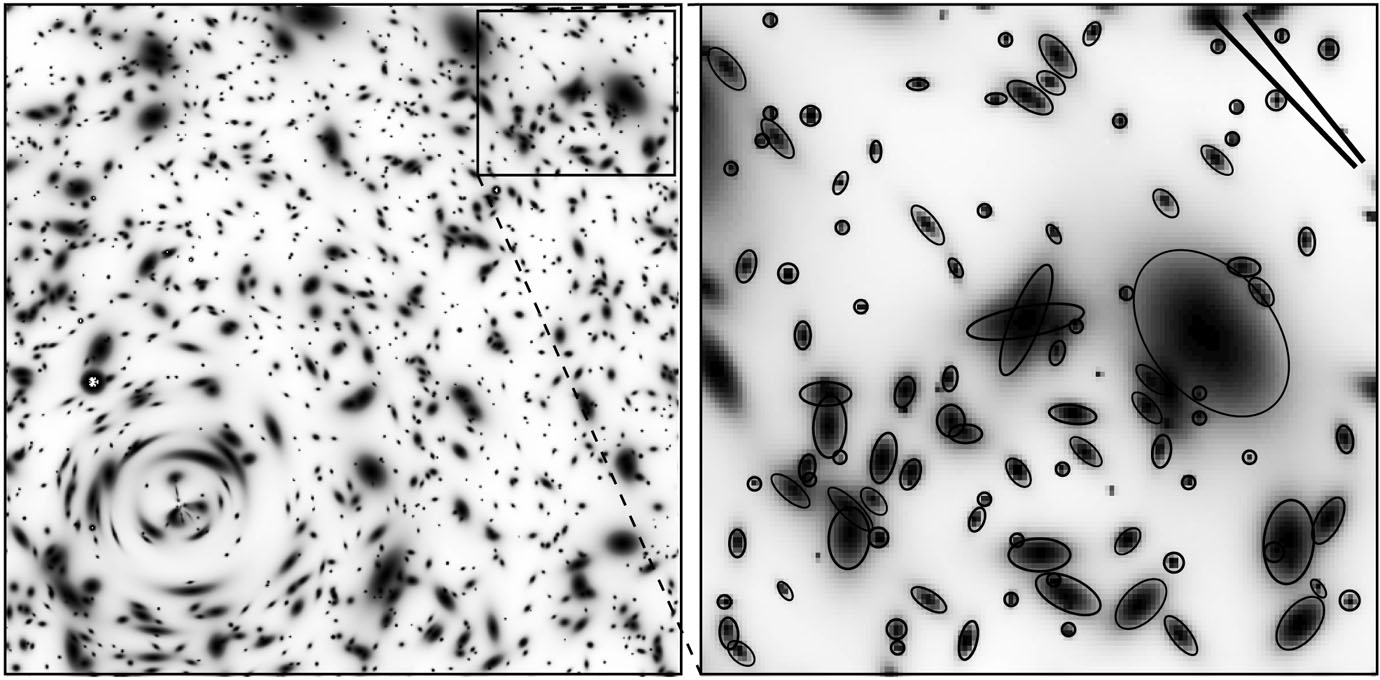}
\end{center}
\caption{\textit{Left panel}: Simulation of a gravitational lens with a
case of strong lensing in the lower left corner at an average redshift of
one.  \textit{Right panel}: Enhancement of the upper right section of the
left panel, depicting
lensing in the weak regime. The contours are ellipses derived from the
quadrupole moment of each galaxy image. The two bars in the upper right
corner show the average shear, where the lower one represents the expected
shear and the upper one the actual shear. The deviation stems from shot
noise due to the intrinsic ellipticity and random orientation of the
galaxies. Taken from \citet{Mellier1998}.}
\label{fig:2-lensing}
\end{figure}

When light passes through a gravitational potential, its trajectory is bent
in a way described by general relativity. This process, dubbed ``lensing'',
can alter the shape, size and brightness of the original image. We usually
distinguish between strong, weak and micro lensing. The mechanism of micro
lensing relies on small objects of low magnitude being the lens, such as
brown or white dwarfs, neutron stars, planets and so on, that transit a
bright source and temporarily increase the source's brightness on time
scales of several seconds to several years \citep{Paczynski1996}. The other
two cases are static on those time scales, since more massive lenses like
galaxies or clusters are involved, which means that the distances involved
are several orders of magnitude larger. Strong lensing occurs when multiple
images of the same galaxy from behind a super massive object can be
observed, as seen in the bottom left corner of the left panel in
fig.~\ref{fig:2-lensing}.

\begin{figure}[htbp]
\begin{center}
\includegraphics[width=\textwidth]{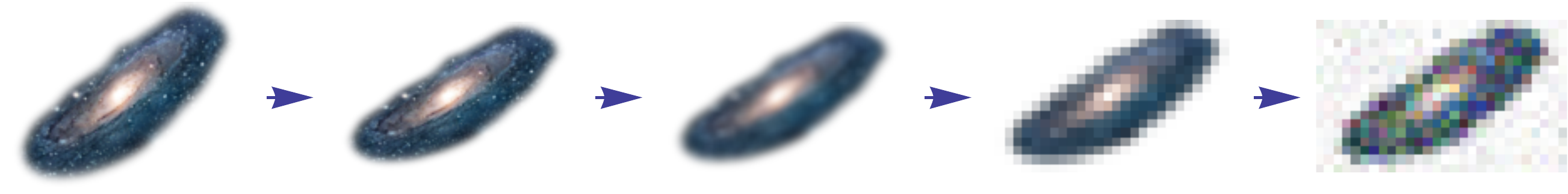}
\end{center}
\vspace{-.5cm}
\small\hspace{1cm}a)\hfill b)\hfill c)\hfill d)\hfill e)\hspace{1cm}
\caption{A schematic illustration of the process of weak lensing detection,
demonstrating its difficulty.
a) The original source galaxy. b) Weak lensing distorts the shape and adds a
shear to the image. c) The image is convolved by the telescopes point spread
function and, in
case of ground based surveys, the atmosphere. d) The effect of the finite
resolution of the detector. e) Additional noise is applied to the image. The
challenge now is to infer b) from e). (Image of M31 by courtesy of Robert
Gendler.)}
\label{fig:cosmicshear}
\end{figure}

While strong and micro lensing are relatively rare, weak lensing is an
effect that can be observed over large areas of the sky (see upper right
corner of the right panel in fig.~\ref{fig:2-lensing}). Its power lies in
statistical analysis, since weakly lensed objects are only slightly
distorted and impossible to detect individually. Only the average distortion
field reveals the cosmological information that we are looking for. As it
turns out, weak lensing changes the ellipticity of galaxies (in a first
order approximation), but the intrinsic ellipticity always dominates the
shape. We need to gather large enough samples and then
subtract the noise, which is relatively simple if the magnitude and
direction of the intrinsic ellipticity is uncorrelated to the signal. This
is unfortunately not entirely the case due to tidal effects
(\textit{intrinsic alignment}), and this is only one of the many challenges
that weak lensing harbors. The most obvious one is illustrated in
fig.~\ref{fig:cosmicshear}. Others include photometric redshift errors,
calibration errors and uncertainties in power spectrum theory. A lot of
these systematic errors can be accounted for by introducing several
nuisance parameters~\citep{Bernstein2009}. The trade-off is that a high
number of nuisance parameters diminish the merit of the Fisher matrix
formalism, as there are degeneracies to be expected, so we will only account
for photometric redshift error and the intrinsic ellipticity in this thesis.

An expression describing the shear can be obtained by perturbing the
Friedmann-Lema\^itre-Robertson-Walker
metric in the Newtonian gauge
\begin{equation}
 \dd s^2 = -(1+2\Psi)\dd t^2 + a^2(t)(1+2\Phi)(\dd x_1^2 + \dd x_2^2 + \dd
 x_3^2)\,,
 \label{eq:2-newt-gauge}
\end{equation}
where $\Psi$ and $\Phi$ are scalar gravitational potentials. We can then
solve the geodesic equation for a light ray
\begin{equation}
\frac{\dd k^\mu}{\dd \lambda}+ \Gamma_{\alpha\beta}^\mu k^\alpha k^\beta =
0\,,
\end{equation}
which can be rewritten in our case as
\begin{equation}
\frac{\dd^2}{\dd r^2}(r\theta_i) =  \Phi_{,i}-\Psi_{,i}\,.
\end{equation}
Thus, we obtain the lensing equation
\begin{equation}
\theta_i = \theta_{0i} + \int_0^r \dd r'\left(1-\frac{
r'}{r}\right) \phi_{,i}(r' \theta_{01},  r' \theta_{02},r')
\end{equation}
with the lensing potential $\psi = \Phi-\Psi$. Here $r$ is the radial
comoving coordinate, $\theta_i=x_i/r$ is the angle of the light ray with
respect to the $r$-axis, and $(x_1,x_2)$ are displacement coordinates
perpendicular to the $r$-axis.

\begin{figure}[htbp]
\begin{center}
\scalebox{.75}{\input{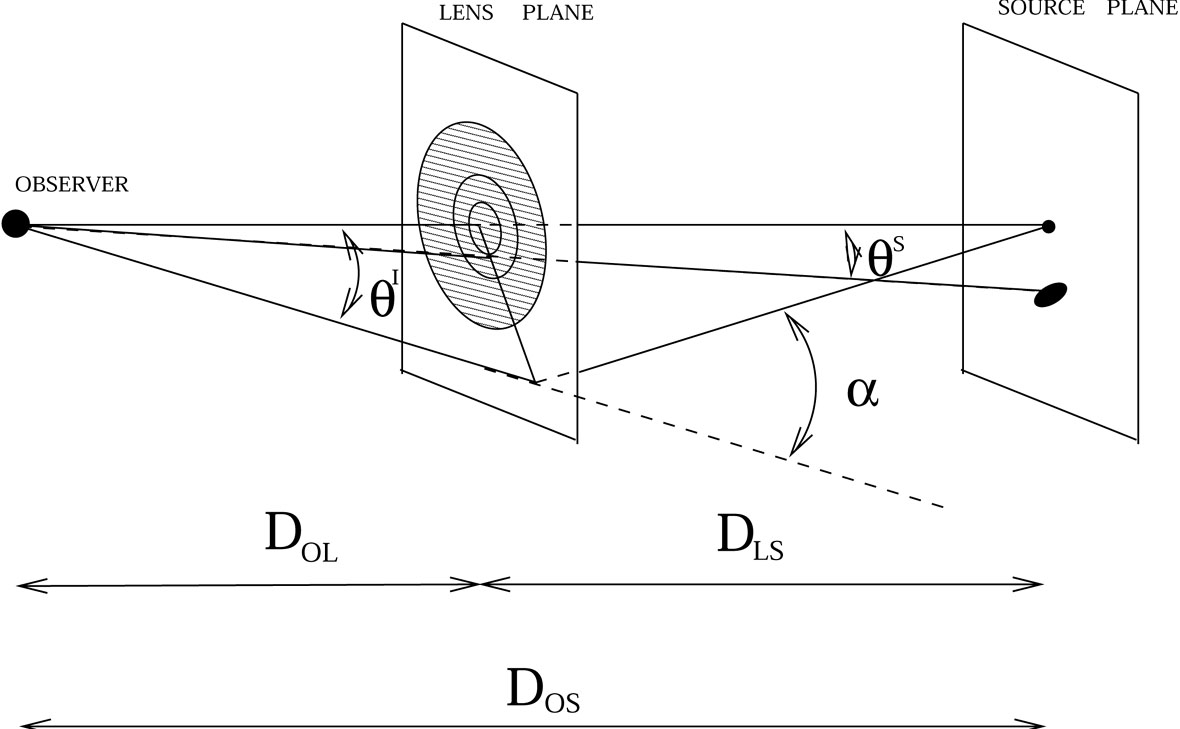}}
\end{center}
\caption{Description of the lensing configuration. Note that comoving
distances cannot be added trivially, i.e. $r_s \neq r_l+r_{ls}$.}
\label{fig:2-descrip}
\end{figure}

The distortion of a source image at distance $r_s$ to first order is a
linear transformation described by the symmetric matrix
\begin{equation}
A_{ij}\equiv \frac{\partial \theta_{si}}{\partial \theta_{0j}} =
\delta_{ij} + D_{ij}
\label{}
\end{equation}
with 
\begin{equation}
D_{ij}(r_s) = \int_0^{r_s}\dd r' \left( 1-\frac{ r'}{r_s}
\right) r' \psi_{,ij} = \left( \begin{array}{cc}-\kappa_\mathrm{wl}
-\gamma_1 &
-\gamma_2\\ -\gamma_2 & -\kappa_\mathrm{wl}+\gamma_1\end{array}\right)\,.
\end{equation}
Here we introduced the convergence
\begin{equation}
\kappa_\mathrm{wl} = -\frac{1}{2}\int_0^{r_s}\dd  r' \left(
1-\frac{ r'}{r_s}\right) r'(\psi_{,11}+\psi_{,22})\,,
\label{eq:2-conv}
\end{equation}
which is a measure for the magnification of the image, and the shear
\begin{align}
\gamma_1& = -\frac{1}{2}\int_0^{r_s}\dd  r' \left(
1-\frac{ r'}{r_s}\right) r'(\psi_{,11}-\psi_{,22})\,,\\
\gamma_2& = -\int_0^{r_s}\dd r'\left( 1-\frac{r'}{r_s}
\right) r' \psi_{,12}\,,
\end{align}
which describes the distortion. These quantities will become important when
we want to describe the cosmic shear by statistical means, in particular its
power spectrum.

We can measure the ellipticity of real astronomical images by computing
their quadrupole moment, which is defined as
\begin{equation}
q_{ij} = \int I(\vec \theta)\theta_i\theta_j\dd^2\theta \,,
\end{equation}
where $I(\vec\theta)$ is the luminous intensity of the galaxy image with its
center at $\vec \theta =0$. The complex ellipticity is then given by
\begin{equation}
\chi\equiv\frac{q_{11}-q_{22}+2iq_{12}}{q_{11}+q_{22}}\,.
\end{equation}
It is straight forward to show that according to this definition the complex
ellipticity for an ellipse with semiaxes $a$ and $b$, rotated by an angle
$\phi$, is
\begin{equation}
\chi = \frac{a^2-b^2}{a^2+b^2}e^{2i\phi}\,.
\end{equation}
To first order, a weak lens distorts a spherical object into an ellipse with
a simple relation between the shear and the complex ellipticity, namely
\begin{equation}
\chi = 2(\gamma_1 + i\gamma_2)\,.
\end{equation}
We can use this to compute the power spectra $P_{\gamma_i}(\ell)$, which
describe the auto-correlation of the shear field at the multipole $\ell$,
the adjoint variable to $\theta$. It can be
shown that the convergence power spectrum $P$ is related to the matter power
spectrum and can be written to first order
as a linear combination $P = P_E = c_1P_{\gamma_1}+c_2P_{\gamma_2}$. Another
linear combination, $P_B = c_1P_{\gamma_1}-c_2P_{\gamma_2}$, must vanish.
These are usually referred to as the electric and magnetic part of the
shear field. Thus, the convergence power spectrum contains all the
information while the magnetic part provides a good check for systematics.

But, as mentioned before,  not all galaxies appear circular even if their
intrinsic shape was a perfectly circular disc and there was no gravitational
lensing distorting our view of the sky, as they are randomly oriented and
thus, viewed from the side, look like ellipses.
This intrinsic ellipticity, denoted by $\gamma_\mathrm{int}$, can be
reflected in a noise term that is added to the power spectrum, if we assume
that the noise is uncorrelated to the weak lensing signal\footnote{As
mentioned before, because of tidal effects this is not necessarily the
case.}:
\begin{equation}
P = P_\mathrm{signal} + \gamma_\mathrm{int}^2 n^{-1}\,,
\end{equation}
where $n$ is the average galaxy density (A\&T, ch. 14.4).

The shear field is a vector field of dimension two. To access the full power
of weak lensing, we must also include the third dimension, which can be done
by considering the redshift of each source galaxy. Weak lensing works
linearly, so we can add up the transformation matrices of all galaxies and
find for the full transformation matrix
\begin{equation}
\mathcal D_{ij} = \int_0^\infty \dd r' n(r') D_{ij}(r')\,,
\label{eq:2-trans}
\end{equation}
where $n(r)\dd r$ is the number of galaxies in a shell with width $\dd r$
and the galaxy distribution function $n(r)$ is normalized to unity. Making
use of the fact that for any real, integrable functions $f(x)$ and $g(x,y)$
the identity 
\begin{equation}
\int_0^\infty \dd x f(x) \int_0^x \dd y g(x,y) = \int_0^\infty \dd y
\int_y^\infty \dd x f(x) g(x,y)
\end{equation}
holds, we can rewrite eq.~(\ref{eq:2-trans}) to get
\begin{equation}
\mathcal D_{ij} =  \int_0^\infty \dd r w(r) \psi_{,ij}\,.
\end{equation}
Here we abbreviated the weight function as
\begin{equation}
w(r) \equiv \int_r^\infty\dd r'\left( 1-\frac{r}{r'} \right)r n(r')\,.
\end{equation}
Thus, the total convergence field from eq.~(\ref{eq:2-conv}) now reads
\begin{equation}
\kappa_\mathrm{wl} = -\frac{1}{2}\int_0^\infty \dd r w(r)
\left(\psi_{,11}+\psi_{,22}\right)\,.
\label{eq:2-totconv}
\end{equation}

By grouping all observed
galaxies into redshift bins, we are able to extract more information out of
the given data by calculating the power spectrum in each each redshift bin
as well as their cross correlation. This is called \textit{power spectrum
tomography}~\citep{Hu1999}. We shall cover this in more detail in
section~\ref{sec:tomography}.
 
\section{Power spectra} 

Almost every book and every lecture on modern cosmology begins with the
cosmological principle: the assumption that the Universe is, simply put,
homogeneous and isotropic. Now obviously the average density, say,
$100\unit{m}$ below the University Square in Heidelberg is quite different
than the average density $100\unit{m}$ above, and it is still quite
different at the midpoint between the Sun and Alpha Centauri -- we
say that the Universe has \textit{structure}. But if we look at the Universe
at sufficiently large scales of $\gtrsim 1\unit{Gpc}$, the overall success
of the standard model confirms our initial assumption~\citep{Komatsu2011}.
While there are research groups who investigate the possibility of genuine
large scale inhomogeneities~\citep{Barausse2005,Kolb2005}, we remain
confident that there is good reason to believe that the cosmological
principle applies to the Universe as a whole.

The matter distribution in tour Universe consists mostly of structures at
different scales, which can be roughly classified as voids,
super clusters, clusters, groups, galaxies, solar systems, stars, planets,
moons, asteroids and so on. This ``lumpiness'' is inherently random, but
like all randomness it can be characterized by an underlying set of well
defined rules by using statistics. The matter distribution in our Universe
is one sample drawn from an imaginary ensemble of Universes with an
according probability distribution function (PDF), and we are now facing the
challenge of determining that PDF.

We require our models to make precise predictions regarding the structure of
matter. Naturally, no model will be able to predict the distance between the
Milky Way and Andromeda for instance, but it should be able to predict how
likely it is for two objects of that size to have the separation that they
have. To quantify and measure the statistics of the irregularities in the
matter distribution in the cosmos, we define the \textit{two-point
correlation function} $\xi$ as the average of the relative excess number of
pairs found at a given distance.  That is, if $n_i,\, i\in\{1,2\},$ denotes
the number of point-like objects (nucleons, or galaxies, or galaxy clusters)
found in some volume element $\dd V_i$ at position $\vec r_i$ with a sample
average $\langle n_i\rangle=n_0 \dd V_i$, we can write the average number of
pairs found in $\dd V_1$ and $\dd V_2$ as
\begin{equation}
\langle n_1 n_2 \rangle = n_0^2 \dd V_1 \dd V_2 [1+\xi(\vec r_1,\vec r_2)],
\label{eq:2-pairs}
\end{equation}
where $n_0=N/V$ is the mean number density. If the matter distribution was
truly random, then any two volume elements would be uncorrelated, which
means that $\xi$ would vanish and the average number of pairs would simply
be the product of the average amount of objects in each volume element. If
$\xi$ is positive (negative), then we say those two volume elements are
correlated (anti-correlated).

Assuming a statistically homogeneous Universe, $\xi$ can only depend on the
difference vector $\vec r_{12} = \vec r_2-\vec r_1$. Further assuming
statistical isotropy allows $\xi$ to only depend on the distance $r$ between
$\vec r_1$ and $\vec r_2$.  Hence, we denote the two-point correlation
function as $\xi(r)$.  Solving eq.~(\ref{eq:2-pairs}) leads to
\begin{equation}
\xi(|\vec r_2-\vec r_1|) = \frac{\langle n_1n_2\rangle}{n_0^2 \dd V_1 \dd
V_2} - 1 = \langle \delta_m(\vec r_1) \delta_m(\vec r_2) \rangle,
\label{eq:2-corr}
\end{equation}
which can be easily checked by plugging in the definition of the density
contrast
\begin{equation} 
\delta_m\left( \vec r_i \right) = \frac{n_i - \langle n_i \rangle}{\langle
n_i \rangle}.
\label{eq:2-contrast}
\end{equation}
Thus, the correlation function is often written as the average over all
possible positions
\begin{equation}
\xi(|\vec r|) = 
\frac{1}{V}\int_V \delta_m(\vec x)\delta_m(\vec x+\vec r) \dd^3 x.
\label{eq:2-corr2}
\end{equation}

Another important tool that will later prove to be invaluable is the
\textit{power spectrum}, which is in a cosmological context the square of
the Fourier transform of a perturbation variable (up to some normalization
constant). For instance, the matter
power spectrum is defined as
\begin{equation}
P_m(\vec k) \equiv V \tilde \delta_m(\vec k) \tilde \delta_m^*(\vec k)=
\frac{1}{V}\left|\int \delta_m(\vec x) 
e^{-i\vec k\cdot\vec x} \dd^3x\right|^2.
\label{}
\end{equation}
Rewriting the norm yields
\begin{equation}
P_m(\vec k) = \frac{1}{V}\int\delta_m(\vec x)\delta_m(\vec y) e^{-i\vec k
\cdot (\vec x - \vec y)}\dd^3x\dd^3y
\end{equation}
or, if we substitute $\vec r = \vec x-\vec y$
\begin{equation}
P_m(\vec k) = \frac{1}{V} \int \delta_m(\vec x)\delta_m(\vec x - \vec r)
e^{-i\vec k \vec r} \dd^3x \dd^3r
 = \int \xi(r)e^{-i\vec k \vec r} \dd^3r\,,
\end{equation}
where we used eq.~(\ref{eq:2-corr2}) in the last step. Hence the power
spectrum is the Fourier transform of the two-point correlation
function.\footnote{This is the Wiener-Khinchin Theorem.} 

The power spectrum is amongst the cosmologist's favorite tool to describe
our Universe in a meaningful way. It is often applied for instance to the
matter distribution, the anisotropies of the cosmic microwave background
radiation or the shear field of weak lensing. 

The convergence power spectrum describing the cosmic shear of galaxy images
can be expressed in terms of the matter power spectrum. Since the matter
distribution is 3-dimensional, but the convergence is a function of the
2-dimensional sky, we need Limber's theorem to relate those two. It states
that the power spectrum for a projection 
\begin{equation}
F(\theta_x,\theta_y) = \int_0^{\infty} w(r) f(\theta_x r,\theta_y r, r) \dd
r\,,
\end{equation}
where $w(r)$ is a weight function (normalized to unity), turns out to be
\begin{equation}
P(q) = \int_0^\infty \dd r \frac{w(r)^2}{r^2}p\left( \frac{q}{r} \right)\,,
\label{}
\end{equation}
where $p(k)$ is the power spectrum of $f$. This theorem is directly
applicable to the total convergence field in eq.~(\ref{eq:2-totconv}), so
that its power spectrum becomes
\begin{align}
P_{\kappa_\mathrm{wl}}(q) &= \frac{1}{4}\int_0^\infty \dd
r\frac{w(r)^2}{r^2} P_{\Sigma_i \psi_{,ii}}\left( \frac{q}{r} \right)\\
&= \frac{1}{4}\int_0^\infty \dd z \frac{W(z)^2}{H(z)}P_{\Sigma_i
\psi_{,ii}}\left( \frac{q}{r} \right)
\end{align}
with 
\begin{equation}
W(z) \equiv \frac{w(r(z))}{r(z)}
\end{equation}
being the window function. All we need to know now is the power spectrum of
$\psi_{,ij}$, which is simply
\begin{equation}
P_{\psi_{,ij}} = k_i^2 k_j^2 |\tilde\psi|^2
\end{equation}
because the Fourier transform of $\psi_{,ij}$ is $-k_ik_j\tilde \psi$. Then
we can plug in the Poisson equation in Fourier space, which is
\begin{equation}
k^2\tilde\psi = 3a^2H(a)^2\Omega(a) \delta_m(a)
\end{equation}
(in the absence of anisotropic stress, i.e. $\psi=2\Phi$) to obtain
\begin{equation}
P_{\Sigma_i \psi_{,ii}}(k) = k^4\tilde\psi = 9 H(a)^4 \Omega_m^2/(1+z)^4
P_m(k)\,.
\end{equation}
Putting it all together, and replacing $q$ with the multipole $\ell/\pi$,
finally yields the power spectrum for the convergence,
\begin{equation}
P(\ell) = \frac{9H_0^3}{4}\int_0^\infty
\frac{W(z)^2E(z)^3\Omega_m(z)^2}{(1+z)^4}P_m\left(
\frac{\ell}{\pi r(z)}\right)\dd z\,.
\label{eq:2-convps}
\end{equation}
More details can be found in A\&T (ch.  4.11, 14.4) or \citet{Hu2004}.

\begin{figure}[htpb]
\begin{center}
\includegraphics[width=9cm]{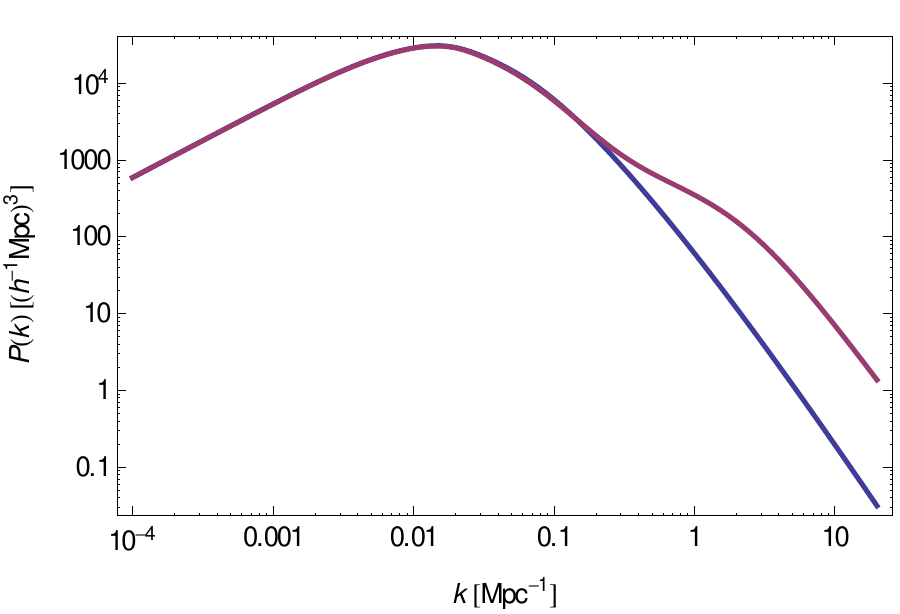}
\includegraphics[width=9cm]{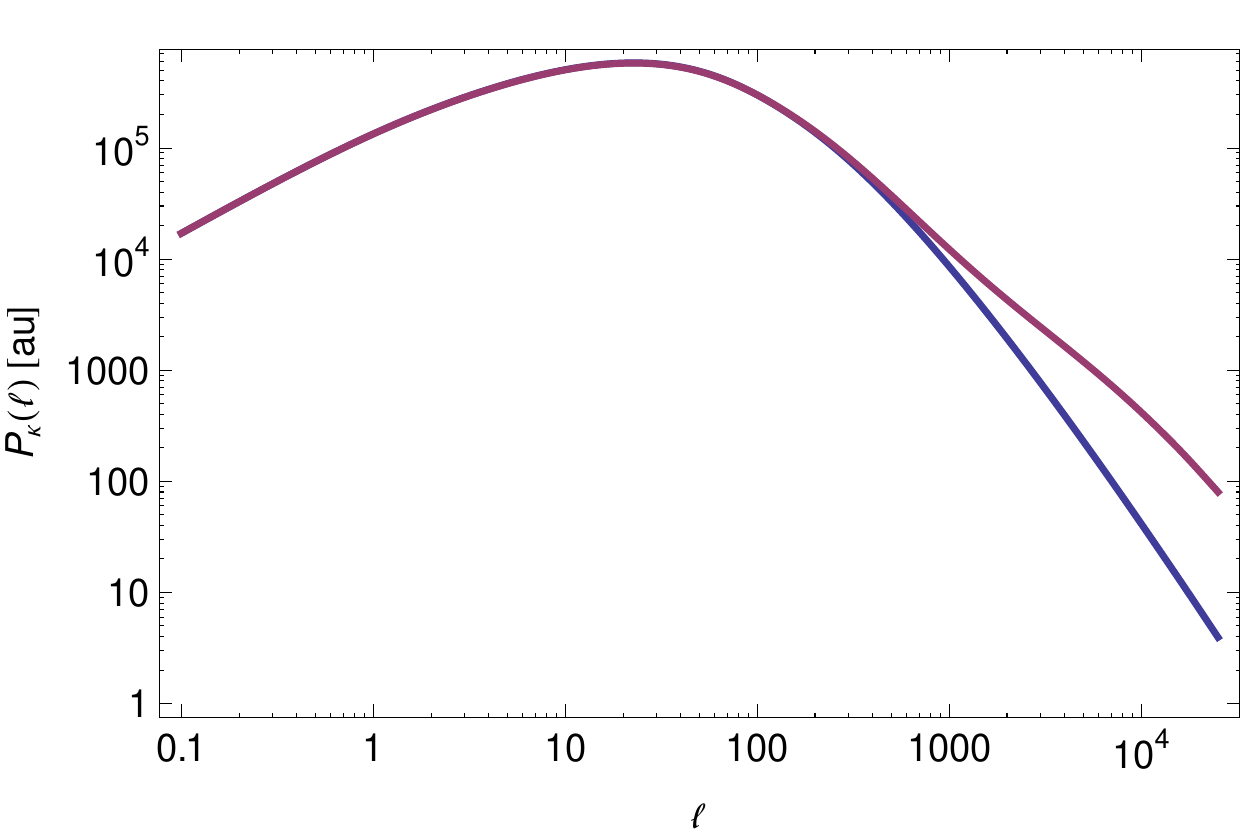}
\end{center}
\caption{\textit{Top panel:} Comparison between a matter power spectrum
with (purple) and without (blue) non-linear corrections in arbitrary units.
The fitting formulae were taken from \citet{Eisenstein1999} and
\citet{Smith2003}.
Non-linear corrections become important only at wave numbers
$k>0.1\unit{Mpc}^{-1}$. Note that $k$ is not in units of $h\unit{Mpc}^{-1}$
in this case. \textit{Bottom panel:} Comparison between a convergence
power spectrum from eq.~(\ref{eq:2-convps}) based on the linear and
non-linear matter power spectrum (no redshift binning).} \label{fig:2-ps}
\end{figure}

In fig.~\ref{fig:2-ps} we can see the matter power spectrum derived
from linear perturbation theory as well as non-linear corrections.
Fig.~\ref{fig:2-ps} shows the convergence power spectrum, derived from
the linear and non-linear matter power spectrum according to
eq.~(\ref{eq:2-convps}).

\section{Fisher matrix formalism} 

The last tool we will need is the powerful theory of Bayesian statistics.
In physics we define a theory by a set of parameters encapsulated in the
vector $\vec \theta$. The ultimate goal is to deduce the value of
$\vec\theta$ by performing a series of experiments that yield the data $\vec
x$. If we were given the values of $\vec \theta$, we could calculate the
probability density function $f(\vec x, \vec \theta)$ and thus predict the
probability $P$ of getting a realization of the data $\vec x$ given the
theory $\vec \theta$, which is usually denoted by $P(\vec x|\vec \theta)$.
But since we rarely are given the exact theory, all we can do is draw
samples $\vec x_i$ of an unknown PDF by conducting experiments in order to
estimate the parameters $\vec\theta$. In a {\textit{frequentist}} approach,
we would take those parameters as the true ones that maximize the so-called
\textit{likelihood function}
\begin{equation}
\mathcal L(\vec \theta,\vec x_i) \equiv \prod\limits_i f(\vec x_i,\vec
\theta)\,,  \label{eq:2-pdf}
\end{equation}
which is nothing but the joined PDF, with $\vec \theta$ now being
interpreted as a variable and $\vec x$ as a fixed parameter. But in
cosmology, we often have prior knowledge about the parameters from other
observations or theories that we would like to account for. For this, we
need Bayes' theorem:
\begin{equation} 
P(\vec \theta| \vec x,I) = \frac{ P(\vec \theta|I) P(\vec x|\vec
\theta,I)}{P(\vec x|I)}\,.
\label{eq:2-bayes}
\end{equation}
In this \textit{Bayesian} approach, we turn the situation around so that we
are asking the question: \textit{What is the probability of the theory,
given the observed data?}  Note that we included the background information
$I$ in every term even though it is hardly relevant to the derivation. It is
to remind ourselves that we always assume some sort of prior knowledge, even
when we do not realize it.  For instance, we might implicitly assume that
the Copernican principle holds true, or that the Universe is described by
the Einstein field equations.

Here the prior knowledge (or just \textit{prior}) is given by $P(\vec
\theta|I)$, and $P(\vec x|I)$ is the marginal probability of the data $\vec
x$ which acts as a normalization factor, requiring that $\int \mathcal
L(\vec \theta,\vec x_i)\dd^n  \theta = 1$.
$P(\vec x|\vec \theta,I)$ is identical to the likelihood, which is given
by the model. This could be for instance a Gaussian PDF
\begin{equation}
\prod\limits_{i=1}^n\frac{1}{\sigma_i \sqrt{2\pi}} \exp\left( -
\frac{(\hat{x}_i - x_i(\vec \theta))^2}{2\sigma_i^2} \right)\,.
\label{eq:2-gauss}
\end{equation}
The left hand side of
eq.~(\ref{eq:2-bayes}) is then called the
\textit{posterior}~\citep[p. 42]{Hobson2009a}. We now have all the
ingredients to calculate the posterior probability of the parameters given
the data. This quantity is important for computing the so-called evidence,
which is defined by the integral over the numerator in
eq.~(\ref{eq:2-bayes})
and needed for model selection.

Unfortunately, the likelihood is often not a simple Gaussian. Finding the
values of $\vec \theta$ that maximize the PDF given by
eq.~(\ref{eq:2-bayes}) can be computationally demanding, since a (na\"ive)
algorithm for finding the extremum scales exponentially with the number of
dimensions of the parameter space. This will becomes a problem when
considering 30 parameters as we will later in this thesis. It is not unusual
for Bayesian statistics to suffer from the ``curse of
dimensionality''\citep{Bellman2003}. One way to overcome this curse is by
using Markov chain Monte Carlo methods. In the case of
a likelihood that is not a multivariate Gaussian, we can simply assume that
it can at least be approximated by one. This is not an unreasonable
assumption, since the logarithm of the likelihood can always be Taylor
expanded up to the second order around its maximum, denoted by $\hat{\vec
\theta}$:
\begin{equation}
\ln\mathcal L(\vec \theta) \approx \ln \mathcal L(\hat{\vec\theta}) +
\frac{1}{2}\sum\limits_{i,j} \left.\frac{\partial^2 \ln \mathcal L(\vec
\theta)}{\partial\theta_i \partial\theta_j}\right|_{\vec
\theta=\hat{\vec\theta}}(\theta_i-\hat\theta_i)(\theta_j-\hat\theta_j)\,.
\label{}
\end{equation}
No first order terms appear since they vanish in a maximum by definition.
It is now quite useful to define the \textit{Fisher matrix} as the negative
of the Hessian of the likelihood,
\begin{equation} 
F_{ij} \equiv -\left.
\frac{\partial^2\ln\mathcal L(\vec
\theta)}{\partial\theta_i\partial\theta_j}\right|_{\vec
\theta=\hat{\vec\theta}}\,.\label{eq:2-fm}
\end{equation}
This definition is not very useful when we need to find the
maximum of a multi-dimensional function and then compute the derivatives
numerically
anyway, since many expensive evaluations is what we wanted to avoid in the
first place.  However, since this thesis deals with constraints by
\textit{future} surveys, using
some fiducial model which we already know the expected outcome of and thus
the peaks of the likelihood.  This is the reason why the Fisher matrix is a
perfect tool in assessing the merit of future cosmological surveys. 

Let us consider a survey in which we measure some set of observables $\vec
x$ (along with their respective standard errors $\vec \sigma$)
whose theoretical values $\hat{\vec x}(\vec \theta)$ depend on our model.
These observables could be the same quantity at different redshifts, i.e.
$x_i = x(z_i)$, or different quantities all together, e.g. $\vec
x=(z_\mathrm{CMB},c_s,\Omega_k)$, or a mixture of both. Assuming a Gaussian
PDF we can write down the likelihood as 
\begin{equation}
\mathcal L(\vec\theta) \propto \exp\left(-\frac12 \sum\limits_i
\frac{(x_i-\hat{ x_i}(\vec\theta))^2}{\sigma_i^2}\right)\,.
\end{equation}
Using the definition in eq.~(\ref{eq:2-fm}) we get for the Fisher matrix
\begin{equation}
F_{ij} = - \sum\limits_k \left.\frac{1}{\sigma_k^2}\frac{\partial^2 \hat
x_k(\vec \theta)}{\partial\theta_i\partial\theta_j}\right|_{\vec
\theta=\hat{\vec\theta}} \,.
\end{equation}

\subsection{Calculation rules}
\label{calcrules}

We will not go into a full derivation of the rules, as they can be found
in~A\&T~(ch. 13.3) and are quite straight forward, but rather simply state
them here.

\paragraph{Fixing a parameter.} If we want to know what the Fisher matrix
would be given that we knew one particular parameter $\theta_i$ precisely,
we simply remove the $i$-th row and column of the Fisher matrix.

\paragraph{Marginalizing over a parameter.} If, on the other hand, we want
to disregard a particular parameter $\theta_i$, we remove $i$-th row
and column from the inverse of the Fisher matrix (the so-called
\textit{correlation matrix}) and invert again afterwards. If we are only
interested in exactly one parameter $\theta_i$, then we cross out all other
rows and columns until the correlation matrix only has one entry left. Thus,
we arrive at the important result
\begin{equation}
\sigma(\theta_i)^2= (\mat F^{-1})_{ii}\,.
\end{equation}
This implies that the Fisher matrix must be positive definite, as it must be
as the Hessian in a maximum.

\paragraph{Combination of Fisher matrices.} Including priors in our analysis
is extremely simple in the Fisher matrix formalism, since all we need to do
is add the Fisher matrix:
\begin{equation}
\mat F' = \mat F + \mat F^\mathrm{prior}\,.
\end{equation}
This only holds if both matrices have been calculated with the same
underlying fiducial model, i.e. the maximum likelihood is the same. The only
difficulty lies in ensuring that the parameters of the model belonging to
each matrix are identical and line up properly. If one matrix covers
additional parameters, then the other matrix must be extended with rows and
columns of zeros accordingly.

\paragraph{Parameter transformation.} Often a particular parameterization of
a model is not unique and there exists a transformation of parameters
\begin{equation}
\vec \theta' = \vec\theta'(\vec \theta)\,,
\end{equation}
which might happen when combining Fisher matrices from different sources.
Then the Fisher matrix transforms like a tensor, i.e.
\begin{equation}
\mat F' = \mat J^T \mat F \mat J\,,
\end{equation}
where $\mat J$ is the Jacobian matrix
\begin{equation}
J_{ij} = \frac{\partial \theta_i}{\partial \theta_j'}\,,
\end{equation}
which does not necessarily need to be a square matrix. If it is not,
however, note that the new Fisher matrix will be degenerate and using it
only makes sense when combining it with at least one matrix from a different
source.

%% file: graphics/lensing-description.tex
\ifx\du\undefined
  \newlength{\du}
\fi
\setlength{\du}{15\unitlength}
\begin{tikzpicture}
\pgftransformxscale{1.000000}
\pgftransformyscale{-1.000000}
\definecolor{dialinecolor}{rgb}{0.000000, 0.000000, 0.000000}
\pgfsetstrokecolor{dialinecolor}
\definecolor{dialinecolor}{rgb}{1.000000, 1.000000, 1.000000}
\pgfsetfillcolor{dialinecolor}
\pgfsetlinewidth{0.100000\du}
\pgfsetdash{}{0pt}
\pgfsetdash{}{0pt}
\pgfsetmiterjoin
\pgfsetbuttcap
\definecolor{dialinecolor}{rgb}{1.000000, 1.000000, 1.000000}
\pgfsetfillcolor{dialinecolor}
\fill (14.000000\du,5.000000\du)--(19.000000\du,8.000000\du)--(19.000000\du,19.000000\du)--(14.000000\du,16.000000\du)--cycle;
\definecolor{dialinecolor}{rgb}{0.000000, 0.000000, 0.000000}
\pgfsetstrokecolor{dialinecolor}
\draw (14.000000\du,5.000000\du)--(19.000000\du,8.000000\du)--(19.000000\du,19.000000\du)--(14.000000\du,16.000000\du)--cycle;
\pgfsetlinewidth{0.100000\du}
\pgfsetdash{}{0pt}
\pgfsetdash{}{0pt}
\pgfsetmiterjoin
\pgfsetbuttcap
\definecolor{dialinecolor}{rgb}{1.000000, 1.000000, 1.000000}
\pgfsetfillcolor{dialinecolor}
\fill (29.000000\du,5.000000\du)--(34.000000\du,8.000000\du)--(34.000000\du,19.000000\du)--(29.000000\du,16.000000\du)--cycle;
\definecolor{dialinecolor}{rgb}{0.000000, 0.000000, 0.000000}
\pgfsetstrokecolor{dialinecolor}
\draw (29.000000\du,5.000000\du)--(34.000000\du,8.000000\du)--(34.000000\du,19.000000\du)--(29.000000\du,16.000000\du)--cycle;
\pgfsetlinewidth{0.100000\du}
\pgfsetdash{}{0pt}
\pgfsetdash{}{0pt}
\pgfsetbuttcap
{
\definecolor{dialinecolor}{rgb}{0.000000, 0.000000, 0.000000}
\pgfsetfillcolor{dialinecolor}
\definecolor{dialinecolor}{rgb}{0.000000, 0.000000, 0.000000}
\pgfsetstrokecolor{dialinecolor}
\draw (2.000000\du,12.000000\du)--(18.000000\du,16.000000\du);
}
\pgfsetlinewidth{0.100000\du}
\pgfsetdash{{1.000000\du}{1.000000\du}}{0\du}
\pgfsetdash{{0.300000\du}{0.300000\du}}{0\du}
\pgfsetbuttcap
{
\definecolor{dialinecolor}{rgb}{0.000000, 0.000000, 0.000000}
\pgfsetfillcolor{dialinecolor}
\definecolor{dialinecolor}{rgb}{0.000000, 0.000000, 0.000000}
\pgfsetstrokecolor{dialinecolor}
\draw (18.980581\du,16.196116\du)--(28.000000\du,18.000000\du);
}
\definecolor{dialinecolor}{rgb}{0.000000, 0.000000, 0.000000}
\pgfsetstrokecolor{dialinecolor}
\node[anchor=west] at (1.000000\du,11.000000\du){Observer};
\definecolor{dialinecolor}{rgb}{0.000000, 0.000000, 0.000000}
\pgfsetstrokecolor{dialinecolor}
\node[anchor=west] at (18.000000\du,6.000000\du){Lens plane};
\definecolor{dialinecolor}{rgb}{0.000000, 0.000000, 0.000000}
\pgfsetstrokecolor{dialinecolor}
\node[anchor=west] at (32.000000\du,6.000000\du){Source plane};
\pgfsetlinewidth{0.100000\du}
\pgfsetdash{{\pgflinewidth}{0.060000\du}}{0cm}
\pgfsetdash{{\pgflinewidth}{0.200000\du}}{0cm}
\pgfsetbuttcap
{
\definecolor{dialinecolor}{rgb}{0.000000, 0.000000, 0.000000}
\pgfsetfillcolor{dialinecolor}
\definecolor{dialinecolor}{rgb}{0.000000, 0.000000, 0.000000}
\pgfsetstrokecolor{dialinecolor}
\draw (2.000000\du,12.000000\du)--(17.074442\du,13.507444\du);
}
\pgfsetlinewidth{0.100000\du}
\pgfsetdash{}{0pt}
\pgfsetdash{}{0pt}
\pgfsetbuttcap
{
\definecolor{dialinecolor}{rgb}{0.000000, 0.000000, 0.000000}
\pgfsetfillcolor{dialinecolor}
\pgfsetarrowsstart{stealth}
\pgfsetarrowsend{stealth}
\definecolor{dialinecolor}{rgb}{0.000000, 0.000000, 0.000000}
\pgfsetstrokecolor{dialinecolor}
\pgfpathmoveto{\pgfpoint{23.490228\du}{17.098160\du}}
\pgfpatharc{50}{-42}{1.034542\du and 1.034542\du}
\pgfusepath{stroke}
}
\pgfsetlinewidth{0.100000\du}
\pgfsetdash{}{0pt}
\pgfsetdash{}{0pt}
\pgfsetbuttcap
{
\definecolor{dialinecolor}{rgb}{0.000000, 0.000000, 0.000000}
\pgfsetfillcolor{dialinecolor}
\pgfsetarrowsstart{stealth}
\pgfsetarrowsend{stealth}
\definecolor{dialinecolor}{rgb}{0.000000, 0.000000, 0.000000}
\pgfsetstrokecolor{dialinecolor}
\pgfpathmoveto{\pgfpoint{11.399874\du}{14.350109\du}}
\pgfpatharc{50}{-32}{1.827551\du and 1.827551\du}
\pgfusepath{stroke}
}
\pgfsetlinewidth{0.100000\du}
\pgfsetdash{}{0pt}
\pgfsetdash{}{0pt}
\pgfsetbuttcap
{
\definecolor{dialinecolor}{rgb}{0.000000, 0.000000, 0.000000}
\pgfsetfillcolor{dialinecolor}
\pgfsetarrowsstart{stealth}
\pgfsetarrowsend{stealth}
\definecolor{dialinecolor}{rgb}{0.000000, 0.000000, 0.000000}
\pgfsetstrokecolor{dialinecolor}
\pgfpathmoveto{\pgfpoint{21.999913\du}{13.950085\du}}
\pgfpatharc{46}{-45}{1.328125\du and 1.328125\du}
\pgfusepath{stroke}
}
\pgfsetlinewidth{0.100000\du}
\pgfsetdash{}{0pt}
\pgfsetdash{}{0pt}
\pgfsetbuttcap
{
\definecolor{dialinecolor}{rgb}{0.000000, 0.000000, 0.000000}
\pgfsetfillcolor{dialinecolor}
\pgfsetarrowsstart{stealth}
\pgfsetarrowsend{stealth}
\definecolor{dialinecolor}{rgb}{0.000000, 0.000000, 0.000000}
\pgfsetstrokecolor{dialinecolor}
\draw (2.000000\du,20.000000\du)--(16.500000\du,20.000000\du);
}
\pgfsetlinewidth{0.100000\du}
\pgfsetdash{}{0pt}
\pgfsetdash{}{0pt}
\pgfsetbuttcap
{
\definecolor{dialinecolor}{rgb}{0.000000, 0.000000, 0.000000}
\pgfsetfillcolor{dialinecolor}
\pgfsetarrowsstart{stealth}
\pgfsetarrowsend{stealth}
\definecolor{dialinecolor}{rgb}{0.000000, 0.000000, 0.000000}
\pgfsetstrokecolor{dialinecolor}
\draw (16.500000\du,20.000000\du)--(31.500000\du,20.000000\du);
}
\pgfsetlinewidth{0.100000\du}
\pgfsetdash{}{0pt}
\pgfsetdash{}{0pt}
\pgfsetbuttcap
{
\definecolor{dialinecolor}{rgb}{0.000000, 0.000000, 0.000000}
\pgfsetfillcolor{dialinecolor}
\pgfsetarrowsstart{stealth}
\pgfsetarrowsend{stealth}
\definecolor{dialinecolor}{rgb}{0.000000, 0.000000, 0.000000}
\pgfsetstrokecolor{dialinecolor}
\draw (2.000000\du,23.000000\du)--(31.450000\du,22.950000\du);
}
\definecolor{dialinecolor}{rgb}{0.000000, 0.000000, 0.000000}
\pgfsetstrokecolor{dialinecolor}
\node[anchor=west] at (16.450000\du,23.700000\du){$r_s$};
\definecolor{dialinecolor}{rgb}{0.000000, 0.000000, 0.000000}
\pgfsetstrokecolor{dialinecolor}
\node[anchor=west] at (9.000000\du,20.000000\du){};
\definecolor{dialinecolor}{rgb}{0.000000, 0.000000, 0.000000}
\pgfsetstrokecolor{dialinecolor}
\node[anchor=west] at (9.000000\du,21.000000\du){$r_l$};
\definecolor{dialinecolor}{rgb}{0.000000, 0.000000, 0.000000}
\pgfsetstrokecolor{dialinecolor}
\node[anchor=west] at (24.000000\du,21.000000\du){$r_{ls}$};
\pgfsetlinewidth{0.100000\du}
\pgfsetdash{{1.000000\du}{1.000000\du}}{0\du}
\pgfsetdash{{0.300000\du}{0.300000\du}}{0\du}
\pgfsetbuttcap
{
\definecolor{dialinecolor}{rgb}{0.498039, 0.498039, 0.498039}
\pgfsetfillcolor{dialinecolor}
\definecolor{dialinecolor}{rgb}{0.498039, 0.498039, 0.498039}
\pgfsetstrokecolor{dialinecolor}
\draw (16.500000\du,6.500000\du)--(16.500000\du,17.500000\du);
}
\pgfsetlinewidth{0.100000\du}
\pgfsetdash{{0.300000\du}{0.300000\du}}{0\du}
\pgfsetdash{{0.300000\du}{0.300000\du}}{0\du}
\pgfsetbuttcap
{
\definecolor{dialinecolor}{rgb}{0.498039, 0.498039, 0.498039}
\pgfsetfillcolor{dialinecolor}
\definecolor{dialinecolor}{rgb}{0.498039, 0.498039, 0.498039}
\pgfsetstrokecolor{dialinecolor}
\draw (14.000000\du,10.500000\du)--(19.000000\du,13.500000\du);
}
\pgfsetlinewidth{0.100000\du}
\pgfsetdash{{0.300000\du}{0.300000\du}}{0\du}
\pgfsetdash{{0.300000\du}{0.300000\du}}{0\du}
\pgfsetbuttcap
{
\definecolor{dialinecolor}{rgb}{0.498039, 0.498039, 0.498039}
\pgfsetfillcolor{dialinecolor}
\definecolor{dialinecolor}{rgb}{0.498039, 0.498039, 0.498039}
\pgfsetstrokecolor{dialinecolor}
\draw (31.500000\du,6.500000\du)--(31.500000\du,17.500000\du);
}
\pgfsetlinewidth{0.100000\du}
\pgfsetdash{{0.300000\du}{0.300000\du}}{0\du}
\pgfsetdash{{0.300000\du}{0.300000\du}}{0\du}
\pgfsetbuttcap
{
\definecolor{dialinecolor}{rgb}{0.498039, 0.498039, 0.498039}
\pgfsetfillcolor{dialinecolor}
\definecolor{dialinecolor}{rgb}{0.498039, 0.498039, 0.498039}
\pgfsetstrokecolor{dialinecolor}
\draw (29.000000\du,10.500000\du)--(34.000000\du,13.500000\du);
}
\definecolor{dialinecolor}{rgb}{0.498039, 0.498039, 0.498039}
\pgfsetstrokecolor{dialinecolor}
\node[anchor=west] at (29.000000\du,11.000000\du){};
\definecolor{dialinecolor}{rgb}{0.498039, 0.498039, 0.498039}
\pgfsetstrokecolor{dialinecolor}
\node[anchor=west] at (16.500000\du,8.000000\du){$x_1$};
\definecolor{dialinecolor}{rgb}{0.498039, 0.498039, 0.498039}
\pgfsetstrokecolor{dialinecolor}
\node[anchor=west] at (19.000000\du,13.000000\du){};
\definecolor{dialinecolor}{rgb}{0.498039, 0.498039, 0.498039}
\pgfsetstrokecolor{dialinecolor}
\node[anchor=west] at (17.500000\du,12.500000\du){$x_2$};
\definecolor{dialinecolor}{rgb}{0.498039, 0.498039, 0.498039}
\pgfsetstrokecolor{dialinecolor}
\node[anchor=west] at (31.500000\du,8.000000\du){$x_1$};
\definecolor{dialinecolor}{rgb}{0.498039, 0.498039, 0.498039}
\pgfsetstrokecolor{dialinecolor}
\node[anchor=west] at (32.500000\du,12.500000\du){$x_2$};
\definecolor{dialinecolor}{rgb}{0.000000, 0.000000, 0.000000}
\pgfsetstrokecolor{dialinecolor}
\node[anchor=west] at (11.855100\du,13.975000\du){$\vec\theta_0$};
\definecolor{dialinecolor}{rgb}{0.000000, 0.000000, 0.000000}
\pgfsetstrokecolor{dialinecolor}
\node[anchor=west] at (23.900000\du,16.575000\du){$\vec\alpha$};
\definecolor{dialinecolor}{rgb}{0.000000, 0.000000, 0.000000}
\pgfsetstrokecolor{dialinecolor}
\node[anchor=west] at (22.480100\du,13.225000\du){$\vec\theta_s$};
\pgfsetlinewidth{0.100000\du}
\pgfsetdash{}{0pt}
\pgfsetdash{}{0pt}
\pgfsetbuttcap
{
\definecolor{dialinecolor}{rgb}{0.000000, 0.000000, 0.000000}
\pgfsetfillcolor{dialinecolor}
\definecolor{dialinecolor}{rgb}{0.000000, 0.000000, 0.000000}
\pgfsetstrokecolor{dialinecolor}
\draw (18.997785\du,15.933481\du)--(33.000000\du,15.000000\du);
}
\pgfsetlinewidth{0.100000\du}
\pgfsetdash{{\pgflinewidth}{0.200000\du}}{0cm}
\pgfsetdash{{\pgflinewidth}{0.200000\du}}{0cm}
\pgfsetbuttcap
{
\definecolor{dialinecolor}{rgb}{0.000000, 0.000000, 0.000000}
\pgfsetfillcolor{dialinecolor}
\definecolor{dialinecolor}{rgb}{0.000000, 0.000000, 0.000000}
\pgfsetstrokecolor{dialinecolor}
\draw (19.020485\du,13.647144\du)--(33.000000\du,15.000000\du);
}
\pgfsetlinewidth{0.100000\du}
\pgfsetdash{}{0pt}
\pgfsetdash{}{0pt}
\pgfsetbuttcap
{
\definecolor{dialinecolor}{rgb}{0.000000, 0.000000, 0.000000}
\pgfsetfillcolor{dialinecolor}
}
\definecolor{dialinecolor}{rgb}{0.000000, 0.000000, 0.000000}
\pgfsetstrokecolor{dialinecolor}
\draw (31.500000\du,12.000000\du)--(33.000000\du,15.000000\du);
\pgfsetlinewidth{0.100000\du}
\pgfsetdash{}{0pt}
\pgfsetmiterjoin
\pgfsetbuttcap
\definecolor{dialinecolor}{rgb}{0.000000, 0.000000, 0.000000}
\pgfsetfillcolor{dialinecolor}
\pgfpathmoveto{\pgfpoint{31.500000\du}{12.000000\du}}
\pgfpathcurveto{\pgfpoint{31.522361\du}{11.988820\du}}{\pgfpoint{31.555902\du}{12.000000\du}}{\pgfpoint{31.567082\du}{12.022361\du}}
\pgfpathcurveto{\pgfpoint{31.578262\du}{12.044721\du}}{\pgfpoint{31.567082\du}{12.078262\du}}{\pgfpoint{31.544721\du}{12.089443\du}}
\pgfpathcurveto{\pgfpoint{31.522361\du}{12.100623\du}}{\pgfpoint{31.488820\du}{12.089443\du}}{\pgfpoint{31.477639\du}{12.067082\du}}
\pgfpathcurveto{\pgfpoint{31.466459\du}{12.044721\du}}{\pgfpoint{31.477639\du}{12.011180\du}}{\pgfpoint{31.500000\du}{12.000000\du}}
\pgfusepath{fill}
\definecolor{dialinecolor}{rgb}{0.000000, 0.000000, 0.000000}
\pgfsetstrokecolor{dialinecolor}
\pgfpathmoveto{\pgfpoint{31.500000\du}{12.000000\du}}
\pgfpathcurveto{\pgfpoint{31.522361\du}{11.988820\du}}{\pgfpoint{31.555902\du}{12.000000\du}}{\pgfpoint{31.567082\du}{12.022361\du}}
\pgfpathcurveto{\pgfpoint{31.578262\du}{12.044721\du}}{\pgfpoint{31.567082\du}{12.078262\du}}{\pgfpoint{31.544721\du}{12.089443\du}}
\pgfpathcurveto{\pgfpoint{31.522361\du}{12.100623\du}}{\pgfpoint{31.488820\du}{12.089443\du}}{\pgfpoint{31.477639\du}{12.067082\du}}
\pgfpathcurveto{\pgfpoint{31.466459\du}{12.044721\du}}{\pgfpoint{31.477639\du}{12.011180\du}}{\pgfpoint{31.500000\du}{12.000000\du}}
\pgfusepath{stroke}
\pgfsetlinewidth{0.100000\du}
\pgfsetdash{}{0pt}
\pgfsetmiterjoin
\pgfsetbuttcap
\definecolor{dialinecolor}{rgb}{0.000000, 0.000000, 0.000000}
\pgfsetfillcolor{dialinecolor}
\pgfpathmoveto{\pgfpoint{33.000000\du}{15.000000\du}}
\pgfpathcurveto{\pgfpoint{32.977639\du}{15.011180\du}}{\pgfpoint{32.944098\du}{15.000000\du}}{\pgfpoint{32.932918\du}{14.977639\du}}
\pgfpathcurveto{\pgfpoint{32.921738\du}{14.955279\du}}{\pgfpoint{32.932918\du}{14.921738\du}}{\pgfpoint{32.955279\du}{14.910557\du}}
\pgfpathcurveto{\pgfpoint{32.977639\du}{14.899377\du}}{\pgfpoint{33.011180\du}{14.910557\du}}{\pgfpoint{33.022361\du}{14.932918\du}}
\pgfpathcurveto{\pgfpoint{33.033541\du}{14.955279\du}}{\pgfpoint{33.022361\du}{14.988820\du}}{\pgfpoint{33.000000\du}{15.000000\du}}
\pgfusepath{fill}
\definecolor{dialinecolor}{rgb}{0.000000, 0.000000, 0.000000}
\pgfsetstrokecolor{dialinecolor}
\pgfpathmoveto{\pgfpoint{33.000000\du}{15.000000\du}}
\pgfpathcurveto{\pgfpoint{32.977639\du}{15.011180\du}}{\pgfpoint{32.944098\du}{15.000000\du}}{\pgfpoint{32.932918\du}{14.977639\du}}
\pgfpathcurveto{\pgfpoint{32.921738\du}{14.955279\du}}{\pgfpoint{32.932918\du}{14.921738\du}}{\pgfpoint{32.955279\du}{14.910557\du}}
\pgfpathcurveto{\pgfpoint{32.977639\du}{14.899377\du}}{\pgfpoint{33.011180\du}{14.910557\du}}{\pgfpoint{33.022361\du}{14.932918\du}}
\pgfpathcurveto{\pgfpoint{33.033541\du}{14.955279\du}}{\pgfpoint{33.022361\du}{14.988820\du}}{\pgfpoint{33.000000\du}{15.000000\du}}
\pgfusepath{stroke}
\pgfsetlinewidth{0.100000\du}
\pgfsetdash{}{0pt}
\pgfsetdash{}{0pt}
\pgfsetbuttcap
{
\definecolor{dialinecolor}{rgb}{0.000000, 0.000000, 0.000000}
\pgfsetfillcolor{dialinecolor}
}
\definecolor{dialinecolor}{rgb}{0.000000, 0.000000, 0.000000}
\pgfsetstrokecolor{dialinecolor}
\draw (16.500000\du,12.000000\du)--(18.000000\du,16.000000\du);
\pgfsetlinewidth{0.100000\du}
\pgfsetdash{}{0pt}
\pgfsetmiterjoin
\pgfsetbuttcap
\definecolor{dialinecolor}{rgb}{0.000000, 0.000000, 0.000000}
\pgfsetfillcolor{dialinecolor}
\pgfpathmoveto{\pgfpoint{16.500000\du}{12.000000\du}}
\pgfpathcurveto{\pgfpoint{16.523408\du}{11.991222\du}}{\pgfpoint{16.555595\du}{12.005852\du}}{\pgfpoint{16.564373\du}{12.029260\du}}
\pgfpathcurveto{\pgfpoint{16.573151\du}{12.052669\du}}{\pgfpoint{16.558521\du}{12.084855\du}}{\pgfpoint{16.535112\du}{12.093633\du}}
\pgfpathcurveto{\pgfpoint{16.511704\du}{12.102411\du}}{\pgfpoint{16.479518\du}{12.087781\du}}{\pgfpoint{16.470740\du}{12.064373\du}}
\pgfpathcurveto{\pgfpoint{16.461962\du}{12.040964\du}}{\pgfpoint{16.476592\du}{12.008778\du}}{\pgfpoint{16.500000\du}{12.000000\du}}
\pgfusepath{fill}
\definecolor{dialinecolor}{rgb}{0.000000, 0.000000, 0.000000}
\pgfsetstrokecolor{dialinecolor}
\pgfpathmoveto{\pgfpoint{16.500000\du}{12.000000\du}}
\pgfpathcurveto{\pgfpoint{16.523408\du}{11.991222\du}}{\pgfpoint{16.555595\du}{12.005852\du}}{\pgfpoint{16.564373\du}{12.029260\du}}
\pgfpathcurveto{\pgfpoint{16.573151\du}{12.052669\du}}{\pgfpoint{16.558521\du}{12.084855\du}}{\pgfpoint{16.535112\du}{12.093633\du}}
\pgfpathcurveto{\pgfpoint{16.511704\du}{12.102411\du}}{\pgfpoint{16.479518\du}{12.087781\du}}{\pgfpoint{16.470740\du}{12.064373\du}}
\pgfpathcurveto{\pgfpoint{16.461962\du}{12.040964\du}}{\pgfpoint{16.476592\du}{12.008778\du}}{\pgfpoint{16.500000\du}{12.000000\du}}
\pgfusepath{stroke}
\pgfsetlinewidth{0.100000\du}
\pgfsetdash{}{0pt}
\pgfsetmiterjoin
\pgfsetbuttcap
\definecolor{dialinecolor}{rgb}{0.000000, 0.000000, 0.000000}
\pgfsetfillcolor{dialinecolor}
\pgfpathmoveto{\pgfpoint{18.000000\du}{16.000000\du}}
\pgfpathcurveto{\pgfpoint{17.976592\du}{16.008778\du}}{\pgfpoint{17.944405\du}{15.994148\du}}{\pgfpoint{17.935627\du}{15.970740\du}}
\pgfpathcurveto{\pgfpoint{17.926849\du}{15.947331\du}}{\pgfpoint{17.941479\du}{15.915145\du}}{\pgfpoint{17.964888\du}{15.906367\du}}
\pgfpathcurveto{\pgfpoint{17.988296\du}{15.897589\du}}{\pgfpoint{18.020482\du}{15.912219\du}}{\pgfpoint{18.029260\du}{15.935627\du}}
\pgfpathcurveto{\pgfpoint{18.038038\du}{15.959036\du}}{\pgfpoint{18.023408\du}{15.991222\du}}{\pgfpoint{18.000000\du}{16.000000\du}}
\pgfusepath{fill}
\definecolor{dialinecolor}{rgb}{0.000000, 0.000000, 0.000000}
\pgfsetstrokecolor{dialinecolor}
\pgfpathmoveto{\pgfpoint{18.000000\du}{16.000000\du}}
\pgfpathcurveto{\pgfpoint{17.976592\du}{16.008778\du}}{\pgfpoint{17.944405\du}{15.994148\du}}{\pgfpoint{17.935627\du}{15.970740\du}}
\pgfpathcurveto{\pgfpoint{17.926849\du}{15.947331\du}}{\pgfpoint{17.941479\du}{15.915145\du}}{\pgfpoint{17.964888\du}{15.906367\du}}
\pgfpathcurveto{\pgfpoint{17.988296\du}{15.897589\du}}{\pgfpoint{18.020482\du}{15.912219\du}}{\pgfpoint{18.029260\du}{15.935627\du}}
\pgfpathcurveto{\pgfpoint{18.038038\du}{15.959036\du}}{\pgfpoint{18.023408\du}{15.991222\du}}{\pgfpoint{18.000000\du}{16.000000\du}}
\pgfusepath{stroke}
\pgfsetlinewidth{0.100000\du}
\pgfsetdash{}{0pt}
\pgfsetdash{}{0pt}
\pgfsetbuttcap
{
\definecolor{dialinecolor}{rgb}{0.000000, 0.000000, 0.000000}
\pgfsetfillcolor{dialinecolor}
\definecolor{dialinecolor}{rgb}{0.000000, 0.000000, 0.000000}
\pgfsetstrokecolor{dialinecolor}
\draw (19.000000\du,12.000000\du)--(31.500000\du,12.000000\du);
}
\pgfsetlinewidth{0.100000\du}
\pgfsetdash{}{0pt}
\pgfsetdash{}{0pt}
\pgfsetbuttcap
{
\definecolor{dialinecolor}{rgb}{0.000000, 0.000000, 0.000000}
\pgfsetfillcolor{dialinecolor}
\definecolor{dialinecolor}{rgb}{0.000000, 0.000000, 0.000000}
\pgfsetstrokecolor{dialinecolor}
\draw (16.500000\du,12.000000\du)--(2.000000\du,12.000000\du);
}
\node[anchor=west] at (32.480100\du,15.225000\du)
{\includegraphics[width=1cm,height=.8cm]{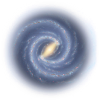}};
\end{tikzpicture}

%% file: chapter3.tex
\chapter[Constraints by future surveys]{Constraints on cosmological
parameters by future dark energy surveys}

With the knowledge of the observational specifications of future weak
lensing surveys, we can use the Fisher matrix formalism to forecast the
errors that said surveys will place on cosmological parameters. In
particular, we want to expand the list of those parameters with values of
the Hubble parameter and growth function at certain redshifts. An invaluable
resource for this chapter is the book by A\&T (ch. 14)

\section{Power spectrum tomography} \label{sec:tomography}

\begin{figure}[htpb]
\begin{center}
\includegraphics[width=9cm]{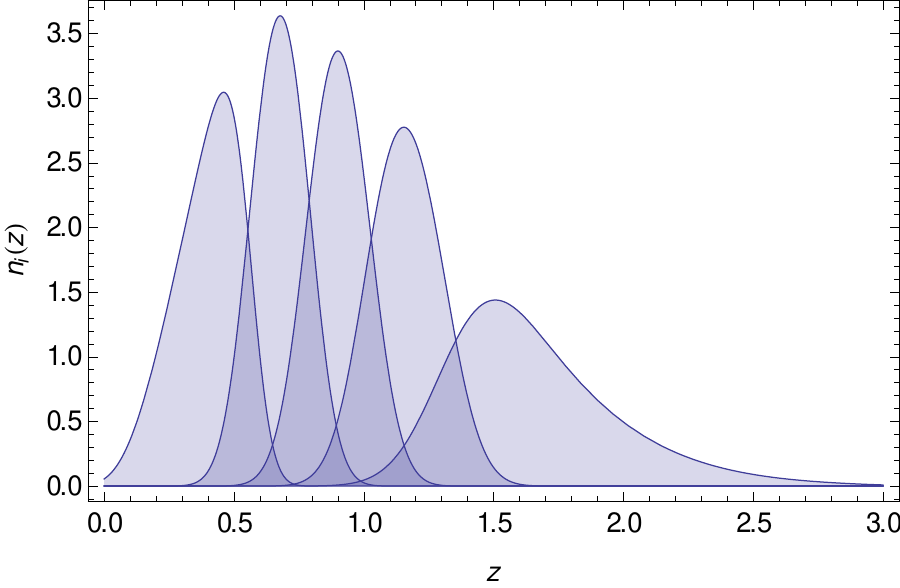}
\end{center}
\caption{Normalized galaxy density functions for each bin with $\calN=5$,
convolved with a Gaussian to account for photometric redshift measurement
errors.} \label{fig:3-ngal}
\end{figure}

It has been shown that dividing up the distribution of lensed galaxies into
redshift bins and measuring the convergence power spectrum in each of these
bins as well as their cross-correlation can increase the amount of
information extracted from weak lensing surveys \citep{Hu1999,Huterer2002}.
This means on one hand that we need to have additional redshift information
on the lensed galaxies, but on the other that we are possibly rewarded with
gaining knowledge about the evolution of dark energy parameters.  Note that
the purpose of the redshift bins in this thesis is two-fold, and power
spectrum tomography is only one of them. The other is the linear
interpolation of $H(z)$ and $G(z)$, where the centers of the redshift bins
act as supporting points, making our analysis independent of assumptions
about the growth function by any particular model. We choose to divide the
redshift space into $\calN$ bins such that each redshift bin contains
roughly the same amount of galaxies according to the galaxy density function
$n(z)$, i.e.
\begin{equation}
\int_{z_{i-1}}^{z_{i}} n(z)\dd z = \frac1\calN
\int_0^{\infty}n(z)\dd z 
\end{equation}
for each $i$. We infer the values of the $z_i$ via a series
of successive numerical integrations with $z_0=0$ and $z_\calN=3$ (see
fig.~ref{fig:2-zbins}). 
A common parametrization of $n(z)$ is 
\begin{equation}
n(z) \propto z^\alpha e^{-\left( {z}/{z_p} \right)^\beta}, 
\end{equation}
with $\alpha=2$ and $\beta=\frac32$. Here $z_p$ is related to the median
redshift $z_m=1.412 z_p$ \citep{Amara2008}.
\begin{figure}[htpb]
\begin{center}
\includegraphics[width=9cm]{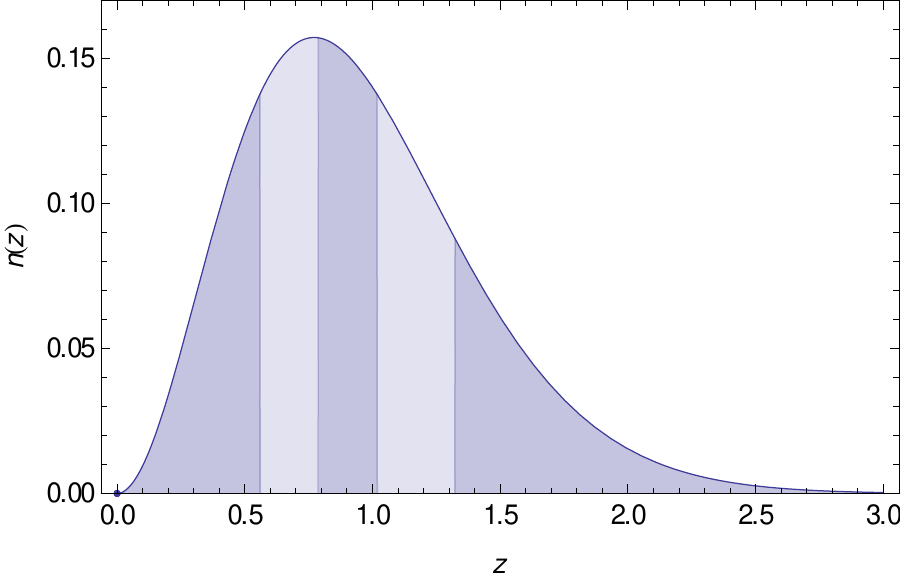}
\end{center}
\caption{The area under the curve of the galaxy density function
is the same in each redshift bin for $\calN=5$.}
\label{fig:2-zbins}
\end{figure}
We need a galaxy density function for each bin, and the naïve choice
would be
\begin{equation}
\hat n_i(z) = \left\{ 
\begin{array}{ll}
n(z) & z_{i-1}<z\leq z_i\,,\\
0 & \mathrm{else.}
\end{array}
\right.
\end{equation}
But to account for redshift measurement uncertainties, we will convolve 
$\hat n(z)$ with the probability distribution of the measured redshift
$z_\mathrm{ph}$ given the real value $z$:
\begin{equation}
n_i(z) = \int_0^\infty\dd  z'
\hat n_i( z') P_i(z_\mathrm{ph}= z'|z)  \,.
\end{equation}
Note that the integral vanishes if $ z'$ lies outside the
$i$th bin, so the limits can be adjusted to the according finite values.
This convolution reflects the fact that we cannot assign a galaxy to a
particular bin by measuring the photometric redshift with finite precision.
The probability distribution is here modeled by Gaussian, i.e.
\begin{equation}
P_i(z_\mathrm{ph}|z) = \frac{1}{\sqrt{2\pi}\sigma_i} \exp\left(-\frac{\left(
z_\mathrm{ph}-z \right)^2}{2 \sigma_i^2}\right)
\end{equation} 
with $\sigma_i = \Delta_z(1+(z_{i-1}+z_i)/2)$ (see \cite{Ma2006} for
details).  Finally, we normalize each function to unity, such that
\begin{equation}
\int_0^\infty n_i(z)\dd z = 1\,.
\end{equation}
The resulting functions can be visualized as in fig. \ref{fig:3-ngal}.

\section{The matter power spectrum}\label{sec:nonlinear}

Future weak lensing surveys will supply us with the convergence power
spectrum in which the cosmological information we are seeking is imprinted.
However, for now we will have to rely on simulated data. As we will see in
section~\ref{sec:convps}, the convergence power spectrum depends on the
matter power spectrum which can be simulated by a
fitting formula derived by \citet{Eisenstein1999}. Non-linear corrections
need to be accounted for, since we are assuming an angular galaxy density of
$35\unit{arcmin}^{-2}$, which corresponds to a multipole up to a order of
magnitude of
\begin{equation}
\lg\left(\sqrt{35 \unit{arcmin}^{-2}}\right)\approx 4\,.
\label{eq:3-ellmax}
\end{equation}
As we saw in fig.~\ref{fig:2-ps}, non-linear corrections are required for
$\ell\gtrsim10^3$, and for this we need to use the results by
\citet{Smith2003}.  Here we will give a brief overview of the essential
results found in the two papers cited above, using our notation and some
simplifications (no massive neutrinos, flat cosmology). 

\subsection{Fitting formulas}\label{subsec:fitting} 

It should be stressed that the growth function $D(z)$ in Eisenstein's paper
differs from ours by a factor of $a=1/(1+z)$, so $D(z) = a G(z)$. A precise
definition of the growth function will be given in
section.~\ref{sec:growth}. They also define $\Theta_{2.7} \equiv
T_\mathrm{CMB}/2.7\unit{K}$. Only in this section will we differentiate
between the linear and the non-linear power spectrum. Later on, the
expression ``power spectrum'' or $P_m(k)$ will always imply that non-linear
corrections have been applied.

Using the definition of the growth function, the matter power spectrum in
the linear regime can be written as (cf. eq.~(\ref{eq:3-mps2})) 
\begin{equation}
P_\mathrm{L}(k) \propto k^{n_s} \frac{G(z)^2}{(1+z)^2} T(z,k)^2\,,
\label{eq:3-mps1}
\end{equation}
where $n_s$ is called the scalar spectral index. The transfer function
$T(k)$ can now be fitted by a series of functions as follows.
\begin{equation}
T(k) = \frac{L B(k)}{L+C q_\mathrm{eff}^2}\,,
\end{equation}
with (we assume that there are no massive neutrinos, in which case $B(k)$
equals unity)
\begin{align}
L &= \ln(e+1.84\beta_c\sqrt{\alpha_\nu} q_\mathrm{eff})\\
C&= 1.44+\frac{325}{1+60.5 q_\mathrm{eff}^{1.11}}
\end{align}
where
\begin{align}
q_\mathrm{eff} &= \frac{k \Theta^2_{2.7}}{\Gamma_\mathrm{eff}
\unit{Mpc}^{-1}}\\
\Gamma_\mathrm{eff} &= \omega_m \left( \sqrt{\alpha_\nu} +
\frac{1-\sqrt{\alpha_\nu}}{1+(0.34ks)^4} \right)\\
\end{align}
and 
\begin{multline}
\alpha_\nu = \frac{\omega_c}{\omega_m}\frac{5-2(p_c +
p_m)}{5-4p_m} \left( 1-0.553 \omega_b + 0.126\omega_b^3\right)
(1+y_d)^{p_m-p_c}\\
\times \left\{ 1 + \frac{p_c-p_{cb}}{2}\left[
1+\frac{1}{(3-4p_c)(7-4p_m)} \right](1-y_d)^{-1} \right\}\,.
\end{multline}
Here, we used the abbreviations
\begin{align}
p_x &= \frac{1}{4}(5-\sqrt{1+24\omega_x})\\
y_d & = \frac{1+z_\mathrm{eq}}{1+z_d}
\end{align}
with
\begin{align}
z_\mathrm{eq} &= 2.50 \times 10^4 \omega_m \Theta_{2.7}^{-4}\\
z_d &= 1291 \frac{ \omega_m^{0.251}}{1+0.659\omega_m^{0.828}}
(1+b_1\omega_m^{b_2})\\
b_1 &= 0.313\omega_m^{-0.419}(1+0.607\omega_m^{0.674})\\
b_2 &= 0.238\omega_m^{0.223}\,.
\end{align}
By definition, the power spectrum needs to be normalized such that
\begin{equation}
\sigma_8^2 = \int_0^\infty \frac{\dd k}{k} \frac{k^3}{2\pi^2}
P_\mathrm{L}(k)|\tilde W_8(k)|^2\,,
\end{equation}
where $\tilde W_R(k)$ is the Fourier transform of the real-space window
function, in this case a spherical top hat of radius $R$ (in Mpc), i.e.
\begin{align}
W_R(r) &\propto \left\{
\begin{array}{ll}1& \mathrm{if}\:\:r \leq R\\ 
0 & \mathrm{otherwise}
\end{array} 
\right.\\
\tilde W_R(k) &= \frac{3}{(kR)^3}(\sin kR -kR \cos kR)\,.
\end{align}

\subsection{Non-linear corrections}

When applying non-linear corrections, we use the dimensionless power
spectrum, which is defined by
\begin{equation}
\Delta_\mathrm{L}^2(k) = \frac{4\pi k^3}{(2\pi)^3 } P_\mathrm{L}(k) \,,
\end{equation}
and similarly for other indices. It turns out that the non-linear power
spectrum can be decomposed into a quasi-linear term and contributions from
self-correlations
\begin{equation}
\Delta_\mathrm{NL}^2 = \Delta^2_\mathrm{Q}(k) + \Delta^2_\mathrm{H}(k)\,,
\end{equation}
which are given by
\begin{align}
\Delta_\mathrm{Q}^2(k) &= \Delta_\mathrm{L}^2(k)\left(
\frac{(1+\Delta_\mathrm{L}^2(k))^{\beta_n}}
{1+\alpha_n\Delta_\mathrm{L}^2(k)} \right)\exp(-y/4-y^2/8)\\
\Delta_\mathrm{H}^2(k) &= \frac{ {\Delta_\mathrm{H}^2}'(k)}{1+\mu_n
y^{-1}+\nu y^{-2}}\,,
\end{align}
with $y \equiv k/k_\sigma$ and
\begin{equation}
{\Delta^2_\mathrm{H}}'=\frac{a_n y^{3f_1(\Omega_m)}}{1+b_n
y^{f_2(\Omega_m)}+(c_n f_3(\Omega_m)y)^{3-\gamma_n}}\,.
\end{equation}
In these equations, $k_\sigma$ is defined via
\begin{equation}
\sigma(R_\mathrm{G})^2 \equiv \int_\infty^\infty \Delta_\mathrm{L}^2(k)
\exp(-k^2 R_\mathrm{G})^2 \dd \ln k 
\end{equation}
by
\begin{equation}
\sigma(k_\sigma^{-1}) = 1\,,
\end{equation}
while the effective index is
\begin{equation}
n = -3- \left. \frac{\dd \ln \sigma(R)^2}{\dd\ln R}\right|_{\sigma=1}
\end{equation}
and the spectral curvature is
\begin{equation}
C = \left.\frac{\dd^2\ln\sigma(R)^2}{\dd\ln R}\right|_{\sigma=1}\,.
\end{equation}
The best fit yielded the following values for the coefficients:
\begin{align}
\lg a_n &= 1.4861+1.8369n+1.6762n^2+0.7940n^3+0.167n^4-0.6206C\nonumber\\
\lg b_n &= 0.9463+0.9466n+0.3084n^2-0.9400C\\
\lg c_n &= -0.2807+0.6669n+0.3214n^2-0.0793C\\
\alpha_n &= 1.3884+0.3700n-0.1452n^2\\
\beta_n &= 0.8291+0.9854n+0.3401n^2\\
\gamma_n &= 0.8649+0.2989n+0.1631C\\
\lg \mu_n &= -3.5442+0.908n\\
\lg \nu_n &= 0.9589 + 1.2857n\,.
\end{align}
Also, the functions $f_i$ in a flat universe are given by
\begin{align}
f_1(\Omega_m) &= \Omega^{-0.0307}_m\\
f_2(\Omega_m) &= \Omega^{-0.0585}_m\\
f_3(\Omega_m) &= \Omega^{0.0743}_m\,.
\end{align}

\subsection{Parameterized post-Newtonian formalism}

We shall allow another degree of freedom in our equations that stems from
scalar theories in more than four dimensions. The Gauss-Bonnet theorem in
differential geometry connects the Euler characteristic of a two-dimensional
surface with the integral over its curvature. In general relativity, it
gives rise to a term with unique properties: It is the most general term
that, when added to the Einstein-Hilbert action in more than four
dimensions, leaves the field equation a second order differential equation.
In four dimensions, the equation of motion does not change at all under this
generalization, unless we are working in the context of a scalar-tensor
theory, in which the Gauss-Bonnet term couples to the scalar part and
modifies the equation of motion. For our purposes, this will effectively
make Newton's gravitational constant $G$ a variable, denoted by $G_*(\eta)$
(for details, see \cite{Amendola2006} and references therein).  This is
called the \textit{parameterized post-Newtonian} (PPN) formalism.

By defining
\begin{equation}
Q \equiv \frac{G_*}{G}\,,
\end{equation}
one can write the Poisson equation in Fourier space as
\begin{equation}
k^2 \phi = -4\pi G a^2 Q \rho_m \Delta_m\,,
\end{equation}
where $\Delta_m$ accounts for comoving density perturbations of matter. If
we admit anisotropic stress, then the two scalar gravitational potentials do
not satisfy $\Psi=-\Phi$, and we parameterize this by
\begin{equation}
\Psi \equiv -(1+\eta(k,a))\Phi\,.
\end{equation}
With weak lensing, only the combination
\begin{equation}
\Sigma \equiv Q(1+\eta/2)
\end{equation}
appears in the convergence power spectrum in a very simple way, where we can
just replace the matter power spectrum $P_m(k)$ by $\Sigma P_m(k)$.
Obviously, in the standard $\Lambda$CDM model, we have $\Sigma=1$, so we
allow it to vary in time and parameterize it as
\begin{equation}
\Sigma(a) =1+\Sigma_0 a\,,
\end{equation}
such that initially it looks like the standard model and then gradually
diverges from there \citep{Amendola2008}. Finally, we add $\Sigma_0$ to our
list of cosmological parameters with a fiducial value of 0.

\section{The Fisher matrix for the convergence power spectrum}
\label{sec:convps}

Building on the methods outlined in chapter 2, we can now compute the weak
lensing Fisher matrix. A full derivation of its expression can be found in
A\&T, ch.  14.4 or \cite{Hu1999}, with the final result being
\begin{equation}
F_{\alpha\beta} = f_\mathrm{sky} \sum_\ell 
\frac{(2\ell+1)\Delta\ell}{2} 
\sum\limits_{ijkm}\frac{\partial P_{ij}(\ell)}{\partial
\theta_\alpha}C^{-1}_{jk}\frac{\partial P_{km}(\ell)}{\partial
\theta_\beta}C^{-1}_{mi}\,.
\label{eq:3-fm}
\end{equation}
For this we need a multitude of cosmological functions that will be defined
in this section from the bottom up. First of all, note that the sum should
go over all multipoles $\ell$ inside an interval
$\ell_\mathrm{min}..\ell_\mathrm{max}$ that is determined by the fractional
survey size $f_\mathrm{sky}$ and the angular galaxy density $n_\vartheta$.
However, this poses a computational challenge. Hence, we rather sum over
bands with width $\Delta\ell$ while keeping in mind that one multipole
$\ell$ contains $(2\ell+1)$ modes \citep{Hu2004}. The multipole intervals
will be logarithmically spaced and the choice $\lg\Delta\ell$ shall be
justified in section~\ref{sec:lgell}.

$C_{jk}$ is the covariance matrix, given by
\begin{equation}
C_{jk}=P_{jk} + \delta_{jk}\gamma_\mathrm{int}^2 n^{-1}_j\,,
\end{equation}
where $\gamma_\mathrm{int}$ is the shot noise due to the intrinsic
ellipticity and $n_j$ is the number of galaxies per steradians belonging to
the $j$-th bin:
\begin{equation} 
n_j = 3600 \left( \frac{180}{\pi} \right)^2 
n_\vartheta\int_{0}^\infty n_j(z)\dd z,
\label{}
\end{equation}
where $n_\vartheta$ is the galaxy density per $\unit{arcmin}^2$.
The convergence spectrum $P_{ij}$ depends on the matter power spectrum
$P_{m}$ and is given by (now with the PPN parameter)
\begin{equation}
P_{ij}(\ell) = \frac{9}{4}\int_0^\infty
\frac{W_i(z)W_j(z)H^3(z)\Omega_m(z)^2}{(1+z)^4}\Sigma(z) P_{m}\left(
\frac{\ell}{\pi r(z)}\right)\dd z\,,
\label{eq:3-convspec}
\end{equation}
which can be slightly simplified, by using
$\Omega_m(z)=\Omega_m\times(1+z)^3/E(z)^2$, to
\begin{equation}
P_{ij}(\ell) = \frac{9H_0^3}{4}\Omega_m^2\int_0^\infty
\frac{W_i(z)W_j(z)(1+z)^2}{E(z)}\Sigma(z) P_{m}\left(
\frac{\ell}{\pi r(z)}\right)\dd z\,.
\end{equation}
Here, $H(z)$ denotes the Hubble parameter, which is
given by the first Friedmann equation (\eq{eq:2-friedmann1}) as
\begin{equation} 
H(z) = H_0 \sqrt{ \Omega_m(1+z)^3 + (1-\Omega_m)
\exp(f_w(z))}\,,
\label{eq:3-e(z)}
\end{equation}
and $E(z)\equiv H(z)/H_0$ is its dimensionless equivalent. The function
\begin{equation}
f_w(z)=3\int_0^z
\frac{1+w(z)}{1+ z'}\dd z'\,.
\end{equation}
describes how the dark energy density scales with the scale factor
and is the solution of the continuity
equation~(\ref{eq:2-conti}).
The equation-of-state ratio $w(z)\equiv \rho/p$ is from 
now on assumed to take the common CPL parameterization\footnote{Relevant
literature often uses $w_a$ instead of $w_1$.}
\citep{Chevallier2001,Linder2003}
\begin{equation}
w(z) = w_0 + w_1 \frac{z}{1+z}\,,
\label{eq:3-w(z)}
\end{equation}
in which case the above function can be calculated analytically and takes
the form
\begin{equation}
f_w(z) = 3 \left((w_0+w_1+1) \ln
   (z+1)-\frac{w_1 z}{z+1}\right)\,.
\end{equation}
\begin{figure}[htb]
\begin{center}
\includegraphics[width=9cm]{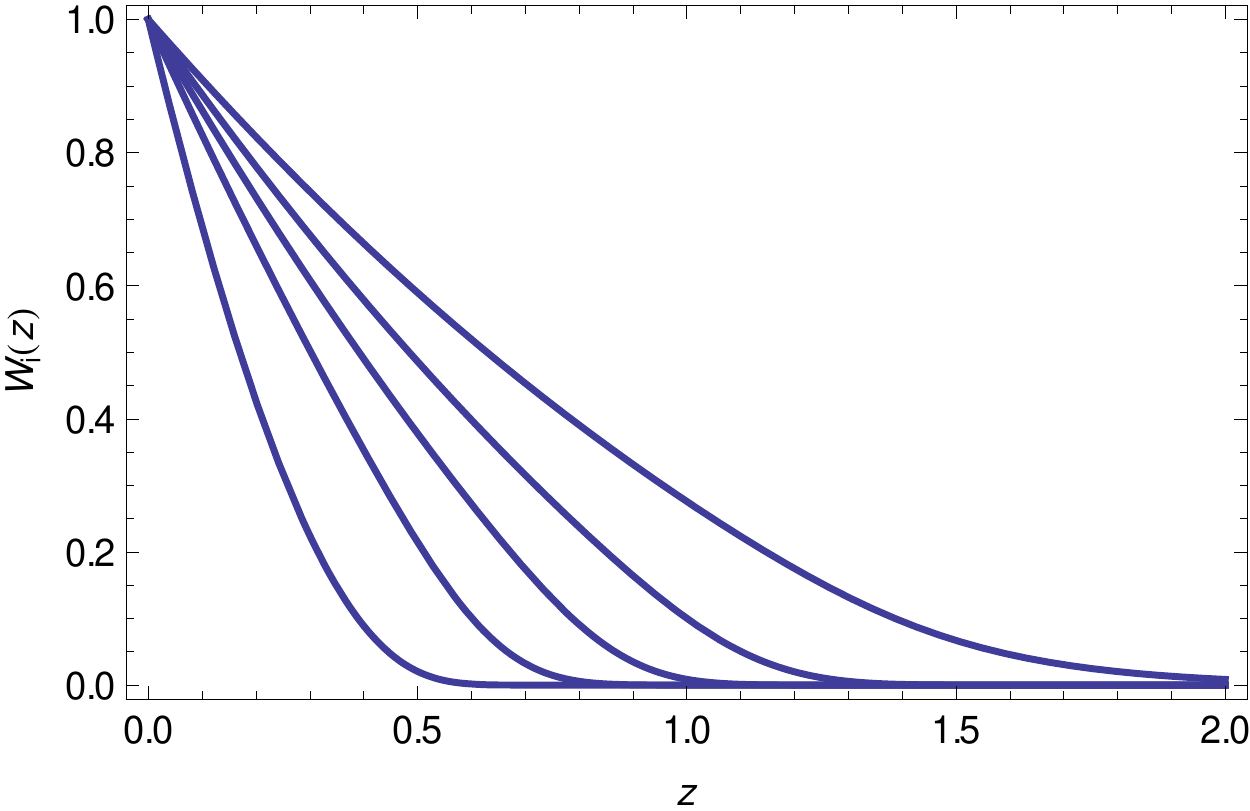}
\end{center}
\caption{The window functions $W_i(z)$ for five redshift bins.}
\label{fig:3-wind}
\end{figure}
One of the most crucial functions being used in this calculation is the
window function
\begin{equation}
W_i(z) = \int_z^\infty \frac{\dd z'}{H( z')}\left( 1-\frac{r(z)}{r( z')}
\right) n_i(r( z')),
\label{eq:3-wind}
\end{equation}
where $r(z)$ is the comoving distance at redshift $z$, i.e.
\begin{equation}
r(z) = \int_0^z \frac{\dd z'}{H(z')}\,.
\end{equation}
Lastly, we need the matter power spectrum $P_{m}$ to infer the
convergence power spectrum. For this we use the fitting formulae covered in
section~\ref{sec:nonlinear}.

\section{Survey parameters}  
\label{survey}

The quality of any weak lensing survey is determined by a set of
parameters that includes the fraction of the sky covered\footnote{Since the
Milky Way is blocking roughly half of our sky for deep surveys, even the
most ambitious missions will have $f_\mathrm{sky}=0.5$.}, the median
redshift, and so on. In table \ref{tab:3-survey} all of the parameters that
are used in our simulation are shown. Weak lensing surveys should aim at
wide and deep field coverage. Since these projects are all still in
development, the exact numbers that will be used in the end may vary, and
some are still undecided.
\begin{table}[htbp]
\centering
\begin{tabular}{ccl}
\hline\hline
Parameter & Value & Comment\tabularnewline
\hline
$f_\mathrm{sky}$ & 0.5 & Fraction of sky surveyed \tabularnewline
$z_m$ & 0.9 & Median redshift\tabularnewline
$\Delta_z$ & 0.05 & Relative photometric redshift error\tabularnewline
$n_\vartheta$& 35 & Galaxy density per $\unit{arcmin}^2$ \tabularnewline
$\gamma_\mathrm{int}$ & 0.22 & Intrinsic shear\tabularnewline
\hline
\end{tabular}
\caption{Shown here are typical values of survey parameters determining the
quality of the experiment, which we will use in our calculation
\citep[cf.][]{Huterer2006}.}
\label{tab:3-survey}
\end{table}

\subsection{Ground based surveys: LSST}

The Large Synoptic Survey Telescope,\footnote{\url{http://www.lsst.org/}}
funded partially privately and by the U.S. Department of Energy, is a wide
field survey telescope working in the visible band, located on the Cerro
Pach\'on in Chile. It is currently in the design and development stage and
is scheduled to be fully operational by 2020 with a lifetime of at least ten
years. It will feature a 3.2 Gigapixel camera and a large aperture  with an
8.4m (6.5m effective) primary mirror and cover the entire extra galactic sky
of 20 000 deg$^2$ with a depth up to an apparent magnitude of $r\sim 27.7$
for point sources. Besides weak lensing measurements, the mission will
observe super novae as well as map the Milky Way and small objects in the
solar system (such as asteroids).
This ultra wide and deep survey will provide a good data
source for weak lensing observations.

\subsection{Space based surveys: \texorpdfstring{\sc Euclid}{Euclid},
WFIRST} 

{\sc Euclid}\footnote{
\url{http://sci.esa.int/science-e/www/object/index.cfm?fobjectid=42266}} is
a proposed mission from ESA for a satellite orbiting the second Lagrangian
point as part of their cosmic vision program. It is still in the definition
phase and currently competing with PLATO and Solar Orbiter to be one the two
missions that will be approved by ESA in mid 2011. It is foreseen to be
launched in 2017 with a nominal mission lifetime of five years. This survey
might also cover up to 20 000 deg$^2$. The telescope will be a 1.2m Korsch
operating within the visible and infrared wavelengths with the ability of
detecting galaxies in the redshift range $0.5<z<2$ and therefore provide an
excellent probe for dark energy.

The Wide Field Infrared Survey Telescope
(WFIRST\footnote{\url{http://wfirst.gsfc.nasa.gov}}, formerly known as
JDEM-Omega) is another survey investigating dark energy, funded by NASA and
the U.S. Department of Energy and proposed by the Joint Dark Energy Mission.
In addition to weak lensing and baryon acoustic oscillations, WFIRST will
also probe supernovae with high precision. It is still at a very early stage
and has been postponed to launch at an unknown date, possibly not before
2025, due to budgeting and scheduling issues on a seperate project, the
James Webb Space Telescope \citep{Overbye2011}.


\begin{table}[htb]
\centering
\begin{tabular}{rrrrrp{4cm}}
\hline\hline
Survey & $f_\mathrm{sky}$ & $\Delta_z$ & $z_m$ &
$n_\vartheta/\unit{arcmin}^{-2}$ & Reference\\
\hline 
LSST & $0.37$ & 0.02 & $0.7$ & $30$ & \cite{Huterer2006,Ivezic2008}\\
{\sc Euclid} & 0.5 &  $0.05$& 0.8 & 30-40 & \cite{Euclid2009}\\
WFIRST & $0.25$ & 0.04 & ? & $\geq 30 $ & \cite{Gehrels2010}\\
\hline
\end{tabular}
\caption{Overview of current and planned surveys.}
\label{tab:surveys}
\end{table}

\section{Fiducial model} 

In order to apply the Fisher matrix formalism, we need a fiducial
cosmological model which represents the vector in parameter space where the
likelihood has its maximum. We use
the standard model of cosmology, a flat universe with cold dark matter and a
cosmological constant (called $\Lambda$CDM, which makes six parameters). We
additionally consider some extra parameters which are allowed to vary in a
non-standard way, in particular 
\begin{equation}
\vec \theta = (\omega_m,\omega_b,\tau,n_s,\Omega_m,
w_0,w_1,\gamma,\Sigma_0 ,\sigma_8)\,, \label{eq:3-params}
\end{equation}
where $\omega_m=\Omega_mh^2$ is the reduced fractional matter density,
and $\omega_b$ is defined analogous for baryons. Their numerical
values according to the WMAP 7-year data~\citep{Komatsu2011} can be seen in
table \ref{tab:wmap7}. The growth index $\gamma$ is taken to be
$6/11\approx0.55$ in the standard model~\citep{Wang1998}.

It should be pointed out that the reionization optical depth $\tau$ is
merely a remnant, and none of the functions in the code in later section
depend on $\tau$; however, it was not removed, however, due to the
possibility of future modifications to the code. The inclusion of functions
that depend on $\tau$ will thus be easier,  as long as we remember to cross
out the according row and column in the resulting Fisher matrix.

\begin{table}
\centering
\begin{tabular}{lll}
\hline\hline
Parameter & Fiducial value & Description\\
\hline
$\omega_m$ & $0.1352(36)$ & Reduced matter density today\\
$\omega_b$ & $0.02255(54)$ & Reduced baryon density today\\
$\Omega_m$ & $0.275(11)$ & Matter density today\\
$\tau$ & $0.088(14)$ & Reionization optical depth\\
$n_s$ & $0.968(12)$ & Scalar spectral index\\
$\sigma_8$ & $0.816(24)$ & Fluctuation amplitude at $8\unit{Mpc}/h$\\
\hline
$w_0^\dagger$ & $-1$ & Current equation-of-state ratio\\
$w_1^\dagger$ & $0$ & Higher order equation-of-state ratio\\
$\gamma^\dagger$ & $6/11$ & Growth index\\
$\Sigma_0^\dagger$ & $0$ & Parameterized Post-Newtonian
parameter\\
\hline
\end{tabular}
\caption{Our ten-parameter fiducial model. The second column represents the
WMAP7+BAO+$H_0$ Mean taken from \citet{Komatsu2011}. Quantities with a
dagger are fixed in
the standard flat $\Lambda$CDM model and thus not included here by the WMAP
measurement. Uncertainties in the last two digits are given in parenthesis.
They will be needed in Section~\ref{sec:priors}.}
\label{tab:wmap7}
\end{table}

\comment{
\section{Parametrization via rectangular functions}
The rectangular $\sqcap(x)$ function can be defined in terms of the
Heaviside step function $\Theta(x)$ as
\begin{equation} \sqcap(x) =
\Theta\left(x+\frac12\right)\Theta\left(\frac12-x\right) = \left\{
\begin{array}{ll} 1\quad \mathrm{if}\; |x|\leq\frac12\\ 0\quad \mathrm{else
} \end{array} \right..  \end{equation}
After choosing $\calN+1$ numbers $x_0,x_1,\ldots,x_\calN$ with
$x_i<x_{i+1}$, we can approximate any function $f(x)$ with rectangular
functions so that the function assumes a constant value in the respective
bin:
\begin{equation}
f(x) \approx \sum_{i=1}^\calN \alpha_i \sqcap\left(
\frac{x-(x_{i-1}+x_i)/2}{x_i-x_{i-1}} \right)
\label{}
\end{equation}
where $\alpha_i=f( (x_{i-1}+x_i)/2)$. In case of a cosmological function of
the redshift $z$, this allows us to treat the coefficients $\alpha_i$ as
independent cosmological parameters after choosing some appropiate redshift
bins. 
\begin{figure}[htpb]
\begin{center}
\includegraphics[width=9cm]{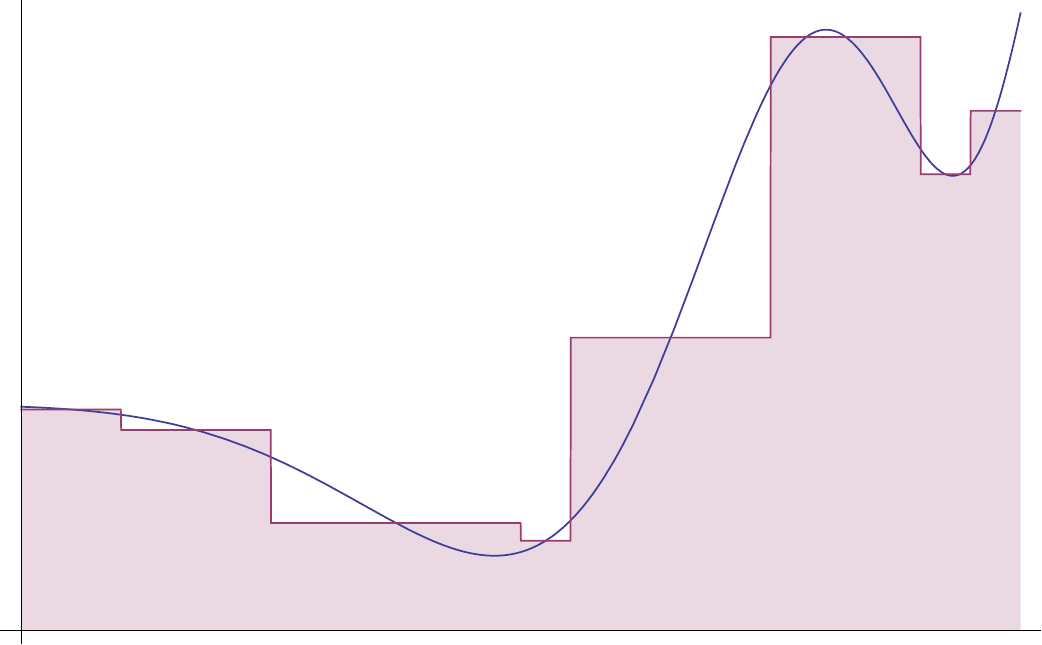}
\end{center}
\caption{A function being approximated by rectangular functions.}
\label{fig:2-rec}
\end{figure}
 }

\section{Model-independent parameterization}

\subsection{Expansion rate}\label{sec:3-hubble}

In order to see how observations can constrain the history of the expansion
rate without assuming a particular parameterization of it, we take the
values of the Hubble parameter at a series of redshifts determined in
section~\ref{sec:tomography} using our fiducial model and build a linearly
interpolated function going through these supporting points. For this, we
define
\begin{equation}
h_i = \ln H(\hat z_i),
\end{equation}
where $\hat z_i$ is at the
center of the $i$-th redshift bin. The use of the logarithm is purely for
convenience. The Hubble parameter itself from now on is a linearly
interpolated function through these points. It is evident that the
dependency of the Hubble parameter on the other cosmological quantities
changes from $H(z;\Omega_m,w_0,w_1)$ to $H(z;h_1,\ldots,h_\calN)$.  The
consequence of this is that all cosmological functions that depend on
$H(z)$ are now treated as functions of the $h_i$s, resulting in $\calN$
additional parameters being added to the fiducial model.

\subsection{Growth function} \label{sec:growth} 

The growth function is defined by 
\begin{equation}
G(a) \equiv \frac{\Phi(a)}{\Phi(a_T)}\,,
 \label{eq:3-growthfunc}
 \end{equation}
where $\Phi$ is the scalar gravitational potential
from~(\ref{eq:2-newt-gauge}) and $a_T$ stands for the transfer epoch, which
typically has a value of $0.03$ (see
for instance A\&T or \citet{Dodelson2003}). Be advised that in relevant
literature the growth function is often defined such that $G(a)$ needs to be
replaced with $aG(a)$ in eq.~(\ref{eq:3-growthfunc}). Also, it is sometimes
referred to as $D$ or $D_1$. The growth function can be identified with the 
matter density contrast over the scale factor, i.e.
\begin{equation}
G(a) = \frac{\delta_m(a)}{a\delta_m(0)}\,, 
\end{equation}
which is given by
\begin{equation}
\delta_m(a) \equiv \frac{\rho(a) }{\langle \rho\rangle}-1\,.
\end{equation}

A practical quantity in cosmology is the so called growth index $\gamma$,
defined by
\begin{equation}
f(a)  \equiv \Omega_m(a)^\gamma\,, \label{eq:3-gammafit}
\end{equation}
where $f$ is the growth rate 
\begin{equation}
f(a)  \equiv  \frac{\dd \ln \delta_m(a)}{\dd \ln
a}\label{eq:3-growthrate}\,. 
\end{equation}
Note that the power law in \eq{eq:3-gammafit} is an empirically found fit
for the growth rate \citep{Wang1998}, which was shown to be accurate to
within 0.2\% of the exact solution \citep{Linder2005}.

After reminding ourselves that
\begin{equation}
\dd \ln a = \frac{\dd a}{a} = (1+z)\dd\left( \frac{1}{1+z} \right) =
-\frac{\dd z}{1+z}\,,
\end{equation}
we can eliminate $\delta_m(a)$ and $f$ from eqs.
(\ref{eq:3-growthfunc} - \ref{eq:3-gammafit}) to find
\begin{equation}
G(z) = G_0 \exp\left({-\int_0^z\frac{\Omega_m( z')^\gamma-1}
{1+ z'}\dd z'}\right)\,,
\label{eq:3-growth}
\end{equation}
where the matter density is 
\begin{equation}
\Omega_m(z) =\frac{\rho_m(z)}{\rho_m(z)+\rho_\Lambda} =
\frac{\Omega_m\times(1+z)^3}{E(z)}\,.
\end{equation}

The only point where the growth function enters the calculation is via the
matter power spectrum, which is given by (see A\&T, p. 75; cf.
eq.~(\ref{eq:3-mps1}))
\begin{equation}
P_m(k,z) = \frac{2\pi^2\delta^2_H}{\Omega_m^2} \left(
\frac{k}{H_0} \right)^{n_s}T(k)^2\frac{G(z)^2}{(1+z)^2} H_0^{-3}.
\label{eq:3-mps2}
\end{equation}
Just as in section \ref{sec:3-hubble} we can now finally define 
\begin{equation}
g_i \equiv \ln G(\hat z_i)
\end{equation}
and replace the original growth function with a linear interpolation through
the $\hat z_i$s. The number of parameters in our model thus increases to 
$10+2\calN$. Naturally, every function that depends on the matter power
spectrum now also becomes a function of the $g_i$s.

\section{The figure of merit}

In order to have a quantifiable gauge of the information content of the
final result, we define the figure of merit (FOM) for the error bars on the
$h_i$s and $g_i$s, respectively, as
\begin{align} 
\FOM(H) \equiv& \sum_{i=1}^\calN \sigma(h_i)^{-2}\,,\\
\FOM(G) \equiv &\sum_{i=1}^\calN \sigma(g_i)^{-2}\,.
\label{eq:3-fom}
\end{align}
It is obvious that the smaller the error bars are, the larger the FOM
becomes. The idea of using this definition is that the FOM should not change
too much if only one error bar is significantly larger than all the other
ones. 

 
While the particular value of the FOM has no deeper meaning, it provides a
convenient tool to easily compare the errors resulting from different sets
of parameters. Note that this definition does not match the
popular definition by the Dark Energy Task Force, which suits a
different purpose \citep{Albrecht2006}.

\section{An implementation in \texorpdfstring{\matha}{Mathematica}} 

To do the actual computation, we use \matha 8.0.1.0, which supplies us with
all necessary numerical procedures.

In an attempt to seperate all the parameters that determine the final
outcome of the computation from the rest of the code, they are all defined
at the beginning of the \matha notebook.
We further
calculate a checksum, a digital fingerprint of those parameters via an
injective\footnote{At least for our purposes.} map from the set of all
parameters onto the set of integers, comparable to a hash function. This
will be useful when we store results in a file when changing the parameters
without having to worry about overwriting previous results.
We can roughly differentiate between physical parameters (such as matter
density, baryon density, etc.) and technical
parameters (the limit of a certain integral, step size in an
interpolating function, etc.). For an overview, we refer to
tables~\ref{tab:3-functions}, \ref{tab:3-physical}, and \ref{tab:3-tech}.

\begin{table}[htb]
\centering
\begin{tabular}{ll}
\hline\hline
\multicolumn{2}{c}{Functions}\\
\hline
\verb|logprint[string]| & Prints messages and saves them to a log file\\
\verb|loadfile[filename]| & Uses pre-calculated expressions if applicaple\\
\verb|ngal[[i]][z]| & $n_i(z)$\\
\verb|hub[z,dbins]| & $H(z;h_1,\dots,h_\calN)$\\
\verb|dist[z,dbins]| & $r(z)$\\
\verb|omgen[z,om,w0,w1]| & $\Omega_m(z)$\\
\verb|growthfastgen[z,gbins]| & $G(z;g_1,\dots,g_\calN)$\\
\verb|wind[i,z,om,dbins]| & $W_i(z)$\\
\verb|pk[k,z,...]| & $P_m(k)$ (Linear)\\
\verb|pknl[k,z,...]| & $P_m(k)$ (Non-linear)\\
\verb|clnngen[i,j,ell,...]| & $P_{ij}(\ell)$\\
\verb|dcdp[pi,i,j,ell,ref]| & $\partial P_{ij}/\partial \theta_{pi}$\\
\hline
\end{tabular}
\caption{Functions as used in the \matha notebook.}
\label{tab:3-functions}
\end{table}

\begin{table}[htb]
\centering
\begin{tabular}{llll}
\hline\hline
\multicolumn{4}{c}{Physical parameters}\\
\hline
\verb|omh2| & $\omega_m$ &
\verb|obh2| & $\omega_b$\\
\verb|tau| & $\tau$ &
\verb|ns| & $n_s$\\
\verb|om| & $\Omega_m$ &
\verb|w0| & $w_0$\\
\verb|w1| & $w_1$ &
\verb|gamma| & $\gamma$\\
\verb|gppn| & $\Sigma_0$ &
\verb|s8| & $\sigma_8$\\
\verb|dbins[[i]]| & $h_i$ & 
\verb|gbins[[i]]| & $g_i$\\
\hline
\end{tabular}
\caption{Variables referring to physical parameters as used for arguments in
functions in the \matha
notebook. Their fiducial value is stored in a variable that has the string
\textit{ref} attached.}
\label{tab:3-physical}
\end{table}

\begin{table}[htb]
\centering
\begin{tabular}{lrp{9cm}}
\hline\hline
\multicolumn{3}{c}{Technical parameters}\\
\hline
\verb|zlimit|&3& Redshift limit after which we assume $n(z)\approx0$\\
\verb|nbins|&5& Number of redshift bins $\calN$\\
\verb|eps|&0.025& Step size for numerical differentiation\\
\verb|zsigma|&4 & Integration length in units of standard deviations\\
\verb|zsmax|& 4& Maximum redshift for interpolated functions\\
\verb|fitstepz|& 0.2 & Step size for interpolated functions\\
\verb|lellmax|& 4 & Decadic logarithm of upper limit of the multipole\\
\verb|lellmin|& 2 & Decadic logarithm of lower limit of the multipole\\
\verb|lstep|&0.2 & Decadic logarithmic step size of the multipole\\
\verb|piv| & & List of parameters that are not fixed\\
\verb|dir|&  & Directory where data is stored\\
\hline
\end{tabular}
\caption{Parameters determining the accuracy of the result and their default
value.}
\label{tab:3-tech}
\end{table}

In the first section of actual computation, we infer the boundaries of the
redshift bins depending only on $\calN$, $\Delta_z$, and the galaxy
distribution function $n(z)$. The definition of the necessary cosmological
functions is straight forward. They often involve the computation of time
intensive numerical integrals, especially the highly used window function.
These functions are interpolated by cubic splines. All functions are
sufficiently ``nice'' (not more than one extremum) so that we do not have to
retreat to more conservative linear splines.

%
In order to save processing time and memory (especially in sequenced
executions of the notebook) we employ a little trick when computing
expensive functions. To avoid
loss of data after unexpected computer crashes, it makes sense to store not
only the final result, i.e. the Fisher matrix, but also intermediate results
on the hard drive. For this, the function \verb|loadfile| evaluates the
expression given in the second argument only if the file given in the first
argument does not exist, and then stores the result in said file and returns
it. If the file does exist, the expression in it will be loaded and
returned. This is applied to all time sensitive functions. All files are
saved in \verb|dir|, which contains the checksum of all parameters, except
for the window functions $W_i(z)$. Because they only depend on $\Omega_m$,
$\calN$, the $h_i$s and the number of the redshift bin, they are stored in a
sibling directory to save computing time when calculating the Fisher matrix
after only, say, $\Delta\lg\ell$ has changed.  The computation of the
interpolating functions of the window functions has also been parallelized
by using the \matha command \verb|ParallelTable|.

Very often we have to interpolate a function depending on several
parameters, such as the window function $W_i(z;\Omega_m, h_1,\dots
h_\calN)$, which generally needs to be evaluated for the derivatives at all
$0<z<3$ for $\Omega_m$, $\Omega_m (1\pm\epsilon)$ and so on. We define the
function such that it is computed only when needed, which can be achieved in
\matha by writing schematically
\begin{verbatim}
fAux[a_] := fAux[a] = 
    Interpolation[Table[{x,f[x,a]},{x,x0,x1,dx}]];
fFast[x,a] := fAux[a][x]; 
\end{verbatim}
The \verb|fFast[x,a]| is a fast, interpolated version of \verb|f[x,a]|,
where \verb|x| is the variable and \verb|a| a parameter. The extra \verb|=|
sign in the first line makes \matha look up the cached value if
there is one, preventing it from making an expensive computation more than
once.

What follows is a relatively straight forward implementation of
sections~\ref{sec:nonlinear} and~\ref{sec:convps}. It is important to note,
however, that all the functions that ultimately depend on the growth
function, i.e. all functions that depend on the linear matter power
spectrum, have been overloaded such that they might take \verb|gbins| as an
argument, in which case the power spectrum uses the interpolated growth
function. The reason for this will become apparent in chapter~4.

Most functions have been carefully defined to avoid warnings by \matha
concerning numerical issues like non-converging integrals. Vanishing
integrals are hard to detect with numerical methods, since a sufficiently
sharp, high peak would require an arbitrary large number of sampling points.
Many functions have been tweaked (if possible) to avoid these warnings. Very
often, \matha also tries to simplify an expression before plugging in actual
values, which causes a lot of warnings and can slow down the computation
significantly. To prevent this, we declared certain arguments as
``numerical'' so that \matha plugs in these values immediately. For instance
the comoving distance is defined as
\begin{verbatim}
    dist[z_?NumericQ, dbins_] := NIntegrate[1/hub[zx, dbins],
        {zx, 0, z}];
\end{verbatim}
This way, \matha will not try to numerically evaluate the integral in a
subsequent definition of a function (which would be futile) before a
specific value for the redshift is given. We found that this significantly
reduces numerical instabilities and improves the overall results.

%% file: chapter4.tex
\chapter{Results}

Before we cover the analysis of the results, it needs to be pointed out that
\textit{ two} Fisher matrices have been calculated. The first Fisher matrix
was computed while using the standard expression in eq.~(\ref{eq:3-growth})
for $G(z)$ and linearly interpolating $H(z)$, and the second Fisher matrix
was computed by linearly interpolating both $H(z)$ and $G(z)$. We will
refer to this as the $H$-\textit{case} and the $G$-\textit{case}
respectively. This introduces a certain degeneracy, since the variables
$h_i$ occur in both the $H$-case and the $G$-case and generally take on
different values. Hence we will denote those variables in the $H$-case as
$\hat h_i$ to avoid confusion, although they will be barely used.

Unless mentioned otherwise, the Fisher matrix has been calculated using the
parameters shown in tab.~\ref{tab:3-tech}, where we cross out the rows and
columns corresponding to $(\tau,w_0,w_1,\gamma,\Sigma_0)$ in the standard
case. The order of our parameters was given in eq.~(\ref{eq:3-params}); here
is a reminder:
\begin{equation}
\vec \theta = (\omega_m,\omega_b,\tau,n_s,\Omega_m,
w_0,w_1,\gamma,\Sigma_0 ,\sigma_8)\,. \label{eq:4-params}
\end{equation}

\begin{figure}[htb]
\begin{center}
\includegraphics[width=9cm]{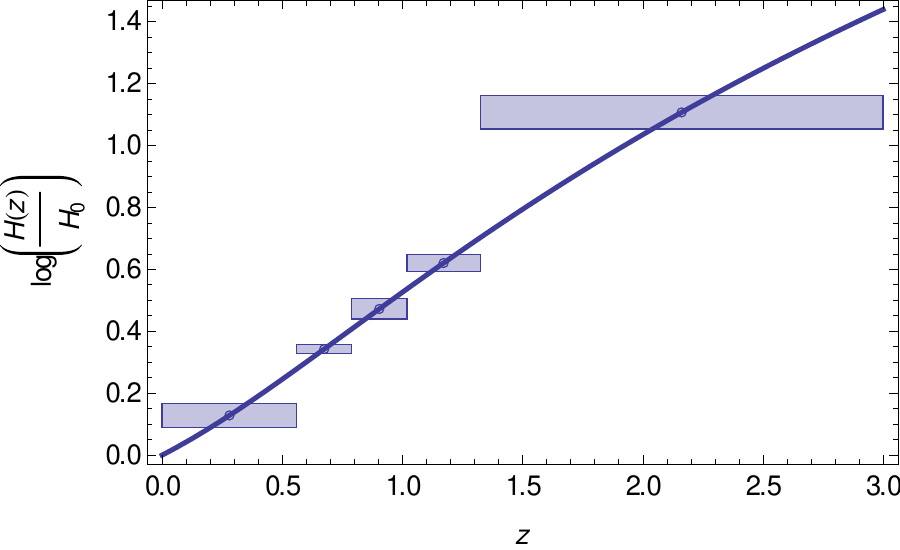}
\includegraphics[width=9cm]{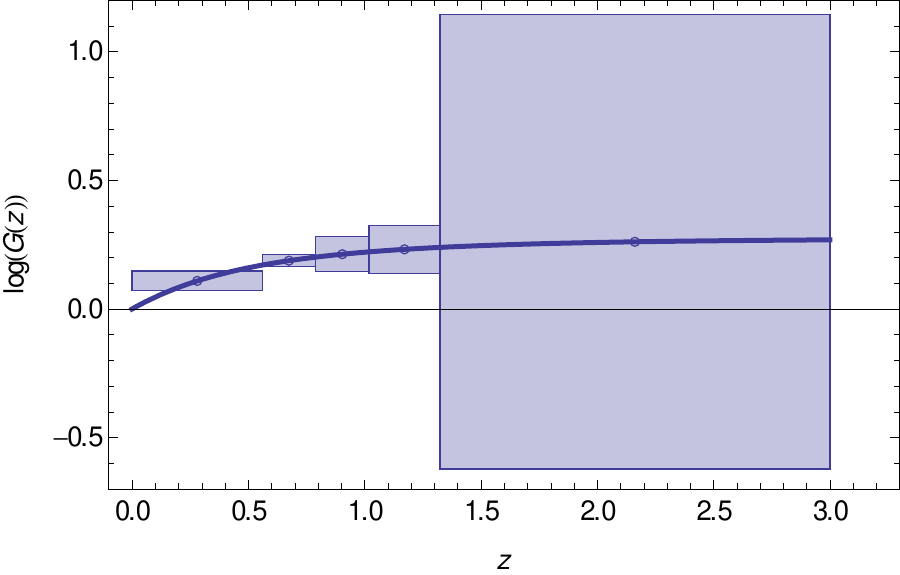}
\caption{A typical plot of the error bars on the $h_i$s (top) and $g_i$s
(bottom) for $\calN=5$
with a figure of merit of 7634 and 2865 respectively. The bins are chosen
such that they all contain the same number of galaxies.}
\label{fig:4-errors}
\end{center}
\end{figure}

From the resulting Fisher matrix,
we can easily extract the errors on each cosmological parameter. A typical
example of plotting the error bars on the Hubble parameter and growth
function can be seen in fig. \ref{fig:4-errors}.

\section{Consistency checks}

Since we naturally cannot compare our results with observations within the
next decade, we have limited means to make sure that they are meaningful and
error free. One way to check is to see whether the matrix satisfies basic
properties, another is to compute intermediate, partial results and compare
them with the numerical results. A first approach would be to visually
compare the derivative of the convergence power spectrum 
in two particular bins with respect to a particular cosmological parameter,
e.g.
\begin{equation}
\frac{\partial P_{12}(\ell)}{\partial g_1}\,,
\end{equation}
with and without interpolated functions. However, this turned out to be
unpractical due to the long runtime of over three days even with the use of
four parallel kernels and reducing the sampling points in the numerical
integration down to 50.

But we can still easily do another check. As the negative of the Hessian of
a function at its maximum, the Fisher matrix needs to be positive definite.
Looking at the eigenvalues of the Fisher matrix, we find in the $H$-case
(descending order)
\begin{multline}
\vec \lambda_H = (1.03\times 10^8,4.87\times 10^6,1.89\times
   10^5,7.29\times 10^4,3.34\times 10^4,\\9.11\times
   10^3,6.71\times 10^3,6.57\times 10^2, 3.2\times
   10^2,1.06\times 10^2,\\4.76\times 10^1,1.67\times
   10^1,3.53,1.04\times 10^{-2},1.4\times
   10^{-15})\,.
\label{eq:ev1}
\end{multline}
Note that the last eigenvalue is thirteen orders of magnitude smaller than
the next largest eigenvalue. The corresponding normalized eigenvector is,
while neglecting numbers that are smaller than $10^{-13}$,
\begin{equation}
\vec e_{H,15} = (0, 0, -1.0, 0, 0, 0, 0, 0, 0, 0, 0, 0, 0, 0, 0)\,.
\end{equation}
This reveals that it points only in the $\tau$-direction in parameter space,
as we can see in eq.~(\ref{eq:4-params}). This degeneracy is expected, since
the
convergence power spectrum does not depend on $\tau$. All other eigenvalues
are positive, so this check is passed.

In the $G$-case, the eigenvalues of the Fisher matrix are
\begin{multline}
\vec \lambda_G = (1.24\times 10^8,5.21\times 10^6,6.63\times
   10^5,1.86\times 10^5,8.85\times 10^4,\\4.08\times
   10^4,9.26\times 10^3,7.65\times 10^3,2.96\times
   10^3,1.\times 10^3,\\5.16\times 10^2,1.4\times
   10^2,30.3,7.36,3.56,0.83,\\-1.92\times 10^{-14},5.33\times
   10^{-15},-5.31\times 10^{-15},1.07\times
   10^{-46})\,.
\label{eq:ev2}
\end{multline}
There are two negative eigenvalues among the last four of them, but as they
are at least thirteen orders of magnitude smaller than the next largest
eigenvalue, they are consistent with zero. Again, analyzing the according
eigenvectors while neglecting values below $10^{-13}$ yields
\begin{align}
\vec e_{G,17} & = (0, 0, 0, 0, 0, 0.085, -0.23, 0.97, 0, 0, 0, 0, 0, 0,
0, 0, 0, 0, 0, 0) \\
\vec e_{G,18} & = (0, 0, 0, 0, 0, 0.98, 0.17, -0.047, 0, 0, 0, 0, 0,
0, 0, 0, 0, 0, 0, 0)\\
\vec e_{G,19} & = (0, 0, 0, 0, 0, -0.15, 0.96, 0.24, 0, 0, 0, 0, 0, 0,
0, 0,0, 0, 0, 0) \\
\vec e_{G,20} & = (0, 0, 1.0, 0, 0, 0, 0, 0, 0, 0, 0, 0, 0, 0, 0, 0, 0, 0,
0, 0)\,. 
\end{align}
Comparing again with eq.~(\ref{eq:4-params}) shows that these eigenvectors
lie in the parameter subspace spanned by $\tau$, $w_0$, $w_1$ and $\gamma$,
as we would expect since now the growth function depends on the $g_i$s
instead of the equation-of-state parameters $w_i$ and the growth index
$\gamma$. At this point, the convergence power spectrum does not depend on
any quantity that in turn depends on those parameters, so the Fisher matrix
will be degenerate if we do not cross out the according rows and columns.
Just like in the $H$-case, all remaining eigenvalues are positive.

\section{Improving the figure of merit}

In this section we want to find out how the FOM behaves when we modify
various parameters. This provides us with a way of gauging the
quality of the tomography process, seeing how much information we were able
to extract out of the weak lensing signal. 

\subsection{Impact of the number of redshift bins}

\begin{figure}[htbp]
\begin{center}
\includegraphics[width=9cm]{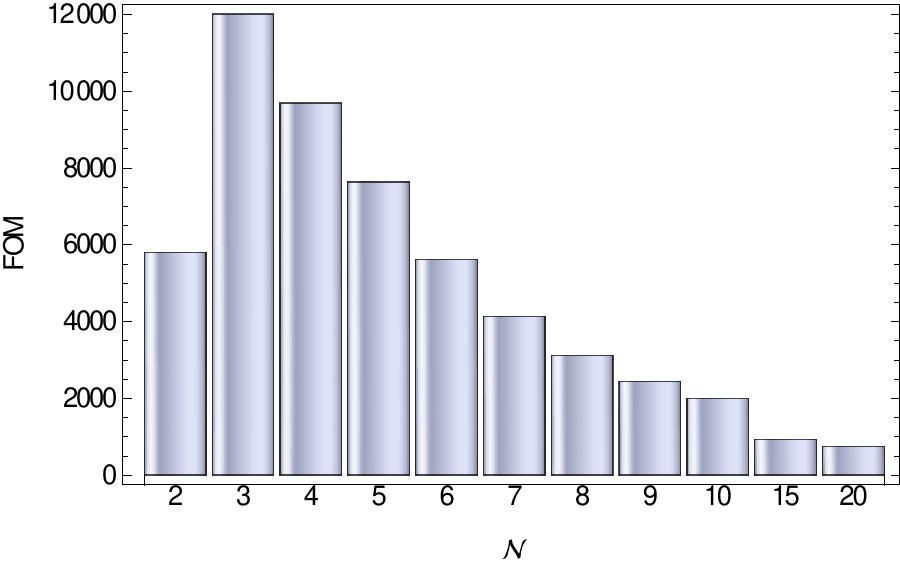}
\end{center}
\caption{Comparing the FOM for various values of $\calN$ in the $H$-case.}
\label{fig:4-h-hist}
\end{figure}

Running the \matha notebook for various values of $\calN$ lets us compare
the FOM for different binning. 
To get optimal results, we let the number of redshift bins $\calN$ run
between 2
and 20, using the values 2, 3, \dots, 10, 15 and 20. In
fig.~\ref{fig:4-h-hist}, we can clearly see a peak at $\calN=3$ in the
$H$-case with a decreasing trend for higher numbers of redshift bins.

\begin{figure}[htb]
\begin{center}
\includegraphics{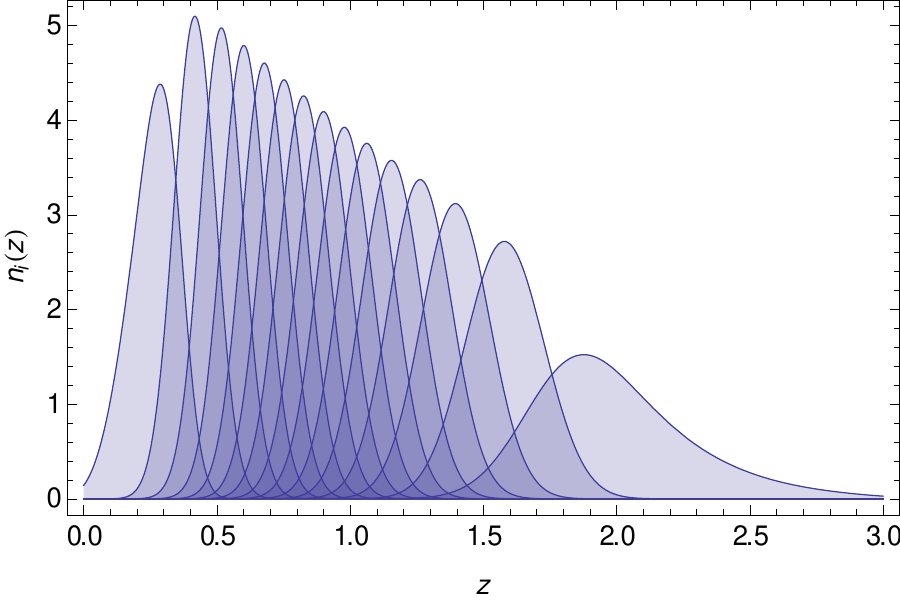}
\end{center}
\caption{The galaxy distribution function in 15 redshift bins, which are
overlapping heavily.}
\label{fig:4-nbins15}
\end{figure}

\begin{figure}[htb]
\begin{center}
\includegraphics[width=9cm]{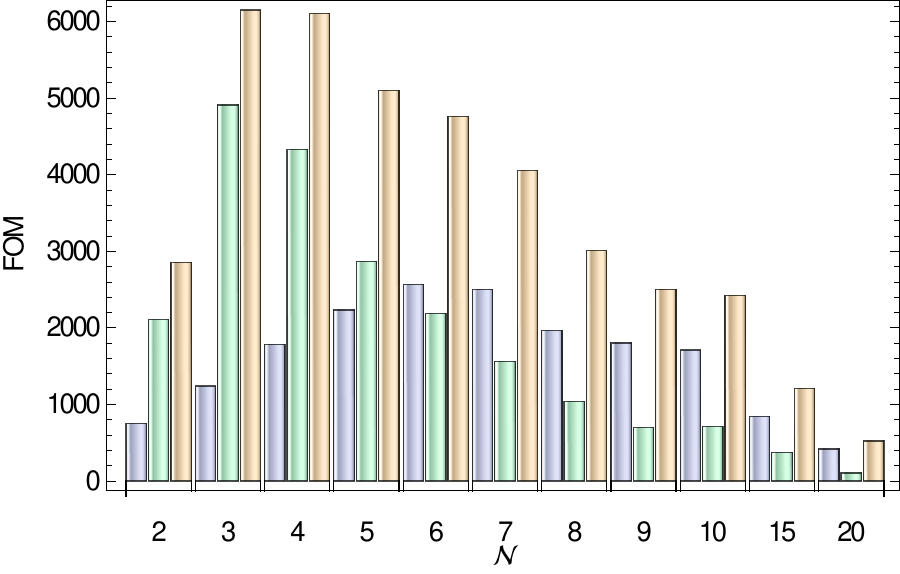}
\end{center}
\caption{In the $G$-case, the FOM in dependence of $\calN$ takes a
slightly different shape. Blue represents $\FOM(H)$, green is $\FOM(G)$
and yellow is the sum of both.}
\label{fig:4-hg-hist}
\end{figure}

Doing the same in the $G$-case yields a slightly different result,
shown in fig. \ref{fig:4-hg-hist}. Here the FOM for $H$ peaks at $\calN=6$
while the FOM for $G$ peaks at $\calN=3$. 

When examining the galaxy distribution function for each bin (see
fig.~\ref{fig:4-nbins15}) in the case of 15 redshift bins, we can see
that with our assumed photometric redshift error they are ``smeared out''
to the point where it becomes impossible to confidently assign a redshift
bin to a given galaxy.

\subsection{\texorpdfstring{On $\ellmax$ and $\Delta{\lg
\ell}$}{On ell max and delta lg ell}}
\label{sec:lgell}

\begin{figure}[htb]
\begin{center}
\includegraphics[width=9cm]{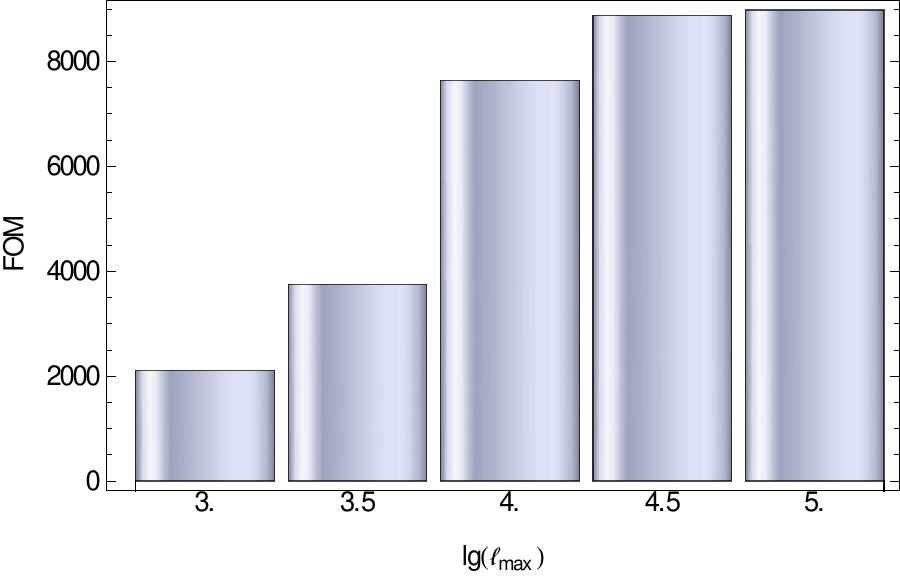}
\end{center}
\caption{When increasing $\ell_\mathrm{max}$, the upper limit of the sum in
eq.~(\ref{eq:3-fm}), the FOM changes, but increases by less than 15\% beyond
$\ell_\mathrm{max}=10^4$. Shown here is the $H$-case, the $G$-case looks
similar.}
\label{fig:4-lellmax}
\end{figure}

It is not obvious how far the sum in eq. (\ref{eq:3-fm}) should go or what
the step size for $\ell$ should be. Estimating a realistic value for the
upper limit $\ellmax$ has been done in eq.~(\ref{eq:3-ellmax}), but it is
interesting to see how the FOM behaves for larger values. 
We therefore try several values for $\ellmax$ and see whether the FOM
approaches some constant or not. The result can be seen in
fig.~\ref{fig:4-lellmax}.
We observe that the FOM only changes slightly for values above 4, as we
would expect due to the finite angular galaxy density.

\begin{figure}[htb]
\begin{center}
\includegraphics[width=9cm]{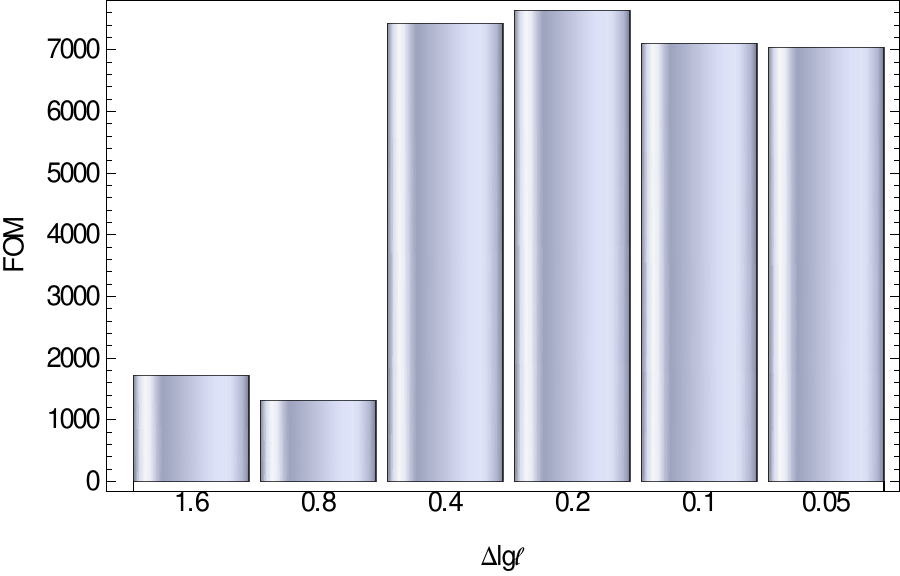}
\end{center}
\caption{The figure of merit for various values of $\Delta{\lg \ell}$ in the
$H$-case.  Going beyond $0.2$ does not change much. The situation is the
same in the $G$-case and thus not shown here.}
\label{fig:4-lellstep}
\end{figure}

Doing the same for some values for $\Delta{\lg \ell}$ at a fixed
$\ell_\mathrm{max}=10^4$ yields the numbers
used in fig. \ref{fig:4-lellstep}. We see that essentially all the
information is being extracted at a logarithmic multipole step size of
$0.2$. This figure roughly matches the result of $0.3$ found in
\cite{Bernstein2009}, where it is argued that the broad window function
``smoothes away fine structure'' in the convergence power spectrum, such
that at some point further increasing of the number of multipole bins will
no longer have beneficial effects. On the other hand, it is
interesting to note that the FOM increases almost by an order of magnitude
by going from $\Delta\lg\ell=0.8$ to $0.4$.

\afterpage{\clearpage}

\section{Fixing various parameters}

\begin{figure}[htb]
\begin{center}
\includegraphics[width=9cm]{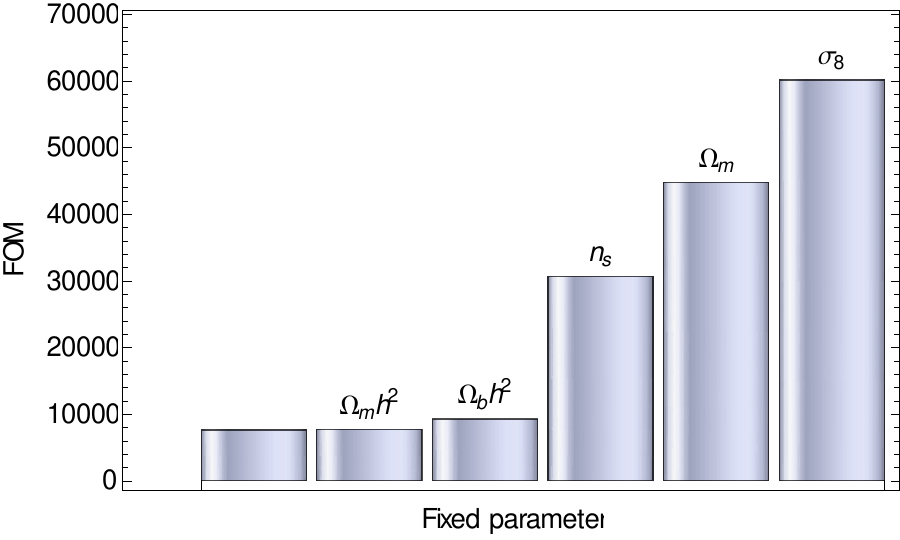}
\end{center}
\caption{Various parameters are being fixed successively ($H$-case). Each
bar is labeled with the parameter that has been fixed on top of the previous
one, such that all of them are fixed on the right side of the diagram.
Fixing the scalar spectral index is most beneficial, as the FOM triples in
the best case.}
\label{fig:fixing}
\end{figure}

In order to determine the parameter which the Fisher matrix is most
sensitive, we will fix one additional parameter after another and analyze
the behavior of the FOM.  Remember, fixing a parameter within the Fisher
matrix formalism is as easy as crossing out the according row and column of
the Fisher matrix. As we can see in fig.~\ref{fig:fixing}, knowing the
scalar spectral index $n_s$ in the $H$-case  as well as possible pays off
the most, as the FOM increases roughly by a factor of three if we know it
exactly.

\begin{figure}[htb]
\begin{center}
\includegraphics[width=9cm]{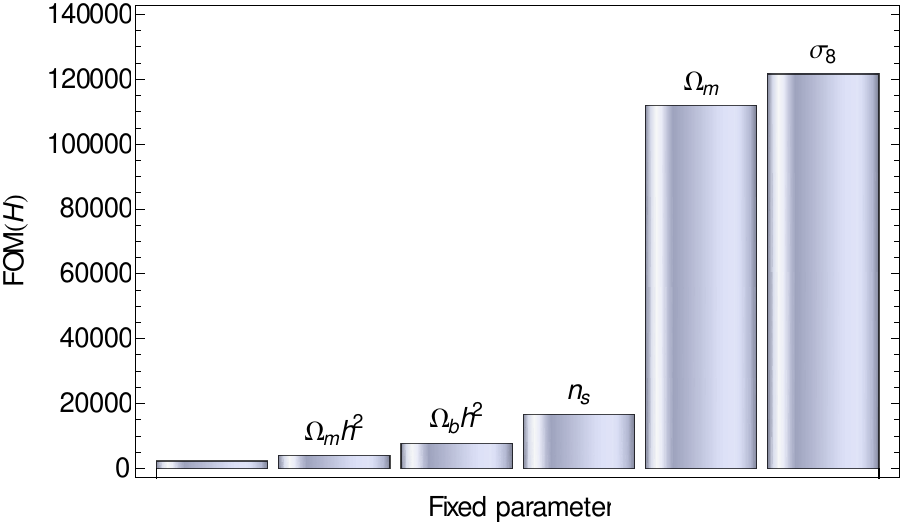}
\includegraphics[width=9cm]{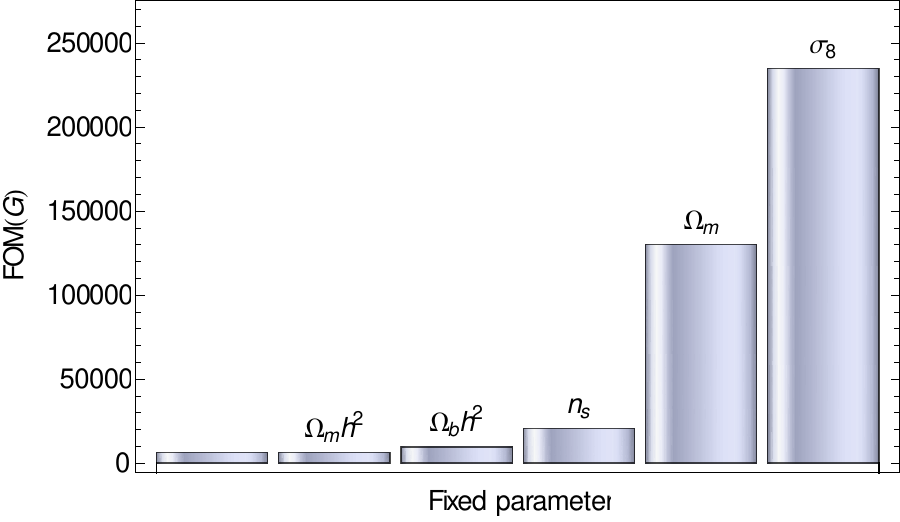}
\end{center}
\caption{Various parameters are, again, being fixed successively ($G$-case).
Here, knowledge about $\Omega_m$ and $\sigma_8$ pays off most for $\FOM(H)$
(\textit{top}) and $\FOM(G)$ (\textit{bottom}) respectively.}
\label{fig:fixing-g}
\end{figure}

On the other hand, in the $G$-case, depicted in fig.~\ref{fig:fixing-g}, it
is most beneficial to know $\Omega_m$ (factor of six) to get the highest
increase in $\FOM(H)$, but knowing $\sigma_8$ exactly gives another factor
of two for $\FOM(G)$.

\section{Uncertainties}

Finally, we can present the uncertainties on the values of $H$ and $G$ at
different redshifts. We use the parameters
$\calN=5$, $\ell_\mathrm{max}=10^4$ and $\Delta{\lg\ell}=0.2$ and find 
the results shown in tables~\ref{tab:4-results} and~\ref{tab:4-resultsH}.
These are derived from the full Fisher matrix as seen in tab.~\ref{tab:fm}
(for the $G$-case). As
expected, the constraints in the $G$-case are not as tight as those in the
$H$-case, since we allow the growth function to vary in a more general way.
\begin{table}[htb]
\centering
\begin{tabular}{lccccc}
\hline\hline
&$\sigma(\omega_m)$ & $\sigma(\omega_b)$ & $\sigma(n_s)$ &
$\sigma(\Omega_m)$ & $\sigma(\sigma_8)$\\
\hline
$H$-case & 0.17 & 0.065& 0.049 & 0.012 & 0.011\\
$G$-case & 0.17 & 0.066 & 0.059 & 0.021 & 0.041\\
\hline
\vspace{.5cm}
\end{tabular}
\begin{tabular}{lccccc}
\hline\hline
$i$ & 1 & 2 & 3 & 4 & 5\\ \hline
$\sigma(\hat h_i)$ & 0.038 & 0.015 & 0.033 & 0.026 & 0.053\\
$\sigma(h_i)$ & 0.044 & 0.036 &0.044 &0.049 &0.48\\
$\sigma(g_i)$ & 0.039 & 0.023 & 0.068 & 0.094 & 0.88\\
\hline
\end{tabular}
\caption{Errors on the cosmological parameters as well as $\hat h_i, h_i,
g_i$ while fixing $\tau, w_0, w_1, \Sigma_0, \gamma$ and keeping $\calN=5$,
$\ell_\mathrm{max}=10^4$, $\Delta{\lg\ell}=0.2$.} \label{tab:4-results}
\end{table}

\begin{table}
\centering
\begin{tabular}{ccccccccc}
\hline\hline
$\sigma(\omega_m)$ & $\sigma(\omega_b)$ & $\sigma(n_s)$ &
$\sigma(\Omega_m)$ & $\sigma(w_0)$ & $\sigma(w_1)$ & $\sigma(\gamma)$ &
$\sigma(\Sigma_0)$ &
$\sigma(\sigma_8)$\\ \hline
0.23 & 0.081 & 0.077 & 0.045 & 3.0 & 9.3 & 0.67 & 0.40 & 0.11\\
\hline
\end{tabular}
\caption{Errors on the cosmological parameters while fixing $\tau$,
marginalizing everything else, and keeping $\calN=5$,
$\ell_\mathrm{max}=10^4$, $\Delta{\lg\ell}=0.2$ in the $H$-case.}
\label{tab:4-resultsH}
\end{table}

\begin{table}[p!]
\small
\begin{flushleft}
\begin{tabular}{rrrrrrr}
\hline
$\Omega_mh^2$ & $\Omega_bh^2$ & $n_s$ & $\Omega_m$ & $\sigma_8$ & $h_1$ &\\
\hline\hline
 5.89991\text{e}6 & -1.29794\text{e}7 &
   3.04031\text{e}6 & 1.77766\text{e}7 &
   8.47979\text{e}6 & -3.10712\text{e}6 & \\
 -1.29794\text{e}7 & 2.85559\text{e}7 &
   -6.68462\text{e}6 & -3.91594\text{e}7 &
   -1.86837\text{e}7 & 6.8399\text{e}6 & \\
 3.04031\text{e}6 & -6.68462\text{e}6 &
   1.57342\text{e}6 & 9.0384\text{e}6 &
   4.30423\text{e}6 & -1.59096\text{e}6 & \\
 1.77766\text{e}7 & -3.91594\text{e}7 &
   9.0384\text{e}6 & 6.56348\text{e}7 &
   3.13299\text{e}7 & -1.06167\text{e}7 & \\
 8.47979\text{e}6 & -1.86837\text{e}7 &
   4.30423\text{e}6 & 3.13299\text{e}7 &
   1.50269\text{e}7 & -5.05321\text{e}6 & \\
 -3.10712\text{e}6 & 6.8399\text{e}6 &
   -1.59096\text{e}6 & -1.06167\text{e}7 &
   -5.05321\text{e}6 & 1.99529\text{e}6 & \ldots \\
 -1.94576\text{e}6 & 4.28365\text{e}6 &
   -9.92991\text{e}5 & -7.29185\text{e}6 &
   -3.45161\text{e}6 & 9.6474\text{e}5 & \\
 -1.16652\text{e}5 & 2.56784\text{e}5 &
   -5.94073\text{e}4 & -4.72584\text{e}5 &
   -2.22597\text{e}5 & 4.63944\text{e}4 & \\
 4.1346\text{e}6 & -9.11132\text{e}6 &
   2.09669\text{e}6 & 1.51661\text{e}7 &
   7.2877\text{e}6 & -2.56814\text{e}6 & \\
 1.91022\text{e}6 & -4.20561\text{e}6 &
   9.75038\text{e}5 & 6.89405\text{e}6 &
   3.2877\text{e}6 & -9.21068\text{e}5 & \\
 4.52229\text{e}4 & -9.9446\text{e}4 &
   2.3215\text{e}4 & 1.77003\text{e}5 &
   8.31533\text{e}4 & -1.48989\text{e}4& \\
\hline
\end{tabular}
\end{flushleft}\vfill
\begin{flushright}
\begin{tabular}{rrrrrr}
\hline
&$h_2$ & $h_3$ & $g_1$ & $g_2$ & $g_3$\\
\hline\hline
 & -1.94576\text{e}6 & -1.16652\text{e}5 &
   4.1346\text{e}6 & 1.91022\text{e}6 &
   4.52229\text{e}4 \\
 & 4.28365\text{e}6 & 2.56784\text{e}5 &
   -9.11132\text{e}6 & -4.20561\text{e}6 &
   -9.9446\text{e}4 \\
 & -9.92991\text{e}5 & -5.94073\text{e}4 &
   2.09669\text{e}6 & 9.75038\text{e}5 &
   2.3215\text{e}4 \\
 & -7.29185\text{e}6 & -4.72584\text{e}5 &
   1.51661\text{e}7 & 6.89405\text{e}6 &
   1.77003\text{e}5 \\
 & -3.45161\text{e}6 & -2.22597\text{e}5 &
   7.2877\text{e}6 & 3.2877\text{e}6 &
   8.31533\text{e}4 \\
 \ldots & 9.6474\text{e}5 & 4.63944\text{e}4 &
   -2.56814\text{e}6 & -9.21068\text{e}5 &
   -1.48989\text{e}4 \\
 & 1.03799\text{e}6 & 7.90972\text{e}4 &
   -1.55933\text{e}6 & -9.6674\text{e}5 &
   -3.22527\text{e}4 \\
 & 7.90972\text{e}4 & 7.59839\text{e}3 &
   -9.07882\text{e}4 & -7.37654\text{e}4 &
   -3.24212\text{e}3 \\
 & -1.55933\text{e}6 & -9.07882\text{e}4 &
   3.61394\text{e}6 & 1.48555\text{e}6 &
   3.17958\text{e}4 \\
 & -9.6674\text{e}5 & -7.37654\text{e}4 &
   1.48555\text{e}6 & 9.18941\text{e}5 &
   3.03478\text{e}4 \\
 & -3.22527\text{e}4 & -3.24212\text{e}3 &
   3.17958\text{e}4 & 3.03478\text{e}4 &
   1.4814\text{e}3\\
   \hline
\end{tabular}
\end{flushright}
\caption{The full Fisher matrix in the $G$-case for three redshift bins.}
\label{tab:fm}
\end{table}

\section{Including priors} \label{sec:priors}

As mentioned in section~\ref{calcrules}, the magic of Fisher lets us easily
include priors from other experiments.  All we need to do is line up the
right rows and columns of two different Fisher matrices and take their sum,
i.e.
\begin{equation}
F_{ij}' = F_{ij} + F_{ij}^\mathrm{prior}.
\label{eq:4-prior}
\end{equation}
We can do this for two measurements of the cosmic microwave background
radiation. The first one is an already completed experiment, namely WMAP,
and the second one a planned mission, {\sc Planck}, both of which use the
same fiducial model as in this thesis.

\subsection{WMAP}

A Gaussian prior, that is a fiducial value $p_i$ with Gaussian error
$\sigma(p_i)$, is represented simply by a Fisher matrix that has
$\sigma(p_i)^{-2}$ as its $i$-th diagonal entry \citep{Albrecht2006}.
Gaussian priors can be obtained from recent observations. We use the data
from the WMAP survey, one of the most precise measurements of cosmological
parameters as of now. The standard deviations can be taken from
table~\ref{tab:wmap7} and---while taking into consideration the order we are
using---the corresponding Fisher matrix thus becomes
\begin{align}
\mat F^\mathrm{WMAP7} = \mathrm{diag}\,&[77160.5, 3.42936\text{e6}, 5102.04,
\nonumber\\
&6944.44, 9291.96, 0, 0, 0, 0,  1736.11, 0, \dots, 0 ]\,.
\end{align}

\subsection{ \texorpdfstring{\sc Planck} {Planck}}

The Fisher matrix $\hat{\mat F}^\mathrm{Planck}$ for the planned {\sc
Planck} mission is shown in table~\ref{tab:fm-planck}. 
\begin{table*}[htb]\scriptsize
\begin{flushleft}
\begin{tabular}{lrrrc}
\hline
 \hline
 & $w_0$ & $w_a$ & $\Omega_\mathrm{DE}$ \\
 \hline
$w_0$ &  .172276e+06 &  .490320e+05 &  .674392e+06 &\\
$w_a$ &  .490320e+05 & .139551e+05 & .191940e+06&\\
$\Omega_\mathrm{DE}$ & .674392e+06 & .191940e+06 & .263997e+07&\\
$\Omega_k$ & $-$.208974e+07 & $-$.594767e+06 & $-$.818048e+07 & \dots \\
$\omega_m$ &  .325219e+07 & .925615e+06 &  .127310e+08 &\\
$\omega_b$ & $-$.790504e+07 &$-$.224987e+07&$-$.309450e+08& \\
$n_S$ & $-$.549427e+05 & $-$.156374e+05 & $-$.215078e+06 & \\
\hline
\end{tabular}
\end{flushleft}
\begin{flushright}
\begin{tabular}{crrrr}
\hline
\hline
& $\Omega_k$ & $\omega_m$ & $\omega_b$ & $n_S$ \\ 
\hline
& $-$.208974e+07 &.325219e+07 & $-$.790504e+07 & $-$.549427e+05 \\ 
& $-$.594767e+06 &.925615e+06 &$-$.224987e+07 &$-$.156374e+05\\ 
&$-$.818048e+07 & .127310e+08 &-.309450e+08 &$-$.215078e+06\\ 
\dots&  .253489e+08& $-$.394501e+08 &  .958892e+08 &  .666335e+06\\ 
& $-$.394501e+08 &.633564e+08 & $-$.147973e+09 & $-$.501247e+06\\ 
& .958892e+08 &$-$.147973e+09  &.405079e+09 & .219009e+07\\
& .666335e+06 &$-$.501247e+06  & .219009e+07  & .242767e+06\\ 
\hline
\end{tabular}
\caption{Fisher matrix $\hat{\mat F}^\mathrm{Planck}$ for ($w_0$, $w_a$,
$\Omega_\mathrm{DE}$, $\Omega_k$, $\omega_m$, $\omega_b$, $n_s$) derived from the
covariance matrix for $(R, l_a, \Omega_b h^2, n_s)$ from {\sc Planck}
\citep{Mukherjee2008}.}
\label{tab:fm-planck}
\end{flushright}
\end{table*}
This matrix needs to be transformed for compatibility with our Fisher
matrix. The parameter transformation is given by
\begin{equation}
\vec \theta'(\vec \theta) =
(\theta_5,\theta_6,\text{const.},\theta_7,1-\theta_3,\theta_1,\theta_2,
\text{const.},\dots,\text{const})\,,
\end{equation}
where $\vec \theta'=(\omega_m,\omega_b,\tau,n_s,\Omega_m, w_0,
w_1,\gamma,\Sigma_0,\sigma_8,h_1,\dots,g_1,\dots)$ is the vector
in parameter space in our formalism, and $\vec
\theta=(w_0,w_a,\Omega_\mathrm{DE},\Omega_k,\omega_m,\omega_b,n_s)$ is the
vector in parameter space in the formalism of \citet{Mukherjee2008}. The
Jacobian matrix thus reads
\begin{equation}
J_{ij}= \left(\frac{\partial \theta'_i}{\partial \theta_j}\right)^T = 
\underbrace{\left(
\begin{array}{cccccccccc}
 0 & 0 & 0 & 0 & 0 & 1 & 0 & 0 & \dots & 0  \\
 0 & 0 & 0 & 0 & 0 & 0 & 1 & 0 & \dots & 0 \\
 0 & 0 & 0 & 0 & -1 & 0 & 0 & 0 & \dots & 0 \\
 0 & 0 & 0 & 0 & 0 & 0 & 0 & 0 & \dots & 0 \\
 1 & 0 & 0 & 0 & 0 & 0 & 0 & 0 & \dots & 0 \\
 0 & 1 & 0 & 0 & 0 & 0 & 0 & 0 & \dots & 0 \\
 0 & 0 & 0 & 1 & 0 & 0 & 0 & 0 & \dots & 0 \\
\end{array}
\right)}_{10+\calN(+\calN)\textrm{ columns}}\,.
\end{equation}
The zeros on the right hand side account for the $\hat h_i$s (or $h_i$s and
$g_i$s), which the original {\sc Planck} Fisher matrix naturally does not
contain.  The Fisher matrix with the {\sc Planck} prior for our use is then
\begin{equation}
\mat F^\mathrm{Planck} = \mat J^T \hat{\mat F}^\mathrm{Planck} \mat J.
\end{equation}

\subsection{Summary}

We can now examine how including priors from WMAP, {\sc Planck}, or both
will affect the constraints on $H(z)$ or $G(z)$. For this, we will choose
$\calN=5$ while marginalizing over all other parameters, although $G(z)$ is
fixed when calculating the $\FOM(H)$. The results are summarized in
table~\ref{tab:priors}.

\begin{table}[htb]
\centering
\begin{tabular}{ccccc}
\hline\hline
  & WL & WL+WMAP7 & WL+{\sc Planck} & WL+{\sc Planck}+WMAP7 \\ 
 \hline
 $\FOM(H)$ & 7636 & 13778 & 25194 & 26096\\
 $\FOM(G)$ & 2865 & 5697 & 4350 & 6805\\
 \hline
\end{tabular}
\caption{Comparison of the FOM when including different priors for the
$H$-case and the $G$-case, respectively.}
\label{tab:priors}
\end{table}

Naturally, the FOM improves in both cases as we include additional priors.
Unexpectedly, the growth function is more constrained by assuming WMAP
priors compared to {\sc Planck} priors. With {\sc Planck} being one
generation ahead of WMAP, it should be the other way around. But if we take
fig.~\ref{fig:fixing-g} into consideration, it is clear that while fixing
$\Omega_m$ is most important for constraining $G$, fixing $\sigma_8$ also
almost doubles the FOM.  Unlike the WMAP prior, the {\sc Planck} prior does
not constrain $\sigma_8$ at all, thus even though {\sc Planck} constrains
$\Omega_m$ much more than WMAP in this parameter configuration, it is not as
important with regard to constraining the growth function. Another
observation we can make is that in either case the FOM does not increase
significantly compared to the next higher value when including the second
prior, i.e. the FOM in the $H$-case, when including WMAP and {\sc Planck},
is almost the same as only including {\sc Planck}. In the $G$-case, the FOM
is almost the same when including both priors as compared to only including
WMAP. This is to be expected, as both experiments probe the cosmic microwave
background radiation, except with different sensitivities. Therefore, no
more cosmological information can be extracted.

\section{Comparison to current constraints}

\begin{figure}[htb]
\begin{center}
\includegraphics{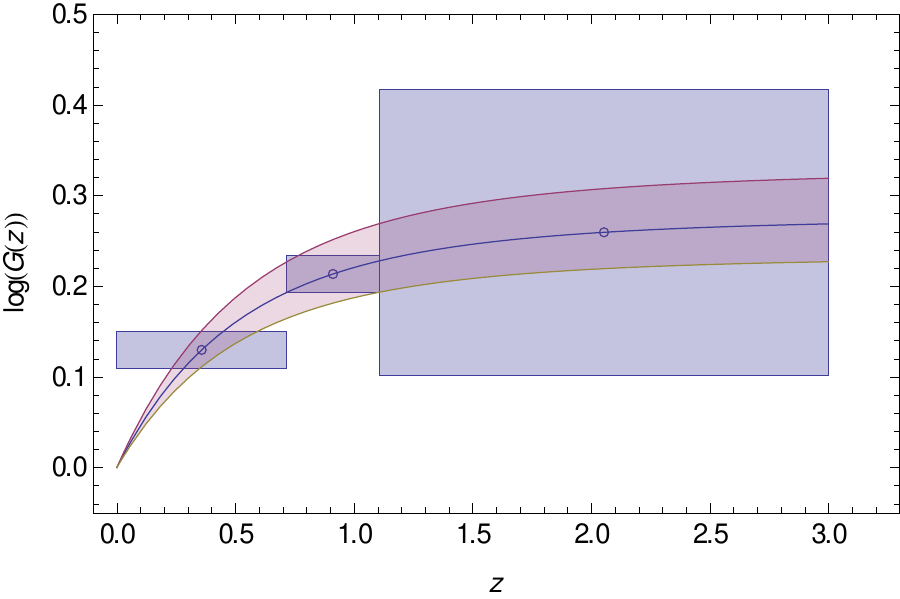}
\end{center}
\caption{Current constraints on the growth function via the growth index by
using X-ray cluster growth data. Shown is the $1\sigma$ confidence band of
the growth funcion.}
\label{fig:current}
\end{figure}

When we try to compare how weak lensing in this context will constrain the
growth function compared to current constrains, we notice that our
constraints are very poor at high redshifts. \cite{Rapetti2010} measured the
growth index via a combination of X-ray cluster growth data with cluster gas
mass fraction, type 1a supernovae, baryon acoustic oscillations and cosmic
microwave background data, where they found that $\gamma =
0.55^{+0.13}_{-0.10}$. As shown in fig.~\ref{fig:current}, these constraints
are not very tight when applied to the growth function (even when assuming
that $\Omega_m$ is known exactly), but the error bars from the Fisher matrix
analysis hardly seem any better, especially at high redshifts. It is
interesting, though, to have a model independent confirmation of our
observations that is comparable with current constraints.

\section{Possibility of ruling out modified gravity models}

\begin{figure}[htb]
\begin{center}
\includegraphics[width=9cm]{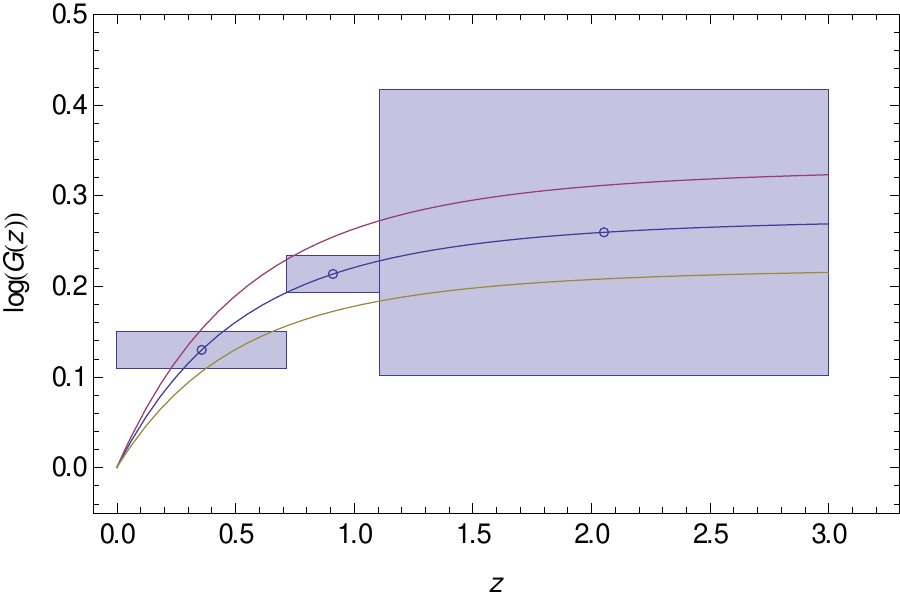}
\includegraphics[width=9cm]{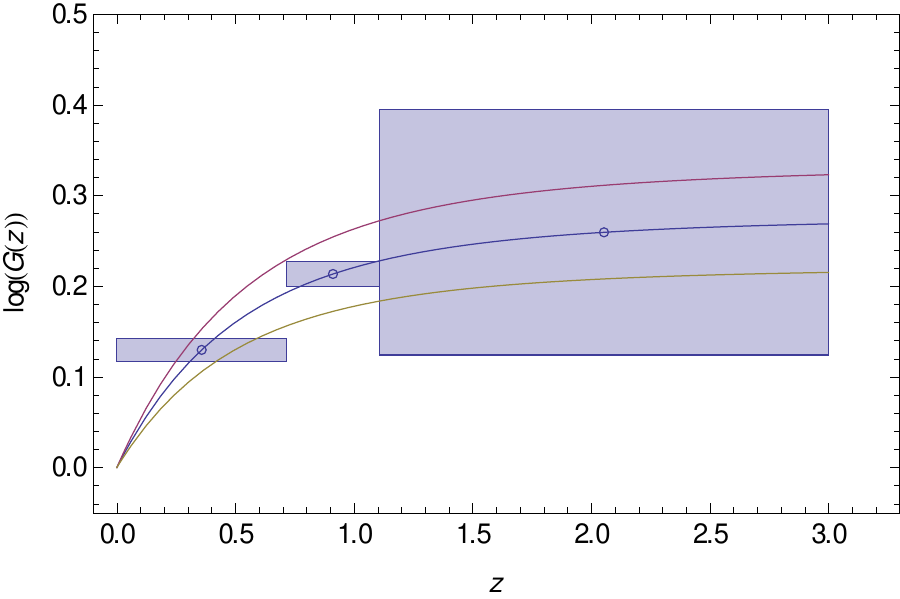}
\end{center}
\caption{The growth function with error bars for the binning with the
highest FOM. Also shown (from the top): The growth function from
eq.~(\ref{eq:3-growth}) for the DGP model $\gamma_\mathrm{DGP}=11/16\approx
0.69$, the standard model $\gamma=6/11\approx0.55$ and the Starobinsky model
with $\gamma_\mathrm{Sta}=0.42$. \textit{Top panel:} Weak lensing only.
\textit{Bottom panel:} Including priors from WMAP and {\sc Planck}.}
\label{fig:growth-best}
\end{figure}

According to the data displayed in fig.~\ref{fig:4-hg-hist}, we get the best
results for the growth function when using three redshift bins, which is why
we shall adopt this number for this and the next section.
Fig.~\ref{fig:growth-best} shows the error bars on $G(z)$ in this case,
along with the growth function as predicted by the Starobinsky model with
$\gamma_\mathrm{Sta}=0.42$ \citep{Fu2010,Starobinsky2007} and the DGP model
with $\gamma_\mathrm{DGP}$ by Dvali, Gabadadze and Porrati
\citep{Gong2008,Dvali2000}. We also see the same diagram with priors from
WMAP and {\sc Planck} included. From the visual, we recognize a slight
disagreement at low redshifts while the error bars at high redshifts are
too large to make any statements about the competing models.

One caveat: The Starobinsky model actually predicts a redshift dependent
growth index. Trying to fit a power law to the growth rate in this case,
i.e.  $f=\Omega_m(z)^\gamma_\mathrm{Sta}$ with a constant $\gamma_{Sta}$,
yields a deviation of up to 10\% at intermediate redshifts, which indicates
that the result cited above is biased. A linear expression for the growth
index, $\gamma_\mathrm{Sta}=\gamma_0 + \gamma_1 z$, is more accurate but
fails at high redshifts with $z>0.5$. However, the result, $\gamma_1\approx
-0.24$ indicates a decreasing growth index \citep{Fu2010}. Thus, in the
case of the Starobinsky model the power law does not seem to be a good fit
and the following discussion should be taken with a grain of salt.

\begin{figure}[htb]
\begin{center}
\includegraphics[width=9cm]{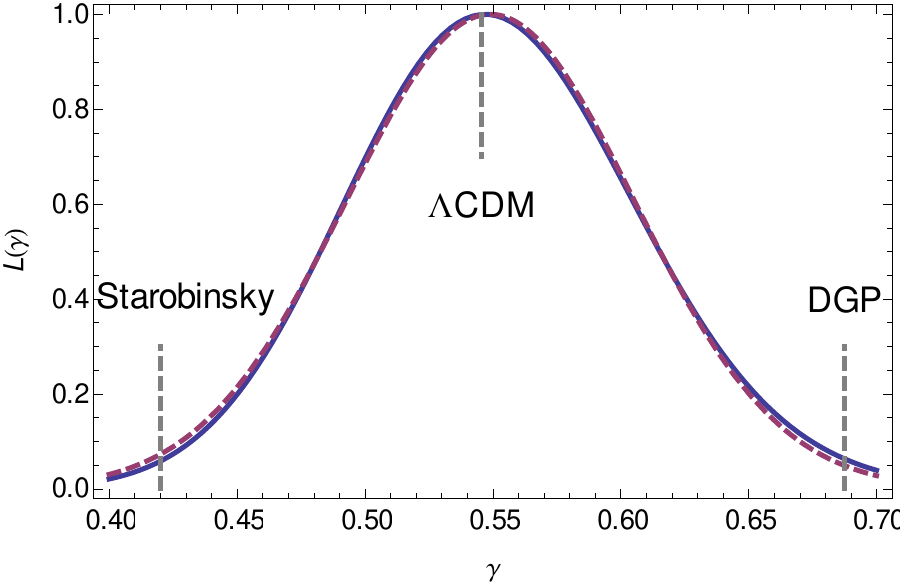}
\includegraphics[width=9cm]{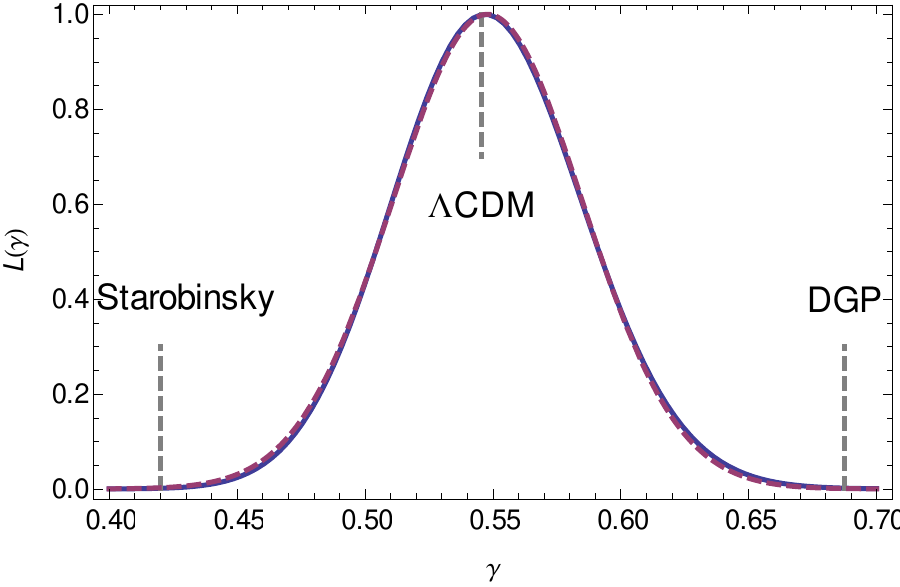}
\end{center}
\caption{The likelihood for the growth index with a fitted Gaussian (dashed
line). Indicated are the predicted values for the Starobinsky, DGP and
standard model.  \textit{Top panel:} Weak lensing only. The Gaussian has a
standard deviation of $\sigma = 0.056$.  \textit{Bottom
panel:} Including priors from WMAP and {\sc Planck}. Here, the standard
deviation is $\sigma = 0.037$.}
\label{fig:likelihood}
\end{figure}

We now want to attempt to quantify the deviation from these two models.
Assuming the errors on the growth function are Gaussian, we can compute the
likelihood function for the growth index $\gamma$ as 
\begin{equation}
\mathcal L(\gamma) = \exp\left( -\frac{1}{2} \sum_{i=1}^{3}
\frac{(G(z_i;\gamma)-g_i)^2}{\sigma(g_i)^2}  \right)
\end{equation}
using the error estimates from the Fisher matrix and keeping $\Omega_m$
fixed. The likelihood is plotted in fig.~\ref{fig:likelihood}. Obviously,
the maximum of the likelihood is found at the fiducial value $\gamma=0.55$.
In this case, the likelihood can be approximated by a Gaussian with good
agreement, and we find a standard deviation of $\sigma_0(\gamma)=0.056$ with
weak lensing only and $\sigma_1(\gamma)=0.037$ while including priors from
WMAP and {\sc Planck}.  Considering that both the DGP model as well as the
Starobinsky model deviate from the fiducial value by at least
$\Delta\gamma=0.13$, they could both be ruled out at the $2\sigma$
confidence level with weak lensing only and at the $3\sigma$ confidence
level with priors. This does not seem to be entirely conclusive, especially
when we consider the uncertainty in the $\gamma_\mathrm{Sta}$ that we
mentioned earlier. However, we need to keep in mind that these constraints
are model independent.

\afterpage{\clearpage}

\section{Problems and outlook}

It is not surprising that the Fisher matrix formalism can only provide a
lower bound on the expected error bars. This bound is called the
Cram\'er-Rao bound, which expresses the fact that the maximum likelihood
estimate constrains the model parameters at best with the value given by the
Fisher matrix \citep{Shao2003}. 
We should also keep in mind that when the errors become too large, the
approximation of the likelihood by a Gaussian may lose its validity.

Besides that, we did not take into account several systematic errors that
might present itself while observing the weak lensing signal, like the
aforementioned intrinsic alignment. Other potential error sources include:
shear calibration  errors, intrinsic galaxy alignment, and non-Gaussian
statistics, some of which are discussed in \citet{Bernstein2009} and
\citet{Albrecht2009}.

%

Regarding the source code of the notebook, while it does satisfy all the
requirements for calculating the Fisher matrix, there certainly is still
substantial room for improvement. One desirable feature would be the
abstraction of cosmological parameters, such that they would not have to be
hard coded into the arguments of any function, making it easier to add or
remove parameters later on. In appendix~\ref{coupledquint}, large parts of
the code were reused, but almost all arguments had to be adjusted manually,
a cumbersome and error prone job. Another issue, that the runtime on a
typical machine is of order several hours to a few days with parallelizing
part of the code, which could perhaps be reduced significantly by examining
the step size of interpolated functions more carefully. Many of them (and
especially the window functions, a very central part of the computation)
only have a single extremum and tend to approach zero after that. Limiting
the evaluation of sampling points specifically to regions where the function
takes non-negligible  values might speed up the computation drastically.
From a more structural point of view, a more modular setup by using \matha
packages would be desirable to avoid duplicate code fragments, which would
make the program more expandable, robust, and easier to use.

%% file: chapter6.tex
\chapter{Conclusion}

We presented and developed the formal basis of the theory of dark energy,
weak lensing, and statistics in cosmology (in particular the Fisher matrix
formalism), on which the remainder of this thesis built. Furthermore, we
described the specifics of the methods of weak lensing tomography using
fitting formulas for the matter power spectrum with non-linear corrections
and how it is used to compute the observable, the convergence power
spectrum. It was demonstrated how we avoided assuming an underlying model
for the Hubble parameter and growth function by using linear interpolation
based on a choice of redshift bins. We showed how the convergence power
spectrum is related to the matter power spectrum and how the matter power
spectrum can be separated into the growth function and the transfer
function. We gave the fitting formula for the transfer function and how the
non-linear corrections are applied. Finally, we outlined how to implement
the computation in \matha and what kind of challenges might arise in doing
so.

Then we analyzed the behavior of the Fisher matrix when varying different
parameters and why it makes sense to use their particular values. We showed
how to include priors from other experiments and how it improves the FOM, a
quantity that we defined as a quality measure. We found that a small number
number of redshift bins is sufficient to extract most of the cosmological
information from the weak lensing signal -- five for best constraints of
$H(z)$ and three for $G(z)$. It turned out that the error bars on the growth
function are most sensitive to the fractional matter density $\Omega_m$ and
power spectrum amplitude $\sigma_8$. The results indicated that the
constraints that we can expect are comparable to current constraints from
several probes combined. 
According to a likelihood analysis, we conclude that future weak lensing
surveys, using this kind of technique on $H(z)$ and $G(z)$, might be able to
rule out the Starobinsky and DGP model for modified gravity only at a
$2\sigma$ level when used by itself. Together with priors from surveys of
the cosmic microwave background like {\sc Planck}, this might improve to a
$3\sigma$ level instead.

It may not have the kind of certainty that meets the ambitious standards of
physics, but this the price we pay for parameterizing cosmological
quantities this way. As a model independent approach it represents an
important check that is not biased towards a model, supplementary to other
measurements.

Regardless, the next decade promises to be an exciting
one for cosmology and dark energy research as new probes and unprecedented
high precision data will become available from {\sc Euclid} and similar
surveys.  Perhaps it is weak lensing what will unravel the mystery that is
dark energy by providing the necessary insights into the evolution of the
growth of structure\dots we can only imagine what the consequences
for physics will be.

%% file: coupled.tex
\chapter{Weak lensing Fisher matrix for coupled
quintessence}\label{coupledquint}

With quintessence being one of the most popular candidates for dark energy
\citep{Copeland2008}, it is interesting to study the coupling between a
scalar field and matter. A simple case would manifest in the conservation
equations with a coupling constant $C$ such that
\begin{align}
T^{\mu\nu}_{(m);\mu} &= -C T_{(m)} \phi^{;\nu}\\
T^{\mu\nu}_{(\phi);\mu}& = C T_{(\phi)}\phi^{;\nu}\,,
\end{align}
where $T_{(m)}^{\mu\nu}$ and $T_{(\phi)}^{\mu\nu}$ describe the
energy-momentum tensor of matter and the scalar field
respectively (see \cite{Amendola1999,DiPorto2008}).  It is convenient to
define a dimensionless coupling constant $\beta$ proportional to $C$, i.e.
\begin{equation}
\beta \equiv \frac{C}{\sqrt{8\pi G}}\,.
\end{equation}
The evolution of the density contrast $\delta(a)$ must then satisfy
\begin{equation}
\delta''(\alpha) + \left( \frac{1}{2} - \frac{\Omega_m'}{2\Omega_m}
\right)\delta'(\alpha) - \frac{3}{2}\Omega_m(1+2\beta^2) \delta(\alpha)=0\,,
\end{equation}
where $\alpha \equiv \ln a$ and $'=\dd/\dd \ln a$. In an attempt to
approximate the solution $\delta(a)$ and considering that $G(z)=\delta(z)$,
one can try to generalize
the solution of the original equation (with $\beta=0$) phenomenologically as
\begin{equation}
G(z) = \exp\left( -\int_0^z
\Omega_m(z')^\gamma(1+c\beta^2)\frac{\dd z'}{1+z} \right)\,.
\end{equation}
\citet{DiPorto2008} found that a well working least square fit to the exact
solution, derived numerically, yields the values 
\begin{equation}
\gamma = 0.56\,,\qquad c=2.1\,.
\end{equation}

Provided that we are given the transfer function for this scenario to plug
into eq.~(\ref{eq:3-mps1}), we can skip the fitting procedure for the matter
power spectrum in
section~\ref{subsec:fitting} and proceed directly with the non-linear
corrections and the convergence power spectrum.
Now we can run the code with a slightly modified growth function and a
fiducial model with the parameters 
\begin{equation}
(\beta, \sigma, \Omega_c, h,\Omega_b,n_s)=(0.01, 0.2, 0.226, 0.703, 0.0451,
0.966)\,,
\end{equation}
where $\sigma$ is the exponent in the potential for the scalar field, which
we take here to be
\begin{equation}
V(\phi) = V_0 \phi^{-\sigma}\,.
\end{equation}

\begin{table}\small
\centering
\begin{tabular}{cccccc}
\hline\hline
$\beta$&$ \sigma$&$ \Omega_c$&$ h$&$\Omega_b$&$n_s$\\
\hline
6.08293\text{e}5 & $-$1.28992\text{e}5 &
   5.48499\text{e}6 & 3.73233\text{e}6 &
   3.3104\text{e}6 & 5.02321\text{e}5 \\
 $-$1.28992\text{e}5 & 2.80677\text{e}4 &
   $-$1.2008\text{e}6 & $-$8.22476\text{e}5 &
   $-$7.44588\text{e}5 & $-$1.044\text{e}5 \\
 5.48499\text{e}6 & $-$1.2008\text{e}6 &
   5.28593\text{e}7 & 3.61948\text{e}7 &
   3.27805\text{e}7 & 4.54416\text{e}6 \\
 3.73233\text{e}6 & $-$8.22476\text{e}5 &
   3.61948\text{e}7 & 2.48333\text{e}7 &
   2.26252\text{e}7 & 3.06263\text{e}6 \\
 3.3104\text{e}6 & $-$7.44588\text{e}5 &
   3.27805\text{e}7 & 2.26252\text{e}7 &
   2.1013\text{e}7 & 2.62304\text{e}6 \\
 5.02321\text{e}5 & $-$1.044\text{e}5 &
   4.54416\text{e}6 & 3.06263\text{e}6 &
   2.62304\text{e}6 & 4.55493\text{e}5\\
\hline
\end{tabular}
\caption{The weak lensing Fisher matrix for coupled quintessence with
$\calN=5$, $\ell_\mathrm{max}=10^4$, $\Delta\lg\ell=0.2$. }
\label{tab:cqfm}
\end{table}

\begin{figure}[htbp]
\begin{center}
\includegraphics[width=\textwidth]{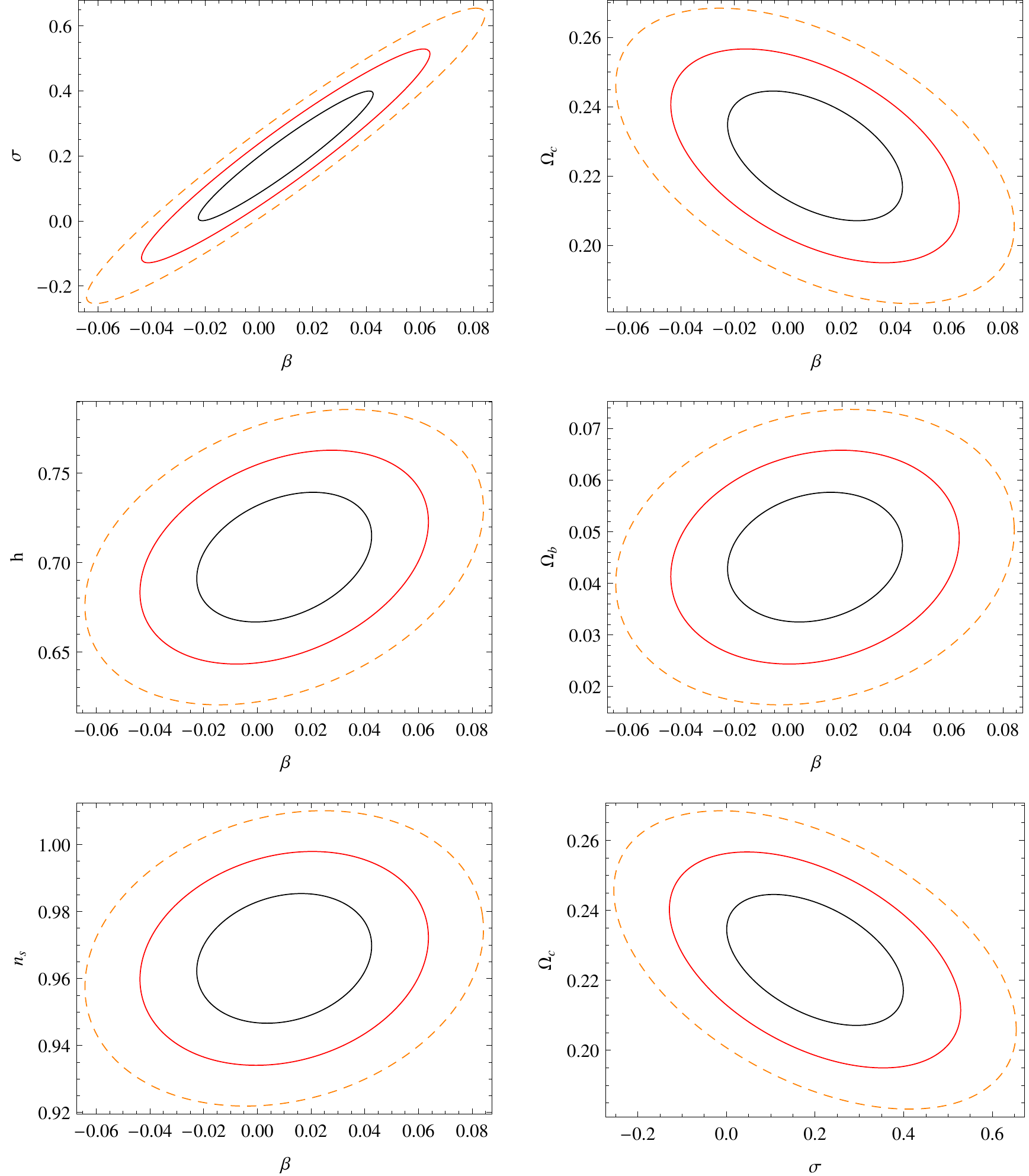}
\end{center}
\caption{Confidence ellipses for dark energy parameters in the coupled
quintessence model. Shown are the ellipses for the 68\%, 95\% and 99\%
confidence level.} 
\label{fig:coup-ellipses1}
\end{figure}

\begin{figure}[htbp]
\begin{center}
\includegraphics[width=\textwidth]{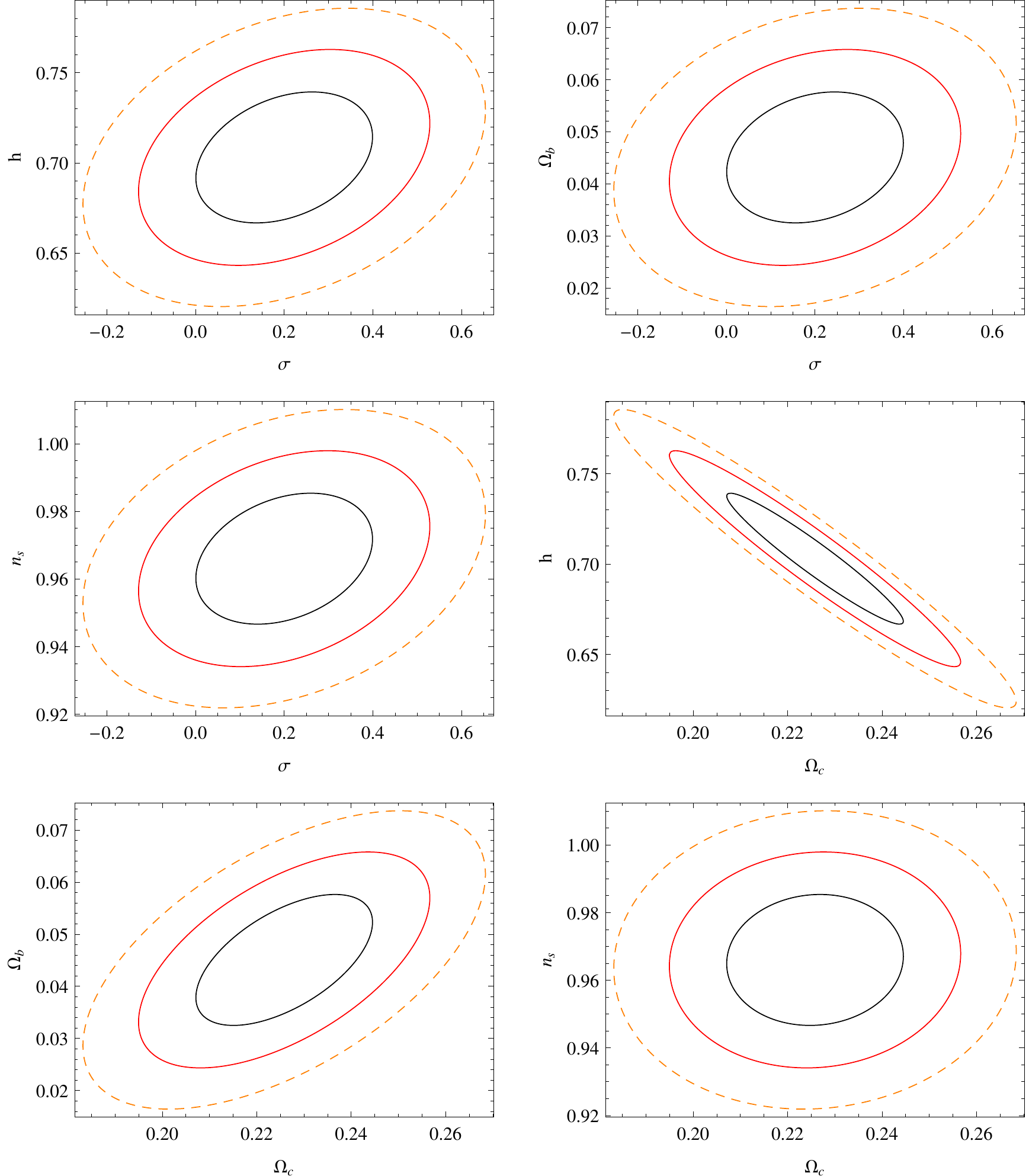}
\end{center}
\caption{Confidence ellipses for dark energy parameters in the coupled
quintessence model (continued).}
\label{fig:coup-ellipses2}
\end{figure}

Using $\calN=5$, $\ell_\mathrm{max}=10^4$, $\Delta\lg\ell=0.2$, we find
the resulting Fisher to be as in tab.~\ref{tab:cqfm}.
We can now show the correlation of all possible parameter pairs by
marginalizing over 
all other parameters such that the correlation matrix takes the form
\begin{equation}
\mat C = \left( \begin{array}{cc}
\sigma_1^2 & \rho\sigma_1\sigma_2\\
\rho\sigma_1\sigma_2 & \sigma_2^2
\end{array}\right)\,.
\end{equation}
Then we can plot ellipses whose semiaxes are proportional to the square
root of the eigenvalues of $\mat C$, i.e.
\begin{equation}
a_{1,2}^2 = \frac{1}{2} \left( \sigma_1^2+\sigma_2^2\mp
\sqrt{\sigma_1^4-2\sigma_1^2\sigma_2^2 + 4\rho^2\sigma_1^2\sigma_2^s +
\sigma_2^4} \right)
\end{equation}
with an angle with respect to the coordinate axes of
\begin{equation}
\tan 2\alpha = \frac{2\rho\sigma_1\sigma_2}{\sigma_1^2-\sigma_2^2}\,.
\end{equation}
The semiaxes need to be scaled by a factor of 1.51, 2.49, 3.44 so that the
area of the ellipsis corresponds to the $1\sigma$, $2\sigma$, $3\sigma$
probability content respectively (A\&T, ch. 13.3). The result can be seen in
fig.~\ref{fig:coup-ellipses1} and~\ref{fig:coup-ellipses2}
\citep{Amendola2011}. These plots enable us to judge how different surveys
can complement each other, e.g. we could overlay ellipses derived from CMB
(if we had it available)
data and see how much the confidence regions overlap. Ideally, we would want
to see thin, perpendicular ellipses.


%% file: thanks.tex
\chapter*{Acknowledgement}

This thesis represents the culmination of a six year long journey through a
wonderful world full of femtoseconds and Gigayears, strings and galaxies,
harmonic oscillators and harmonic oscillators, Germany and New
Mexico, coffee and beer, and many people who I had the pleasure of meeting
during this journey deserve to be mentioned here.

First of all, words cannot express strongly enough my gratitude towards the
love of my life and best friend Kathleen, who was always there for me and
made great sacrifices to do so.

I also wish to thank my parents for their unconditional support over
numerous years, without which studying physics would have been a lot more
difficult and I would not be anywhere near where I am now. Danke!

Of course I also appreciate that Professor Luca Amendola gave me the
opportunity of having an active part in one of the most exciting research
areas of our time, as well as the Institute for Theoretical Physics for
providing an excellent work environment in their beautiful villa. Thanks to
Professor Matthias Bartelmann for agreeing to be the second referee.

Many friends and roommates that I met in Würzburg, Heidelberg and
Albuquerque---too many to name here without taking the risk of forgetting
someone---that I spent endless hours with socializing deserve to be thanked
as well.

For writing this thesis I heavily relied on a number of outstanding free
software tools, for which I thank the many developers of GNU/Linux, \LaTeX,
Vim, subversion, Gimp, KDE and many more.

This research has made use of NASA's Astrophysics Data System.

\pagestyle{empty}
\cleardoublepage

\begingroup
\selectlanguage{\ngerman}
\thispagestyle{empty}
\chapter*{Erklärung}
Ich versichere, dass ich diese Arbeit selbstständig verfasst habe und keine
anderen als die angegebenen Quellen und Hilfsmittel benutzt habe.\\[2cm]
Heidelberg, den 8. Juni 2011 \hspace{3cm}\dotfill\\
\mbox{}\hfill\textit{(Adrian Vollmer)}\hspace{.8cm}
\endgroup

%% file: thesis.bbl
\begin{thebibliography}{59}
\expandafter\ifx\csname natexlab\endcsname\relax\def\natexlab#1{#1}\fi
\expandafter\ifx\csname bibnamefont\endcsname\relax
  \def\bibnamefont#1{#1}\fi
\expandafter\ifx\csname bibfnamefont\endcsname\relax
  \def\bibfnamefont#1{#1}\fi
\expandafter\ifx\csname citenamefont\endcsname\relax
  \def\citenamefont#1{#1}\fi
\expandafter\ifx\csname url\endcsname\relax
  \def\url#1{\texttt{#1}}\fi
\expandafter\ifx\csname urlprefix\endcsname\relax\def\urlprefix{}\fi
\providecommand{\bibinfo}[2]{#2}
\providecommand{\eprint}[2][]{\url{#2}}

\bibitem[{\citenamefont{Albrecht} {}\emph{et~al.}(2009)\citenamefont{Albrecht,
  Amendola, Bernstein, Clowe, Eisenstein, Guzzo, Hirata, Huterer, Kirshner,
  Kolb, and Nichol}}]{Albrecht2009}
\bibinfo{author}{\bibnamefont{Albrecht}, \bibfnamefont{A.}},
  \bibinfo{author}{\bibfnamefont{L.}~\bibnamefont{Amendola}},
  \bibinfo{author}{\bibfnamefont{G.}~\bibnamefont{Bernstein}},
  \bibinfo{author}{\bibfnamefont{D.}~\bibnamefont{Clowe}},
  \bibinfo{author}{\bibfnamefont{D.}~\bibnamefont{Eisenstein}},
  \bibinfo{author}{\bibfnamefont{L.}~\bibnamefont{Guzzo}},
  \bibinfo{author}{\bibfnamefont{C.}~\bibnamefont{Hirata}},
  \bibinfo{author}{\bibfnamefont{D.}~\bibnamefont{Huterer}},
  \bibinfo{author}{\bibfnamefont{R.}~\bibnamefont{Kirshner}},
  \bibinfo{author}{\bibfnamefont{E.}~\bibnamefont{Kolb}}, and
  \bibinfo{author}{\bibfnamefont{R.}~\bibnamefont{Nichol}},
  \bibinfo{year}{2009}, {}\emph{\bibinfo{title}{{Findings of the Joint Dark
  Energy Mission Figure of Merit Science Working Group}}},
  \bibinfo{type}{Technical Report} \bibinfo{number}{2},
  \href{http://arxiv.org/abs/0901.0721}{{\ttfamily {}arXiv:0901.0721}}.

\bibitem[{\citenamefont{Albrecht} {}\emph{et~al.}(2006)\citenamefont{Albrecht,
  Bernstein, Cahn, Freedman, Hewitt, Hu, Huth, Kamionkowski, Kolb, Knox,
  Mather, Staggs} {}\emph{et~al.}}]{Albrecht2006}
\bibinfo{author}{\bibnamefont{Albrecht}, \bibfnamefont{A.}},
  \bibinfo{author}{\bibfnamefont{G.}~\bibnamefont{Bernstein}},
  \bibinfo{author}{\bibfnamefont{R.}~\bibnamefont{Cahn}},
  \bibinfo{author}{\bibfnamefont{W.~L.} \bibnamefont{Freedman}},
  \bibinfo{author}{\bibfnamefont{J.}~\bibnamefont{Hewitt}},
  \bibinfo{author}{\bibfnamefont{W.}~\bibnamefont{Hu}},
  \bibinfo{author}{\bibfnamefont{J.}~\bibnamefont{Huth}},
  \bibinfo{author}{\bibfnamefont{M.}~\bibnamefont{Kamionkowski}},
  \bibinfo{author}{\bibfnamefont{E.~W.} \bibnamefont{Kolb}},
  \bibinfo{author}{\bibfnamefont{L.}~\bibnamefont{Knox}},
  \bibinfo{author}{\bibfnamefont{J.~C.} \bibnamefont{Mather}},
  \bibinfo{author}{\bibfnamefont{S.}~\bibnamefont{Staggs}}, {}\emph{et~al.},
  \bibinfo{year}{2006}, {}\emph{\bibinfo{title}{{Report of the Dark Energy Task
  Force}}}, \bibinfo{type}{Technical Report},
  \href{http://arxiv.org/abs/astro-ph/0609591}{{\ttfamily
  {}arXiv:astro-ph/0609591}}.

\bibitem[{\citenamefont{Amara and Refregier}(2007)}]{Amara2008}
\bibinfo{author}{\bibnamefont{Amara}, \bibfnamefont{A.}}, and
  \bibinfo{author}{\bibfnamefont{A.}~\bibnamefont{Refregier}},
  \bibinfo{year}{2007}, \bibinfo{title}{``{Systematic Bias in Cosmic Shear:
  Beyond the Fisher Matrix}''},
  \href{http://dx.doi.org/10.1111/j.1365-2966.2008.13880.x}{\bibinfo{journal}{%
Monthly Notices of the Royal Astronomical Society}
  \textbf{\bibinfo{volume}{391}}(\bibinfo{number}{1})}, \bibinfo{pages}{11},
  \href{http://arxiv.org/abs/0710.5171v1}{{\ttfamily {}arXiv:0710.5171v1
  [astro-ph]}}.

\bibitem[{\citenamefont{Amendola}(1999)}]{Amendola1999}
\bibinfo{author}{\bibnamefont{Amendola}, \bibfnamefont{L.}},
  \bibinfo{year}{1999}, \bibinfo{title}{``{Coupled Quintessence}''},
  \href{http://dx.doi.org/10.1103/PhysRevD.62.043511}{\bibinfo{journal}{Physic%
al Review D} \textbf{\bibinfo{volume}{62}}(\bibinfo{number}{4})},
  \bibinfo{pages}{43511},
  \href{http://arxiv.org/abs/astro-ph/9908023}{{\ttfamily
  {}arXiv:astro-ph/9908023}}.

\bibitem[{\citenamefont{Amendola} {}\emph{et~al.}(2006)\citenamefont{Amendola,
  Charmousis, and Davis}}]{Amendola2006}
\bibinfo{author}{\bibnamefont{Amendola}, \bibfnamefont{L.}},
  \bibinfo{author}{\bibfnamefont{C.}~\bibnamefont{Charmousis}}, and
  \bibinfo{author}{\bibfnamefont{S.~C.} \bibnamefont{Davis}},
  \bibinfo{year}{2006}, \bibinfo{title}{``{Constraints on Gauss-Bonnet Gravity
  in Dark Energy Cosmologies}''},
  \href{http://dx.doi.org/10.1088/1475-7516/2006/12/020}{\bibinfo{journal}{Jou%
rnal of Cosmology and Astroparticle Physics}
  \textbf{\bibinfo{volume}{2006}}(\bibinfo{number}{12})},
  \href{http://arxiv.org/abs/hep-th/0506137}{{\ttfamily
  {}arXiv:hep-th/0506137}}.

\bibitem[{\citenamefont{Amendola} {}\emph{et~al.}(2008)\citenamefont{Amendola,
  Kunz, and Sapone}}]{Amendola2008}
\bibinfo{author}{\bibnamefont{Amendola}, \bibfnamefont{L.}},
  \bibinfo{author}{\bibfnamefont{M.}~\bibnamefont{Kunz}}, and
  \bibinfo{author}{\bibfnamefont{D.}~\bibnamefont{Sapone}},
  \bibinfo{year}{2008}, \bibinfo{title}{``{Measuring the dark side (with weak
  lensing)}''},
  \href{http://dx.doi.org/10.1088/1475-7516/2008/04/013}{\bibinfo{journal}{Jou%
rnal of Cosmology and Astroparticle Physics}
  \textbf{\bibinfo{volume}{2008}}(\bibinfo{number}{04})}, \bibinfo{pages}{013}.

\bibitem[{\citenamefont{Amendola and Tsujikawa}(2010)}]{Amendola2010}
\bibinfo{author}{\bibnamefont{Amendola}, \bibfnamefont{L.}}, and
  \bibinfo{author}{\bibfnamefont{S.}~\bibnamefont{Tsujikawa}},
  \bibinfo{year}{2010},
  \href{http://www.amazon.com/Dark-Energy-Observations-Luca-Amendola/dp/052151%
6005}{{}\emph{\bibinfo{title}{{Dark Energy: Theory and Observations}}}}
  (\bibinfo{publisher}{Cambridge University Press}), ISBN
  \bibinfo{isbn}{0521516005}.

\bibitem[{\citenamefont{Amendola} {}\emph{et~al.}(2011)\citenamefont{Amendola,
  Valeria, Quercellini, and Vollmer}}]{Amendola2011}
\bibinfo{author}{\bibnamefont{Amendola}, \bibfnamefont{L.}},
  \bibinfo{author}{\bibfnamefont{P.}~\bibnamefont{Valeria}},
  \bibinfo{author}{\bibfnamefont{C.}~\bibnamefont{Quercellini}}, and
  \bibinfo{author}{\bibfnamefont{A.}~\bibnamefont{Vollmer}},
  \bibinfo{year}{2011}, \bibinfo{title}{``{Constraining Coupled Dark Energy
  with Future CMB and Weak Lensing data}''}, \bibinfo{pages}{to be published}.

\bibitem[{\citenamefont{Barausse} {}\emph{et~al.}(2005)\citenamefont{Barausse,
  Matarrese, and Riotto}}]{Barausse2005}
\bibinfo{author}{\bibnamefont{Barausse}, \bibfnamefont{E.}},
  \bibinfo{author}{\bibfnamefont{S.}~\bibnamefont{Matarrese}}, and
  \bibinfo{author}{\bibfnamefont{A.}~\bibnamefont{Riotto}},
  \bibinfo{year}{2005}, \bibinfo{title}{``{Effect of Inhomogeneities on the
  Luminosity Distance-Redshift Relation: Is Dark Energy Necessary in a
  Perturbed Universe?}''},
  \href{http://dx.doi.org/10.1103/PhysRevD.71.063537}{\bibinfo{journal}{Physic%
al Review D} \textbf{\bibinfo{volume}{71}}(\bibinfo{number}{6})},
  \bibinfo{pages}{19}, \href{http://arxiv.org/abs/astro-ph/0501152}{{\ttfamily
  {}arXiv:astro-ph/0501152}}.

\bibitem[{\citenamefont{Bartelmann and Schneider}(1999)}]{Bartelmann2001}
\bibinfo{author}{\bibnamefont{Bartelmann}, \bibfnamefont{M.}}, and
  \bibinfo{author}{\bibfnamefont{P.}~\bibnamefont{Schneider}},
  \bibinfo{year}{1999}, \bibinfo{title}{``{Weak Gravitational Lensing}''},
  \href{http://dx.doi.org/10.1016/S0370-1573(00)00082-X}{\bibinfo{journal}{Phy%
sics Reports} \textbf{\bibinfo{volume}{340}}(\bibinfo{number}{4-5})},
  \bibinfo{pages}{223}, \href{http://arxiv.org/abs/astro-ph/9912508}{{\ttfamily
  {}arXiv:astro-ph/9912508}}.

\bibitem[{\citenamefont{Bellman}(2003)}]{Bellman2003}
\bibinfo{author}{\bibnamefont{Bellman}, \bibfnamefont{R.~E.}},
  \bibinfo{year}{2003},
  \href{http://books.google.com/books?id=fyVtp3EMxasC&pgis=1}{{}\emph{\bibinfo%
{title}{{Dynamic Programming}}}} (\bibinfo{publisher}{Courier Dover
  Publications}), ISBN \bibinfo{isbn}{0486428095}.

\bibitem[{\citenamefont{Bernstein}(2009)}]{Bernstein2009}
\bibinfo{author}{\bibnamefont{Bernstein}, \bibfnamefont{G.~M.}},
  \bibinfo{year}{2009}, \bibinfo{title}{``{Comprehensive Two-Point Analyses of
  Weak Gravitational Lensing Surveys}''},
  \href{http://dx.doi.org/10.1088/0004-637X/695/1/652}{\bibinfo{journal}{The
  Astrophysical Journal} \textbf{\bibinfo{volume}{695}}(\bibinfo{number}{1})},
  \bibinfo{pages}{652}, \href{http://arxiv.org/abs/0808.3400v1}{{\ttfamily
  {}arXiv:0808.3400v1 [astro-ph]}}.

\bibitem[{\citenamefont{Brumfiel}(2007)}]{Brumfiel2007}
\bibinfo{author}{\bibnamefont{Brumfiel}, \bibfnamefont{G.}},
  \bibinfo{year}{2007}, \bibinfo{title}{``{Unseen Universe: A Constant
  Problem.}''},
  \href{http://dx.doi.org/10.1038/448245a}{\bibinfo{journal}{Nature}
  \textbf{\bibinfo{volume}{448}}(\bibinfo{number}{7151})},
  \bibinfo{pages}{245}.

\bibitem[{\citenamefont{Carroll}(2003)}]{Carroll2003}
\bibinfo{author}{\bibnamefont{Carroll}, \bibfnamefont{S.}},
  \bibinfo{year}{2003},
  \href{http://www.amazon.com/Spacetime-Geometry-Introduction-General-Relativi%
ty/dp/0805387323}{{}\emph{\bibinfo{title}{{Spacetime and Geometry: An
  Introduction to General Relativity}}}} (\bibinfo{publisher}{Benjamin
  Cummings}), ISBN \bibinfo{isbn}{0805387323}.

\bibitem[{\citenamefont{Carroll} {}\emph{et~al.}(1992)\citenamefont{Carroll,
  Press, and Turner}}]{Carroll1992}
\bibinfo{author}{\bibnamefont{Carroll}, \bibfnamefont{S.}},
  \bibinfo{author}{\bibfnamefont{W.~H.} \bibnamefont{Press}}, and
  \bibinfo{author}{\bibfnamefont{E.~L.} \bibnamefont{Turner}},
  \bibinfo{year}{1992}, \bibinfo{title}{``{The Cosmological Constant}''},
  \href{http://dx.doi.org/10.1146/annurev.aa.30.090192.002435}{\bibinfo{journa%
l}{Annual Review of Astronomy and Astrophysics}
  \textbf{\bibinfo{volume}{30}}(\bibinfo{number}{1})}, \bibinfo{pages}{499}.

\bibitem[{\citenamefont{Chevallier and Polarski}(2001)}]{Chevallier2001}
\bibinfo{author}{\bibnamefont{Chevallier}, \bibfnamefont{M.}}, and
  \bibinfo{author}{\bibfnamefont{D.}~\bibnamefont{Polarski}},
  \bibinfo{year}{2001}, \bibinfo{title}{``{Accelerating Universes with Scaling
  Dark Matter}''},
  \href{http://dx.doi.org/10.1142/S0218271801000822}{\bibinfo{journal}{Interna%
tional Journal of Modern Physics D} \textbf{\bibinfo{volume}{10}}},
  \bibinfo{pages}{213}, \href{http://arxiv.org/abs/gr-qc/0009008}{{\ttfamily
  {}arXiv:gr-qc/0009008}}.

\bibitem[{\citenamefont{Copeland} {}\emph{et~al.}(2006)\citenamefont{Copeland,
  Sami, and Tsujikawa}}]{Copeland2008}
\bibinfo{author}{\bibnamefont{Copeland}, \bibfnamefont{E.~J.}},
  \bibinfo{author}{\bibfnamefont{M.}~\bibnamefont{Sami}}, and
  \bibinfo{author}{\bibfnamefont{S.}~\bibnamefont{Tsujikawa}},
  \bibinfo{year}{2006}, \bibinfo{title}{``{Dynamics of dark energy}''},
  \href{http://dx.doi.org/10.1142/S021827180600942X}{\bibinfo{journal}{Physics}
  }, \bibinfo{pages}{93}, \href{http://arxiv.org/abs/hep-th/0603057}{{\ttfamily
  {}arXiv:hep-th/0603057}}.

\bibitem[{\citenamefont{{De Felice} and Tsujikawa}(2010)}]{Felice2010}
\bibinfo{author}{\bibnamefont{{De Felice}}, \bibfnamefont{A.}}, and
  \bibinfo{author}{\bibfnamefont{S.}~\bibnamefont{Tsujikawa}},
  \bibinfo{year}{2010}, \bibinfo{title}{``{f(R) Theories}''},
  \bibinfo{journal}{Living Reviews in Relativity}
  \textbf{\bibinfo{volume}{13}}(\bibinfo{number}{3}),
  \bibinfo{pages}{1}\href{http://arxiv.org/abs/1002.4928v2}{{\ttfamily
  {}arXiv:1002.4928v2 [gr-qc]}}.

\bibitem[{\citenamefont{{Di Porto} and Amendola}(2008)}]{DiPorto2008}
\bibinfo{author}{\bibnamefont{{Di Porto}}, \bibfnamefont{C.}}, and
  \bibinfo{author}{\bibfnamefont{L.}~\bibnamefont{Amendola}},
  \bibinfo{year}{2008}, \bibinfo{title}{``{Observational constraints on the
  linear fluctuation growth rate}''},
  \href{http://dx.doi.org/10.1103/PhysRevD.77.083508}{\bibinfo{journal}{Physic%
al Review D} \textbf{\bibinfo{volume}{77}}(\bibinfo{number}{8})},
  \bibinfo{pages}{9}, \href{http://arxiv.org/abs/0707.2686}{{\ttfamily
  {}arXiv:0707.2686}}.

\bibitem[{\citenamefont{Dodelson}(2003)}]{Dodelson2003}
\bibinfo{author}{\bibnamefont{Dodelson}, \bibfnamefont{S.}},
  \bibinfo{year}{2003},
  \href{http://www.amazon.com/Modern-Cosmology-Scott-Dodelson/dp/0122191412}{{%
}\emph{\bibinfo{title}{{Modern Cosmology}}}} (\bibinfo{publisher}{Academic
  Press}), ISBN \bibinfo{isbn}{0122191412}.

\bibitem[{\citenamefont{Douglas}(2003)}]{Douglas2003}
\bibinfo{author}{\bibnamefont{Douglas}, \bibfnamefont{M.~R.}},
  \bibinfo{year}{2003}, \bibinfo{title}{``{The Statistics of String/M Theory
  Vacua}''},
  \href{http://dx.doi.org/10.1088/1126-6708/2003/05/046}{\bibinfo{journal}{Jou%
rnal of High Energy Physics}
  \textbf{\bibinfo{volume}{305}}(\bibinfo{number}{05})}, \bibinfo{pages}{046},
  \href{http://arxiv.org/abs/hep-th/0303194}{{\ttfamily
  {}arXiv:hep-th/0303194}}.

\bibitem[{\citenamefont{Dvali} {}\emph{et~al.}(2000)\citenamefont{Dvali,
  Gabadadze, and Porrati}}]{Dvali2000}
\bibinfo{author}{\bibnamefont{Dvali}, \bibfnamefont{G.}},
  \bibinfo{author}{\bibfnamefont{G.}~\bibnamefont{Gabadadze}}, and
  \bibinfo{author}{\bibfnamefont{M.}~\bibnamefont{Porrati}},
  \bibinfo{year}{2000}, \bibinfo{title}{``{4D Gravity on a Brane in 5D
  Minkowski Space}''},
  \href{http://dx.doi.org/10.1016/S0370-2693(00)00669-9}{\bibinfo{journal}{Phy%
sics Letters B} \textbf{\bibinfo{volume}{485}}(\bibinfo{number}{1-3})},
  \bibinfo{pages}{11}, \href{http://arxiv.org/abs/hep-th/0005016}{{\ttfamily
  {}arXiv:hep-th/0005016}}.

\bibitem[{\citenamefont{Einstein}(1917)}]{Einstein1917}
\bibinfo{author}{\bibnamefont{Einstein}, \bibfnamefont{A.}},
  \bibinfo{year}{1917}, \bibinfo{title}{``{Kosmologische Betrachtungen zur
  allgemeinen Relativit\"{a}tstheorie}''}, \bibinfo{journal}{Sitzungsberichte
  der K\"{o}niglich Preu\ss ischen Akademie der Wissenschaften (Berlin)}
  \textbf{\bibinfo{volume}{1}}, \bibinfo{pages}{142}.

\bibitem[{\citenamefont{Eisenstein and Hu}(1999)}]{Eisenstein1999}
\bibinfo{author}{\bibnamefont{Eisenstein}, \bibfnamefont{D.~J.}}, and
  \bibinfo{author}{\bibfnamefont{W.}~\bibnamefont{Hu}}, \bibinfo{year}{1999},
  \bibinfo{title}{``{Power Spectra for Cold Dark Matter and Its Variants}''},
  \href{http://dx.doi.org/10.1086/306640}{\bibinfo{journal}{The Astrophysical
  Journal} \textbf{\bibinfo{volume}{511}}(\bibinfo{number}{1})},
  \bibinfo{pages}{5}, \href{http://arxiv.org/abs/astro-ph/9710252v1}{{\ttfamily
  {}arXiv:astro-ph/9710252v1}}.

\bibitem[{\citenamefont{Fu} {}\emph{et~al.}(2010)\citenamefont{Fu, Wu, and
  Yu}}]{Fu2010}
\bibinfo{author}{\bibnamefont{Fu}, \bibfnamefont{X.}},
  \bibinfo{author}{\bibfnamefont{P.}~\bibnamefont{Wu}}, and
  \bibinfo{author}{\bibfnamefont{H.}~\bibnamefont{Yu}}, \bibinfo{year}{2010},
  \bibinfo{title}{``{The Growth Factor of Matter Perturbations in f({R})
  Gravity}''},
  \href{http://dx.doi.org/10.1140/epjc/s10052-010-1324-4}{\bibinfo{journal}{The
  European Physical Journal C} \textbf{\bibinfo{volume}{68}}},
  \bibinfo{pages}{271}, \href{http://arxiv.org/abs/1012.2249v1}{{\ttfamily
  {}arXiv:1012.2249v1 [gr-qc]}}.

\bibitem[{\citenamefont{Gehrels}(2010)}]{Gehrels2010}
\bibinfo{author}{\bibnamefont{Gehrels}, \bibfnamefont{N.}},
  \bibinfo{year}{2010}, {}\emph{\bibinfo{title}{{The Joint Dark Energy Mission
  (JDEM) / Omega}}}, \bibinfo{type}{Technical Report},
  \bibinfo{institution}{Goddard Space Flight Center},
  \href{http://arxiv.org/abs/1008.4936v1}{{\ttfamily {}arXiv:1008.4936v1
  [astro-ph.CO]}}.

\bibitem[{\citenamefont{Gong}(2008)}]{Gong2008}
\bibinfo{author}{\bibnamefont{Gong}, \bibfnamefont{Y.}}, \bibinfo{year}{2008},
  \bibinfo{title}{``{Growth Factor Parametrization and Modified Gravity}''},
  \href{http://dx.doi.org/10.1103/PhysRevD.78.123010}{\bibinfo{journal}{Physic%
al Review D} \textbf{\bibinfo{volume}{78}}(\bibinfo{number}{12})},
  \bibinfo{pages}{123010}, \href{http://arxiv.org/abs/0808.1316v2}{{\ttfamily
  {}arXiv:0808.1316v2 [astro-ph]}}.

\bibitem[{\citenamefont{Hobson} {}\emph{et~al.}(2009)\citenamefont{Hobson,
  Jaffe, Liddle, Mukherjee, and Parkinson}}]{Hobson2009a}
\bibinfo{author}{\bibnamefont{Hobson}, \bibfnamefont{M.~P.}},
  \bibinfo{author}{\bibfnamefont{A.~H.} \bibnamefont{Jaffe}},
  \bibinfo{author}{\bibfnamefont{A.~R.} \bibnamefont{Liddle}},
  \bibinfo{author}{\bibfnamefont{P.}~\bibnamefont{Mukherjee}}, and
  \bibinfo{author}{\bibfnamefont{D.}~\bibnamefont{Parkinson}},
  \bibinfo{year}{2009},
  \href{http://www.amazon.com/Bayesian-Methods-Cosmology-Michael-Hobson/dp/052%
1887941}{{}\emph{\bibinfo{title}{{Bayesian Methods in Cosmology}}}}
  (\bibinfo{publisher}{Cambridge University Press}), ISBN
  \bibinfo{isbn}{9780521887946}.

\bibitem[{\citenamefont{Hu}(1999)}]{Hu1999}
\bibinfo{author}{\bibnamefont{Hu}, \bibfnamefont{W.}}, \bibinfo{year}{1999},
  \bibinfo{title}{``{Power Spectrum Tomography with Weak Lensing}''},
  \href{http://dx.doi.org/10.1086/312210}{\bibinfo{journal}{The Astrophysical
  Journal} \textbf{\bibinfo{volume}{522}}(\bibinfo{number}{1})},
  \bibinfo{pages}{L21}, \href{http://arxiv.org/abs/astro-ph/9904153}{{\ttfamily
  {}arXiv:astro-ph/9904153}}.

\bibitem[{\citenamefont{Hu and Jain}(2004)}]{Hu2004}
\bibinfo{author}{\bibnamefont{Hu}, \bibfnamefont{W.}}, and
  \bibinfo{author}{\bibfnamefont{B.}~\bibnamefont{Jain}}, \bibinfo{year}{2004},
  \bibinfo{title}{``{Joint galaxy-lensing observables and the dark energy}''},
  \href{http://dx.doi.org/10.1103/PhysRevD.70.043009}{\bibinfo{journal}{Physic%
al Review D} \textbf{\bibinfo{volume}{70}}(\bibinfo{number}{4})},
  \href{http://arxiv.org/abs/astro-ph/0312395}{{\ttfamily
  {}arXiv:astro-ph/0312395}}.

\bibitem[{\citenamefont{Huterer}(2002)}]{Huterer2002}
\bibinfo{author}{\bibnamefont{Huterer}, \bibfnamefont{D.}},
  \bibinfo{year}{2002}, \bibinfo{title}{``{Weak Lensing and Dark Energy}''},
  \href{http://dx.doi.org/10.1103/PhysRevD.65.063001}{\bibinfo{journal}{Physic%
al Review D} \textbf{\bibinfo{volume}{65}}(\bibinfo{number}{6})},
  \bibinfo{pages}{063001}, \href{http://arxiv.org/abs/1001.1758v3}{{\ttfamily
  {}arXiv:1001.1758v3 [astro-ph.CO]}}.

\bibitem[{\citenamefont{Huterer and Linder}(2007)}]{Huterer2007}
\bibinfo{author}{\bibnamefont{Huterer}, \bibfnamefont{D.}}, and
  \bibinfo{author}{\bibfnamefont{E.}~\bibnamefont{Linder}},
  \bibinfo{year}{2007}, \bibinfo{title}{``{Separating Dark Physics from
  Physical Darkness: Minimalist Modified Gravity versus Dark Energy}''},
  \href{http://dx.doi.org/10.1103/PhysRevD.75.023519}{\bibinfo{journal}{Physic%
al Review D} \textbf{\bibinfo{volume}{75}}(\bibinfo{number}{2})},
  \bibinfo{pages}{023519},
  \href{http://arxiv.org/abs/astro-ph/0608681v2}{{\ttfamily
  {}arXiv:astro-ph/0608681v2}}.

\bibitem[{\citenamefont{Huterer} {}\emph{et~al.}(2006)\citenamefont{Huterer,
  Takada, Bernstein, and Jain}}]{Huterer2006}
\bibinfo{author}{\bibnamefont{Huterer}, \bibfnamefont{D.}},
  \bibinfo{author}{\bibfnamefont{M.}~\bibnamefont{Takada}},
  \bibinfo{author}{\bibfnamefont{G.}~\bibnamefont{Bernstein}}, and
  \bibinfo{author}{\bibfnamefont{B.}~\bibnamefont{Jain}}, \bibinfo{year}{2006},
  \bibinfo{title}{``{Systematic errors in future weak-lensing surveys:
  requirements and prospects for self-calibration}''},
  \href{http://dx.doi.org/10.1111/j.1365-2966.2005.09782.x}{\bibinfo{journal}{%
Monthly Notices of the Royal Astronomical Society}
  \textbf{\bibinfo{volume}{114}}}.

\bibitem[{\citenamefont{Ivezic} {}\emph{et~al.}(2008)\citenamefont{Ivezic,
  Tyson, Allsman, Andrew, Angel, Axelrod, Barr, Becker, Becla, Beldica,
  Blandford, Brandt} {}\emph{et~al.}}]{Ivezic2008}
\bibinfo{author}{\bibnamefont{Ivezic}, \bibfnamefont{Z.}},
  \bibinfo{author}{\bibfnamefont{J.~A.} \bibnamefont{Tyson}},
  \bibinfo{author}{\bibfnamefont{R.}~\bibnamefont{Allsman}},
  \bibinfo{author}{\bibfnamefont{J.}~\bibnamefont{Andrew}},
  \bibinfo{author}{\bibfnamefont{R.}~\bibnamefont{Angel}},
  \bibinfo{author}{\bibfnamefont{T.}~\bibnamefont{Axelrod}},
  \bibinfo{author}{\bibfnamefont{J.~D.} \bibnamefont{Barr}},
  \bibinfo{author}{\bibfnamefont{A.~C.} \bibnamefont{Becker}},
  \bibinfo{author}{\bibfnamefont{J.}~\bibnamefont{Becla}},
  \bibinfo{author}{\bibfnamefont{C.}~\bibnamefont{Beldica}},
  \bibinfo{author}{\bibfnamefont{R.~D.} \bibnamefont{Blandford}},
  \bibinfo{author}{\bibfnamefont{W.~N.} \bibnamefont{Brandt}}, {}\emph{et~al.},
  \bibinfo{year}{2008}, {}\emph{\bibinfo{title}{{LSST: From Science Drivers to
  Reference Design and Anticipated Data Products}}}, \bibinfo{type}{Technical
  Report}, \href{http://arxiv.org/abs/0805.2366v1}{{\ttfamily
  {}arXiv:0805.2366v1 [astro-ph]}}.

\bibitem[{\citenamefont{Kolb} {}\emph{et~al.}(2005)\citenamefont{Kolb,
  Matarrese, Notari, and Riotto}}]{Kolb2005}
\bibinfo{author}{\bibnamefont{Kolb}, \bibfnamefont{E.~W.}},
  \bibinfo{author}{\bibfnamefont{S.}~\bibnamefont{Matarrese}},
  \bibinfo{author}{\bibfnamefont{A.}~\bibnamefont{Notari}}, and
  \bibinfo{author}{\bibfnamefont{A.}~\bibnamefont{Riotto}},
  \bibinfo{year}{2005}, \bibinfo{title}{``{Primordial Inflation Explains Why
  the Universe is Accelerating Today}''},
  \href{http://arxiv.org/abs/hep-th/0503117}{{\ttfamily
  {}arXiv:hep-th/0503117}}.

\bibitem[{\citenamefont{Komatsu} {}\emph{et~al.}(2010)\citenamefont{Komatsu,
  Smith, Dunkley, Bennett, Gold, Hinshaw, Jarosik, Larson, Nolta, Page,
  Spergel, Halpern} {}\emph{et~al.}}]{Komatsu2011}
\bibinfo{author}{\bibnamefont{Komatsu}, \bibfnamefont{E.}},
  \bibinfo{author}{\bibfnamefont{K.~M.} \bibnamefont{Smith}},
  \bibinfo{author}{\bibfnamefont{J.}~\bibnamefont{Dunkley}},
  \bibinfo{author}{\bibfnamefont{C.~L.} \bibnamefont{Bennett}},
  \bibinfo{author}{\bibfnamefont{B.}~\bibnamefont{Gold}},
  \bibinfo{author}{\bibfnamefont{G.}~\bibnamefont{Hinshaw}},
  \bibinfo{author}{\bibfnamefont{N.}~\bibnamefont{Jarosik}},
  \bibinfo{author}{\bibfnamefont{D.}~\bibnamefont{Larson}},
  \bibinfo{author}{\bibfnamefont{M.~R.} \bibnamefont{Nolta}},
  \bibinfo{author}{\bibfnamefont{L.}~\bibnamefont{Page}},
  \bibinfo{author}{\bibfnamefont{D.~N.} \bibnamefont{Spergel}},
  \bibinfo{author}{\bibfnamefont{M.}~\bibnamefont{Halpern}}, {}\emph{et~al.},
  \bibinfo{year}{2010}, \bibinfo{title}{``{Seven-Year Wilkinson Microwave
  Anisotropy Probe ({WMAP}) Observations: Cosmological Interpretation}''},
  \href{http://dx.doi.org/10.1088/0067-0049/192/2/18}{\bibinfo{journal}{The
  Astrophysical Journal Supplement Series}
  \textbf{\bibinfo{volume}{192}}(\bibinfo{number}{2})}, \bibinfo{pages}{57},
  \href{http://arxiv.org/abs/1001.4538v3}{{\ttfamily {}arXiv:1001.4538v3
  [astro-ph.CO]}}.

\bibitem[{\citenamefont{Laureijs}(2009)}]{Euclid2009}
\bibinfo{author}{\bibnamefont{Laureijs}, \bibfnamefont{R.}},
  \bibinfo{year}{2009}, {}\emph{\bibinfo{title}{{Euclid Assessment Study Report
  for the ESA Cosmic Visions}}}, \bibinfo{type}{Technical Report},
  \bibinfo{institution}{ESA},
  \href{http://arxiv.org/abs/0912.0914v1}{{\ttfamily {}arXiv:0912.0914v1
  [astro-ph.CO]}}.

\bibitem[{\citenamefont{Linder}(2003)}]{Linder2003}
\bibinfo{author}{\bibnamefont{Linder}, \bibfnamefont{E.}},
  \bibinfo{year}{2003}, \bibinfo{title}{``{Exploring the Expansion History of
  the Universe}''},
  \href{http://dx.doi.org/10.1103/PhysRevLett.90.091301}{\bibinfo{journal}{Phy%
sical Review Letters} \textbf{\bibinfo{volume}{90}}(\bibinfo{number}{9})},
  \bibinfo{pages}{091301},
  \href{http://arxiv.org/abs/astro-ph/0208512v1}{{\ttfamily
  {}arXiv:astro-ph/0208512v1}}.

\bibitem[{\citenamefont{Linder}(2005)}]{Linder2005}
\bibinfo{author}{\bibnamefont{Linder}, \bibfnamefont{E.}},
  \bibinfo{year}{2005}, \bibinfo{title}{``{Cosmic Growth History and Expansion
  History}''},
  \href{http://dx.doi.org/10.1103/PhysRevD.72.043529}{\bibinfo{journal}{Physic%
al Review D} \textbf{\bibinfo{volume}{72}}(\bibinfo{number}{4})},
  \bibinfo{pages}{043529},
  \href{http://arxiv.org/abs/astro-ph/0507263}{{\ttfamily
  {}arXiv:astro-ph/0507263}}.

\bibitem[{\citenamefont{Ma} {}\emph{et~al.}(2006)\citenamefont{Ma, Hu, and
  Huterer}}]{Ma2006}
\bibinfo{author}{\bibnamefont{Ma}, \bibfnamefont{Z.}},
  \bibinfo{author}{\bibfnamefont{W.}~\bibnamefont{Hu}}, and
  \bibinfo{author}{\bibfnamefont{D.}~\bibnamefont{Huterer}},
  \bibinfo{year}{2006}, \bibinfo{title}{``{Effects of Photometric Redshift
  Uncertainties on Weak-Lensing Tomography}''},
  \href{http://dx.doi.org/10.1086/497068}{\bibinfo{journal}{The Astrophysical
  Journal} \textbf{\bibinfo{volume}{636}}(\bibinfo{number}{1})},
  \bibinfo{pages}{21},
  \href{http://arxiv.org/abs/astro-ph/0506614v2}{{\ttfamily
  {}arXiv:astro-ph/0506614v2}}.

\bibitem[{\citenamefont{Mellier}(1998)}]{Mellier1998}
\bibinfo{author}{\bibnamefont{Mellier}, \bibfnamefont{Y.}},
  \bibinfo{year}{1998}, \bibinfo{title}{``{Probing the Universe with Weak
  Lensing}''},
  \href{http://dx.doi.org/10.1146/annurev.astro.37.1.127}{\bibinfo{journal}{An%
nual Review of Astronomy and Astrophysics} \textbf{\bibinfo{volume}{37}}},
  \bibinfo{pages}{54}, \href{http://arxiv.org/abs/astro-ph/9812172}{{\ttfamily
  {}arXiv:astro-ph/9812172}}.

\bibitem[{\citenamefont{Mukherjee}
  {}\emph{et~al.}(2008)\citenamefont{Mukherjee, Kunz, Parkinson, and
  Wang}}]{Mukherjee2008}
\bibinfo{author}{\bibnamefont{Mukherjee}, \bibfnamefont{P.}},
  \bibinfo{author}{\bibfnamefont{M.}~\bibnamefont{Kunz}},
  \bibinfo{author}{\bibfnamefont{D.}~\bibnamefont{Parkinson}}, and
  \bibinfo{author}{\bibfnamefont{Y.}~\bibnamefont{Wang}}, \bibinfo{year}{2008},
  \bibinfo{title}{``{Planck Priors for Dark Energy Surveys}''},
  \href{http://dx.doi.org/10.1103/PhysRevD.78.083529}{\bibinfo{journal}{Physic%
al Review D} \textbf{\bibinfo{volume}{78}}(\bibinfo{number}{8})},
  \bibinfo{pages}{083529}, \href{http://arxiv.org/abs/0803.1616v1}{{\ttfamily
  {}arXiv:0803.1616v1 [astro-ph]}}.

\bibitem[{\citenamefont{Overbye}(Jan 4, 2011)}]{Overbye2011}
\bibinfo{author}{\bibnamefont{Overbye}, \bibfnamefont{D.}}, \bibinfo{year}{Jan
  4, 2011}, \bibinfo{title}{``Quest for Dark Energy May Fade to Black''},
  \bibinfo{journal}{New York Times} (\bibinfo{number}{Retrieved from nyt.com}),
  \bibinfo{pages}{p. D1}.

\bibitem[{\citenamefont{Paczynski}(1996)}]{Paczynski1996}
\bibinfo{author}{\bibnamefont{Paczynski}, \bibfnamefont{B.}},
  \bibinfo{year}{1996}, \bibinfo{title}{``{Gravitational Microlensing in the
  Local Group}''},
  \href{http://dx.doi.org/10.1146/annurev.astro.34.1.419}{\bibinfo{journal}{An%
nual Review of Astronomy and Astrophysics}
  \textbf{\bibinfo{volume}{34}}(\bibinfo{number}{1})}, \bibinfo{pages}{419},
  \href{http://arxiv.org/abs/astro-ph/9604011v1}{{\ttfamily
  {}arXiv:astro-ph/9604011v1}}.

\bibitem[{\citenamefont{Peebles and Ratra}(2003)}]{Peebles2003}
\bibinfo{author}{\bibnamefont{Peebles}, \bibfnamefont{P.}}, and
  \bibinfo{author}{\bibfnamefont{B.}~\bibnamefont{Ratra}},
  \bibinfo{year}{2003}, \bibinfo{title}{``{The Cosmological Constant and Dark
  Energy}''},
  \href{http://dx.doi.org/10.1103/RevModPhys.75.559}{\bibinfo{journal}{Reviews
  of Modern Physics} \textbf{\bibinfo{volume}{75}}(\bibinfo{number}{2})},
  \bibinfo{pages}{559},
  \href{http://arxiv.org/abs/astro-ph/0207347v2}{{\ttfamily
  {}arXiv:astro-ph/0207347v2}}.

\bibitem[{\citenamefont{Peebles}(1980)}]{Peebles1980}
\bibinfo{author}{\bibnamefont{Peebles}, \bibfnamefont{P.~J.~E.}},
  \bibinfo{year}{1980},
  \href{http://www.amazon.com/Large-Scale-Structure-Universe-Phillip-Peebles/d%
p/0691082405}{{}\emph{\bibinfo{title}{{Large-Scale Structure of the
  Universe}}}} (\bibinfo{publisher}{Princeton University Press}), ISBN
  \bibinfo{isbn}{0691082405}.

\bibitem[{\citenamefont{Perlmutter}
  {}\emph{et~al.}(1998)\citenamefont{Perlmutter, Aldering, Goldhaber, Knop,
  Nugent, Castro, Deustua, Fabbro, Goobar, Groom, Hook, Kim}
  {}\emph{et~al.}}]{Perlmutter1998}
\bibinfo{author}{\bibnamefont{Perlmutter}, \bibfnamefont{S.}},
  \bibinfo{author}{\bibfnamefont{G.}~\bibnamefont{Aldering}},
  \bibinfo{author}{\bibfnamefont{G.}~\bibnamefont{Goldhaber}},
  \bibinfo{author}{\bibfnamefont{R.~A.} \bibnamefont{Knop}},
  \bibinfo{author}{\bibfnamefont{P.}~\bibnamefont{Nugent}},
  \bibinfo{author}{\bibfnamefont{P.~G.} \bibnamefont{Castro}},
  \bibinfo{author}{\bibfnamefont{S.}~\bibnamefont{Deustua}},
  \bibinfo{author}{\bibfnamefont{S.}~\bibnamefont{Fabbro}},
  \bibinfo{author}{\bibfnamefont{A.}~\bibnamefont{Goobar}},
  \bibinfo{author}{\bibfnamefont{D.~E.} \bibnamefont{Groom}},
  \bibinfo{author}{\bibfnamefont{I.~M.} \bibnamefont{Hook}},
  \bibinfo{author}{\bibfnamefont{A.~G.} \bibnamefont{Kim}}, {}\emph{et~al.},
  \bibinfo{year}{1998}, \bibinfo{title}{``{Measurements of {O}mega and {L}ambda
  from 42 High-Redshift Supernovae}''},
  \href{http://dx.doi.org/10.1086/307221}{\bibinfo{journal}{The Astrophysical
  Journal} \textbf{\bibinfo{volume}{517}}(\bibinfo{number}{2})},
  \bibinfo{pages}{21}, \href{http://arxiv.org/abs/astro-ph/9812133}{{\ttfamily
  {}arXiv:astro-ph/9812133}}.

\bibitem[{\citenamefont{Rapetti} {}\emph{et~al.}(2010)\citenamefont{Rapetti,
  Allen, Mantz, and Ebeling}}]{Rapetti2010}
\bibinfo{author}{\bibnamefont{Rapetti}, \bibfnamefont{D.}},
  \bibinfo{author}{\bibfnamefont{S.~W.} \bibnamefont{Allen}},
  \bibinfo{author}{\bibfnamefont{A.}~\bibnamefont{Mantz}}, and
  \bibinfo{author}{\bibfnamefont{H.}~\bibnamefont{Ebeling}},
  \bibinfo{year}{2010}, \bibinfo{title}{``{The Observed Growth of Massive
  Galaxy Clusters - III. Testing General Relativity on Cosmological Scales}''},
  \href{http://dx.doi.org/10.1111/j.1365-2966.2010.16799.x}{\bibinfo{journal}{%
Monthly Notices of the Royal Astronomical Society}
  \textbf{\bibinfo{volume}{406}}}, \bibinfo{pages}{1796},
  \href{http://arxiv.org/abs/0911.1787v2}{{\ttfamily {}arXiv:0911.1787v2
  [astro-ph.CO]}}.

\bibitem[{\citenamefont{Riess} {}\emph{et~al.}(1998)\citenamefont{Riess,
  Filippenko, Challis, Clocchiatti, Diercks, Garnavich, Gilliland, Hogan, Jha,
  Kirshner, Leibundgut, Phillips} {}\emph{et~al.}}]{Riess1998}
\bibinfo{author}{\bibnamefont{Riess}, \bibfnamefont{A.~G.}},
  \bibinfo{author}{\bibfnamefont{A.~V.} \bibnamefont{Filippenko}},
  \bibinfo{author}{\bibfnamefont{P.}~\bibnamefont{Challis}},
  \bibinfo{author}{\bibfnamefont{A.}~\bibnamefont{Clocchiatti}},
  \bibinfo{author}{\bibfnamefont{A.}~\bibnamefont{Diercks}},
  \bibinfo{author}{\bibfnamefont{P.~M.} \bibnamefont{Garnavich}},
  \bibinfo{author}{\bibfnamefont{R.~L.} \bibnamefont{Gilliland}},
  \bibinfo{author}{\bibfnamefont{C.~J.} \bibnamefont{Hogan}},
  \bibinfo{author}{\bibfnamefont{S.}~\bibnamefont{Jha}},
  \bibinfo{author}{\bibfnamefont{R.~P.} \bibnamefont{Kirshner}},
  \bibinfo{author}{\bibfnamefont{B.}~\bibnamefont{Leibundgut}},
  \bibinfo{author}{\bibfnamefont{M.~M.} \bibnamefont{Phillips}},
  {}\emph{et~al.}, \bibinfo{year}{1998}, \bibinfo{title}{``{Observational
  Evidence from Supernovae for an Accelerating Universe and a Cosmological
  Constant}''}, \href{http://dx.doi.org/10.1086/300499}{\bibinfo{journal}{The
  Astronomical Journal} \textbf{\bibinfo{volume}{116}}(\bibinfo{number}{3})},
  \bibinfo{pages}{1009},
  \href{http://arxiv.org/abs/astro-ph/9805201v1}{{\ttfamily
  {}arXiv:astro-ph/9805201v1}}.

\bibitem[{\citenamefont{Sandage}(1961)}]{Sandage1961}
\bibinfo{author}{\bibnamefont{Sandage}, \bibfnamefont{A.}},
  \bibinfo{year}{1961}, \bibinfo{title}{``{The Ability of the 200-INCH
  Telescope to Discriminate Between Selected World Models}''},
  \href{http://dx.doi.org/10.1086/147041}{\bibinfo{journal}{The Astrophysical
  Journal} \textbf{\bibinfo{volume}{133}}}, \bibinfo{pages}{355}.

\bibitem[{\citenamefont{Scranton} {}\emph{et~al.}(2003)\citenamefont{Scranton,
  Connolly, Nichol, Stebbins, Szapudi, Eisenstein, Afshordi, Budavari, Csabai,
  Frieman, Gunn, Johnson} {}\emph{et~al.}}]{Scranton2003}
\bibinfo{author}{\bibnamefont{Scranton}, \bibfnamefont{R.}},
  \bibinfo{author}{\bibfnamefont{A.~J.} \bibnamefont{Connolly}},
  \bibinfo{author}{\bibfnamefont{R.~C.} \bibnamefont{Nichol}},
  \bibinfo{author}{\bibfnamefont{A.}~\bibnamefont{Stebbins}},
  \bibinfo{author}{\bibfnamefont{I.}~\bibnamefont{Szapudi}},
  \bibinfo{author}{\bibfnamefont{D.~J.} \bibnamefont{Eisenstein}},
  \bibinfo{author}{\bibfnamefont{N.}~\bibnamefont{Afshordi}},
  \bibinfo{author}{\bibfnamefont{T.}~\bibnamefont{Budavari}},
  \bibinfo{author}{\bibfnamefont{I.}~\bibnamefont{Csabai}},
  \bibinfo{author}{\bibfnamefont{J.~A.} \bibnamefont{Frieman}},
  \bibinfo{author}{\bibfnamefont{J.~E.} \bibnamefont{Gunn}},
  \bibinfo{author}{\bibfnamefont{D.}~\bibnamefont{Johnson}}, {}\emph{et~al.},
  \bibinfo{year}{2003}, \bibinfo{title}{``{Physical Evidence for Dark
  Energy}''}, \href{http://arxiv.org/abs/astro-ph/0307335}{{\ttfamily
  {}arXiv:astro-ph/0307335}}.

\bibitem[{\citenamefont{Shao}(2003)}]{Shao2003}
\bibinfo{author}{\bibnamefont{Shao}, \bibfnamefont{J.}}, \bibinfo{year}{2003},
  \href{http://www.amazon.com/Mathematical-Statistics-Jun-Shao/dp/0387953825}{%
{}\emph{\bibinfo{title}{{Mathematical Statistics}}}}
  (\bibinfo{publisher}{Springer}), ISBN \bibinfo{isbn}{9780387953823}.

\bibitem[{\citenamefont{Smith} {}\emph{et~al.}(2003)\citenamefont{Smith,
  Peacock, Jenkins, White, Frenk, Pearce, Thomas, Efstathiou, and
  Couchman}}]{Smith2003}
\bibinfo{author}{\bibnamefont{Smith}, \bibfnamefont{R.~E.}},
  \bibinfo{author}{\bibfnamefont{J.~A.} \bibnamefont{Peacock}},
  \bibinfo{author}{\bibfnamefont{A.}~\bibnamefont{Jenkins}},
  \bibinfo{author}{\bibfnamefont{S.~D.~M.} \bibnamefont{White}},
  \bibinfo{author}{\bibfnamefont{C.~S.} \bibnamefont{Frenk}},
  \bibinfo{author}{\bibfnamefont{F.~R.} \bibnamefont{Pearce}},
  \bibinfo{author}{\bibfnamefont{P.~A.} \bibnamefont{Thomas}},
  \bibinfo{author}{\bibfnamefont{G.}~\bibnamefont{Efstathiou}}, and
  \bibinfo{author}{\bibfnamefont{H.~M.~P.} \bibnamefont{Couchman}},
  \bibinfo{year}{2003}, \bibinfo{title}{``{Stable Clustering, the Halo Model
  and Non-linear Cosmological Power Spectra}''},
  \href{http://dx.doi.org/10.1046/j.1365-8711.2003.06503.x}{\bibinfo{journal}{%
Monthly Notices of the Royal Astronomical Society}
  \textbf{\bibinfo{volume}{341}}(\bibinfo{number}{4})}, \bibinfo{pages}{1311},
  \href{http://arxiv.org/abs/astro-ph/0207664v2}{{\ttfamily
  {}arXiv:astro-ph/0207664v2}}.

\bibitem[{\citenamefont{Starobinsky}(2007)}]{Starobinsky2007}
\bibinfo{author}{\bibnamefont{Starobinsky}, \bibfnamefont{A.~A.}},
  \bibinfo{year}{2007}, \bibinfo{title}{``{Disappearing Cosmological Constant
  in f(R) Gravity}''},
  \href{http://dx.doi.org/10.1134/S0021364007150027}{\bibinfo{journal}{JETP
  Letters} \textbf{\bibinfo{volume}{86}}(\bibinfo{number}{3})},
  \bibinfo{pages}{157}, \href{http://arxiv.org/abs/0706.2041}{{\ttfamily
  {}arXiv:0706.2041}}.

\bibitem[{\citenamefont{Susskind}(2005)}]{Susskind2005}
\bibinfo{author}{\bibnamefont{Susskind}, \bibfnamefont{L.}},
  \bibinfo{year}{2005},
  \href{http://www.amazon.com/Cosmic-Landscape-String-Illusion-Intelligent/dp/%
0316155799}{{}\emph{\bibinfo{title}{{The Cosmic Landscape}}}}
  (\bibinfo{publisher}{Little, Brown and Company}), ISBN
  \bibinfo{isbn}{0316155799}.

\bibitem[{\citenamefont{Wang and Steinhardt}(1998)}]{Wang1998}
\bibinfo{author}{\bibnamefont{Wang}, \bibfnamefont{L.}}, and
  \bibinfo{author}{\bibfnamefont{P.~J.} \bibnamefont{Steinhardt}},
  \bibinfo{year}{1998}, \bibinfo{title}{``{Cluster Abundance Constraints for
  Cosmological Models with a Time-varying, Spatially Inhomogeneous Energy
  Component with Negative Pressure}''},
  \href{http://dx.doi.org/10.1086/306436}{\bibinfo{journal}{The Astrophysical
  Journal} \textbf{\bibinfo{volume}{508}}(\bibinfo{number}{2})},
  \bibinfo{pages}{483}.

\bibitem[{\citenamefont{Weinberg}(1989)}]{Weinberg1989}
\bibinfo{author}{\bibnamefont{Weinberg}, \bibfnamefont{S.}},
  \bibinfo{year}{1989}, \bibinfo{title}{``{The Cosmological Constant
  Problem}''},
  \href{http://dx.doi.org/10.1103/RevModPhys.61.1}{\bibinfo{journal}{Reviews of
  Modern Physics} \textbf{\bibinfo{volume}{61}}(\bibinfo{number}{1})},
  \bibinfo{pages}{1}.

\bibitem[{\citenamefont{Weinberg}(2008)}]{Weinberg2008}
\bibinfo{author}{\bibnamefont{Weinberg}, \bibfnamefont{S.}},
  \bibinfo{year}{2008},
  \href{http://www.amazon.com/Cosmology-Steven-Weinberg/dp/0198526822}{{}\emph%
{\bibinfo{title}{{Cosmology}}}} (\bibinfo{publisher}{Oxford University Press,
  USA}), ISBN \bibinfo{isbn}{0198526822}.

\bibitem[{\citenamefont{Zlatev} {}\emph{et~al.}(1999)\citenamefont{Zlatev,
  Wang, and Steinhardt}}]{Zlatev1999}
\bibinfo{author}{\bibnamefont{Zlatev}, \bibfnamefont{I.}},
  \bibinfo{author}{\bibfnamefont{L.}~\bibnamefont{Wang}}, and
  \bibinfo{author}{\bibfnamefont{P.}~\bibnamefont{Steinhardt}},
  \bibinfo{year}{1999}, \bibinfo{title}{``{Quintessence, Cosmic Coincidence,
  and the Cosmological Constant}''},
  \href{http://dx.doi.org/10.1103/PhysRevLett.82.896}{\bibinfo{journal}{Physic%
al Review Letters} \textbf{\bibinfo{volume}{82}}(\bibinfo{number}{5})},
  \bibinfo{pages}{896}, \href{http://arxiv.org/abs/astro-ph/9807002}{{\ttfamily
  {}arXiv:astro-ph/9807002}}.

\end{thebibliography}
